\newlength{\bibsep}
\documentclass[3p,10pt,letterpaper,nonatbib,times]{elsarticle}
\usepackage[utf8]{inputenc}
\DeclareUnicodeCharacter{0306}{\u}
\usepackage{amsfonts}
\usepackage{amssymb}
\usepackage{mathtools}
\usepackage{microtype}
\usepackage{booktabs}
\usepackage{bbm}
\usepackage{todonotes}
\usepackage{verbatim} 
\usepackage{multirow}
\usepackage[displaymath,mathlines,running]{lineno}
\usepackage{longtable}
\usepackage{arydshln}
\usepackage{subfig}
\usepackage{threeparttable}%
\usepackage{multirow}
\usepackage{newtxmath}
\usepackage{tcolorbox} 
\usepackage{xcolor} 
\usepackage{hyperref}
\hypersetup{colorlinks=true, linkcolor=blue, citecolor=blue, urlcolor=blue}
\usepackage{xurl}
\let\oldnl\nl
\newcommand{\nonl}{\renewcommand{\nl}{\let\nl\oldnl}}
\newcommand{\LB}{\left[\begin{MAT}(r){l}}
\newcommand{\RB}{\\ \end{MAT}\right]}
\newcommand{\dQbar}{d\bar{\mathbf{q}}}
\newcommand{\Mbar}{\bar{\mathbf{M}}}
\newcommand{\Abar}{\bar{\mathbf{A}}}

\newcommand{\dLambda}{d\boldsymbol{\lambda}}
\newcommand{\dX}{d\mathbf{x}}

\newcommand{\dP}{d\mathbf{p}}
\newcommand{\dC}{d\mathbf{c}}
\newcommand{\dY}{d\mathbf{y}}
\newcommand{\dZ}{d\mathbf{z}}
\newcommand{\dFF}{d\boldsymbol{\mathfrak{f}}}
\newcommand{\LCP}{LCP$(\dQbar,\Mbar)$}
\newcommand{\Lambdaf}{d\lambda_{\text{f}}^{pq}}
\newcommand{\Lambdas}{d\lambda^{pq}}

\newcommand{\Felastpq}{\mathbf{F}^{\text{elastic, }pq}}
\newcommand{\Dudefpq}{\delta\mathbf{u}^{\text{def,}\,pq}}
\newcommand{\Dxdefpq}{\delta\mathbf{x}^{\text{def,}\,pq}}
\newcommand{\Dthetadefpq}{\delta\boldsymbol{\theta}^{\text{def},\,pq}}

\newcommand{\Melastpq}{\mathbf{M}^{\text{elastic, }pq}}

\newcommand{\CircCirc}{\raisebox{0.7ex}{${\smash{\raisebox{-0.8ex}{$\scriptstyle\circ$}} \atop \smash{\scriptstyle\circ}}$}}
\newcommand{\BulletCirc}{\raisebox{0.7ex}{${\smash{\raisebox{-0.8ex}{$\scriptstyle\bullet$}} \atop \smash{\scriptstyle\circ}}$}}
\newcommand{\CircBullet}{\raisebox{0.7ex}{${\smash{\raisebox{-0.8ex}{$\scriptstyle\circ$}} \atop \smash{\scriptstyle\bullet}}$}}
\newcommand{\BulletBullet}{\raisebox{0.7ex}{${\smash{\raisebox{-0.8ex}{$\scriptstyle\bullet$}} \atop \smash{\scriptstyle\bullet}}$}}
\makeatletter
\def\ps@pprintTitle{%
  \let\@oddhead\@empty
  \let\@evenhead\@empty
}
\begin{document}
%
%
%
\begin{tcolorbox}[
    colback=gray!5, 
    colframe=gray!70, 
    arc=0mm, 
    width=\textwidth,
    boxrule=0.5pt,
    top=2mm, bottom=2mm, left=2mm, right=2mm
]
\small
\textbf{Please cite this article as:} \\
  Matthew R. Kuhn (2025),
  ``Quasi-static loading of granular media as a linear complementarity problem,''
  \emph{Computer Methods in Applied Mechanics and Engineering},
  Vol. 444,
  118117,
  https://doi.org/10.1016/j.cma.2025.118117.
\end{tcolorbox}
\vspace{1em} 
\begin{frontmatter}
\title{\Large Quasi-static loading of granular media as a
       linear complementarity problem}
\author[1]{Matthew R. Kuhn}%
%
%
%
\affiliation[1]{organization={Emeritus Professor,
                              Donald P. Shiley School of Engr.,
                              Univ. of Portland},
                city={Portland},
                state={Oregon},
                country={U.S.A.}}
\begin{abstract}
The discrete quasi-static
response of rate-independent dissipative
granular media is addressed.
Granular systems are conventionally simulated with methods that
are intrinsically dynamic, such as the discrete element (DEM)
and discontinuous deformation (DDA) methods,
with the particles' accelerations and damping being essential
aspects.
In contrast,
quasi-static methods derive from the static stiffness relationships among
the particles.
With frictional contacts, an assembly's stiffness is
incrementally non-linear and dependent on the direction of loading.
The paper resolves this difficulty by casting the
response as a linear complementarity problem (LCP).
The approach benefits from a foundation of past research
on existence, uniqueness, and stability of LCP solutions.
Results of this research are expounded in a concise set of rules.
The LCP of a granular system is derived,
accounting for geometric effects that arise from curvatures
of particles at their contacts,
frictional contact stiffnesses, and displacement constraints on the particles.
A Bott--Duffin generalized inverse
is used for resolving the displacement constraints.
The paper describes eight aberrant
conditions (pathologies) of granular systems,
in the forms of various bifurcations and instabilities.
When placed in the context of an LCP,
the eight conditions are unambiguously defined.
These pathologies include three types of bifurcation: discrete,
continuous-bounded (not yet reported in the literature),
and continuous-unbounded.
Another pathology, not yet reported in the literature, is
an abrupt discontinuous change in movements upon a continuous change
of the data.
Methods and results are illustrated with examples of granular systems.
Results show that instability and
bifurcation are pervasive conditions near
failure; that failure and softening are sensitive to the geometric
contours of the particles at contacts;
and that quasi-static systems can
encounter states that require a dynamic transition to resolve
a lack of paths consistent with further loading. 
\end{abstract}
\begin{keyword}
complementarity \sep granular material \sep bifurcation
\sep stability \sep friction
\end{keyword}
\end{frontmatter}
%
%
\section{\normalsize Introduction}
This paper addresses the quasi-static
response and failure of rate-independent frictional
granular media during loading,
with the granular system being viewed as an
assembly of discrete grains
that interact at their contacts.
The quasi-static regime applies in
many geotechnical and industrial situations
in which deformation and failure
originate from the micro-scale
interactions of particles that are sufficiently slow
that the grains' inertial
forces are much smaller than the forces imparted at contacts.
Although
the conduct of individual particles can derive from simple,
almost trivial, contact rules, their
collective interactions
are exceedingly complex, and when viewed at an intermediate, meso-scale,
the particles' movements appear as a slow tumult of
shifting, rolling, and rotation, advancing in coordinated but
bewildering sequences.
Remarkably, this 
apparent discordant dance results in
a repeatable and consistent progression of the
macro-scale deformation and stress,
to the extent that macro-scale behavior is often expressed
with relatively simple constitutive operators.
Presented herein is a particle-scale stiffness model
that simulates granular assemblies,
with the intent of understanding the
micro-to-macro interplay,
of identifying the underlying
micro-mechanisms of response,
of pin-pointing the origin of aberrant behaviors,
and of understanding the origins of failure.
\par
Current particle-scale methods
can be classified as either intrinsically
static or dynamic and using either explicit or implicit
integration of the motion equations.
The approach developed herein uses an implicit approach to maintain
static equilibrium during a sequence of loading increments.
This approach is \emph{most different} from
the discrete element method (DEM)~---
the most widely used method~--- which
uses explicit integration 
to track the particles' dynamics \cite{Cundall:1979a}.
Unlike implicit methods, stiffness in
DEM is local
to the individual contacts and is not part of
a multi-particle stiffness matrix,
and current particle accelerations are explicitly
integrated to update the particles'
velocities and positions.
Numeric stability with DEM requires damping of some form,
with damping coefficients determined through
trial-and-error or are perfunctorily assigned.
Because the method is intrinsically dynamic and requires
internal damping,
the quasi-static limit is approximated with DEM only
by reducing the loading rate, but
at considerable expense in computation time.
Yet even with prolonged loading, the
DEM relies upon a continual imbalance of
internal forces to propel particles to new positions,
and
after boundary movements are halted, the particles
continue to vibrate until they approximate equilibrium.
\par
Similar to DEM,
the discontinuous deformation analysis (DDA) method is a
dynamic method, but one that uses implicit integration with
an assembly's stiffness matrix to account for
particle interactions that extend beyond
the influence of a particle's immediate neighbors
\cite{Shi:1988a,Ke:1995a}.
Early implementations required
iteration within each time step to approximate
the implicit integration of the motion equations; but
more recently, the DDA equations are solved as a mixed
linear complementarity problem \cite{Zheng:2015a}.
As with DEM,
the method requires viscous damping \cite{Doolin:2004a},
and the quasi-static limit is approximated with slow
loading.
\par
The Contact dynamics (CD) method also uses implicit
integration,
but, unlike DDA,
assumes both rigid-particle and rigid-contact behaviors.
As such, CD
employs non-smooth unilateral constraints
(zero penetration) between rigid particles and the non-smooth
constraint of Coulomb friction at rigid-frictional contacts.
The implicit scheme uses a mass matrix to resolve the
velocity jumps that occur during time steps.
\par
Although particle-scale motions are
tracked with DEM, DDA, and CD methods,
they do not provide acute insight into the underlying,
meso-scale origin of a particle's motion, which
derives from the coordinated movements of the particle
and its neighborhood of particles.
The approach proposed herein, which allows such discernment,
is most similar to the implicit
granular element
method (GEM) of Kishino \cite{Kishino:1988a},
who employed a full stiffness matrix to maintain static
equilibrium throughout an assembly during quasi-static
loading (unlike DEM, DDA, and CD.
See also \cite{Bashir:1991a,Shodja:2003a}).
The primary difficulty of implicit static methods
is that frictional contacts
introduce incremental nonlinearities
in an assembly's stiffness, and in early implementations,
this difficulty was confronted by iterating within each
time step or by carrying forward any unresolved
force imbalance into the
next step.
The equilibrium equations have also been viewed as
the Karush--Kuhn--Tucker (KKT) optimal conditions of
a minimization problem, which is solved with
mathematical programming techniques,
such as the minimization of an energy functional
\cite{Holtzman:2010a},
solving a mixed complementarity problem \cite{Meng:2017a},
solving a cone programming problem \cite{Meng:2019a}, or
using conventional finite element (FEM) methods
referenced to the undeformed configuration
\cite{Krabbenhoft:2018a}.
\par
The quasi-static problem is posed herein
as a linear complementarity problem (LCP).
Rather than arriving at the LCP by way of the
KKT conditions of a minimization problem~--- %
conditions that are necessary but not
sufficient for a solution~--- %
we directly derive the LCP from the
equilibrium equations and contact stiffnesses.
In this manner, we can investigate issues
of uniqueness (or bifurcation),
of existence (or incongruity), of instability, and of other
aberrant conditions, as well as gaining insight
into the particle-scale origins of these issues.
\par
The plan of the paper is as follows.
In Section~\ref{section:model},
three equivalent but alternative formulations, including
the LCP, are
posed for the quasi-static response of a granular assembly.
Conditions on existence, uniqueness, and degeneracy
are reviewed in Section~\ref{sec:exist} and illustrated
with examples in Section~\ref{sec:LCPillus}.
The stiffness of a granular assembly
is derived in Section~\ref{sec:formulate1},
including geometric effects (Section~\ref{sec:GeomH}),
contact stiffness formulations (Section~\ref{sec:linfrict}),
and displacement constraints on the particles
(Section~\ref{sec:FinalForms1}).
Using these results,
the LCP is derived in Section~\ref{sec:formulate2}.
Section~\ref{sec:pathologies} describes eight aberrant
conditions (pathologies) that are available to
granular systems, and these are expounded
for the alternative formulations.
Examples of three granular systems are analyzed in
Section~\ref{sec:examples}, with attention given to their pathologies.
A summary of the derivations,
insights from the examples,
computational performance of the LCP examples,
and a comparison with DEM simulations
are discussed in
Section~\ref{sec:conclusions}.
Appendices provide algorithms for solving LCPs and
identifying pathologies.
\section{\normalsize Three Forms of the Quasi-static Problem}%
\label{section:model}
An assembly of $N$ three-dimensional particles is assumed
in equilibrium,
both in the reference state at
time $t$ and in the displaced
state at $t+dt$.
We also assume full knowledge of the assembly's configuration
at $t$, which includes the following:
the particles' positions and
orientations,
$\mathbf{u}$ and $\boldsymbol{\theta}$;
the applied external forces and moments on the particles,
$\mathbf{b}$ and $\mathbf{w}$; and the conditions of 
(i.e., the forces and the particles' shapes at) the particles' $M$
contacts.
Between times $t$ and $t+dt$,
changes in the external forces and boundary movements cause
the particles to undergo displacements and rotations,
in increments
$d\mathbf{u}$ and $d\boldsymbol{\theta}$.
These movements
are collected in two $3N\times 1$ vectors,
$[d\mathbf{u}]_{3N\times 1}$ and $[d\boldsymbol{\theta}]_{3N\times 1}$,
and for brevity, the two vectors are stacked into a $6N\times 1$ vector
$[d\mathbf{x}]_{6N\times 1}$.
The corresponding applied incremental external forces and moments on
the particles are collected in vectors
$[d\mathbf{b}]_{3N\times 1}$ and $[d\mathbf{w}]_{3N\times 1}$,
which are stacked into a vector
$[d\mathbf{p}]_{6N\times 1}$.
Details on the construction of these vectors
are given in Section~\ref{sec:formulate1}.
\par
The fundamental problem is finding the movements $[\dX]$ produced
by a given
loading $[\dP]$ and subject to displacement constraints,
but we also consider
whether a granular system can succumb to
aberrant behaviors, called ``stiffness pathologies'':
for example,
whether no solution exists (incongruity);
whether multiple solutions exist (bifurcation);
whether the system is 
stable and controllable;
whether the response is sensitive to the loading
or to the particles' arrangement, etc.
Such pathologies are considered in Section~\ref{sec:pathologies}.
%
%
\subsection{Alternative problems}\label{sec:forms}
The incremental problem is posed in three alternative,
equivalent forms, stated in the following paragraphs.
\subsubsection{Problem~1: conventional stiffness problem}%
\label{sec:Problem1}
The conventional structural approach of Problem~1 is
posed as a direct incremental stiffness relationship
between the particles' movements $[\dX]$ and the external loads $[\dP]$,
\begin{linenomath}
\begin{subequations}\label{eq:form1}
\begin{align}
  \text{Problem~1: }& \text{find } \left[d\mathbf{x}\right] \text{such that}
  \\
  & \left[\mathbf{H}\right]
    \left[d\mathbf{x}\right] =
    \left[d\mathbf{p}\right]
  \\
  &\left[\mathbf{C}\right]
    \left[d\mathbf{x}\right] =
    \left[d\mathbf{c}\right]
  \\
  &\left[\mathbf{H}\right] \in
    \left\{
      \left[\mathbf{H}^{i}\right]
      ,\;
      i = 1, 2, \ldots , B
    \right\}
  ,\quad
  \left[d\mathbf{x}\right]
    \in\Omega^{i} \subset \mathbb{R}^{6N}
  \\
  &\bigcup_{i=1}^{B}\Omega^{i}
    =
   \mathbb{R}^{6N}
\end{align}
\end{subequations}
\end{linenomath}
%
where $[\mathbf{H}]$ is the assembly's
$6N\times 6N$ stiffness matrix (Section~\ref{sec:assemstiff})
\cite{Kuhn:2019b}.
The stiffness relation of Eq.~(\ref{eq:form1}b) is
augmented with equality constraints on the particles' displacements,
given in Eq.~(\ref{eq:form1}c) with
an $R\times6N$ constraint matrix
$[\mathbf{C}]$ and imposed displacement increments $[d\mathbf{c}]$.
These constraints are examined in Section~\ref{sec:FinalForms1},
leading to an alternative form of Problem~1, given
in Eq.~(\ref{eq:form1b}).
%
\par
The incremental stiffness matrix $[\mathbf{H}]$
represents a single member of a set of
candidate matrices $[\mathbf{H}^{i}]$
with cardinality $B$,
in which the $i$th matrix applies to a particular cone of
movement directions, $\Omega^{i}$
(Eq.~\ref{eq:form1}d).
That is,
the stiffness relation (\ref{eq:form1}a) is incrementally nonlinear,
with $B$ stiffness branches, $[\mathbf{H}^{i}]$.
Because the material is assumed rate-independent
and operator $[\mathbf{H}]$ is positive-homogeneous of degree~1,
the $\Omega^{i}$ must be cones,
so that if $[\dX]$ is a solution of Eq.~(\ref{eq:form1}),
then $\gamma[\dX]$ is a solution of the associated problem, with
$[\dP]\rightarrow\gamma[\dP]$ and $[\dC]\rightarrow\gamma[\dC]$,
$\forall\gamma\in\mathbb{R}^{+}$.
The $B$ cones only overlap at their individual boundaries
$\partial\Omega^{i}$,
but together,
they cover the full range $\mathbb{R}^{6N}$ of movements
(Eq.~\ref{eq:form1}e).
The continuum analogy of the $\Omega^{i}$
are the tensorial zones of Darve's constitutive operators
\cite{Darve:1982b}.
\par
The multiple branches result from the nonlinear
nature of contact stiffness.
Examples include sliding/non-sliding movements at contacts that have reached the frictional limit
and sliding/non-sliding rotational movements at contacts that have
reached a rolling friction limit.
Only sliding friction is considered herein,
and Section~\ref{sec:linfrict} presents this dissipative mechanism
in relation to the three Problems. 
%
\par
The number of branches, $B$, can be quite large.
If the assembly's particles touch at $M$
contacts and a subset of contacts of size
$M_{\lambda}\le M$ are
\emph{active}, meaning that they have reached a limiting condition
(friction limit, etc.),
with each active contact admitting two stiffnesses
(e.g., sliding or non-sliding),
then $B=2^{M_{\lambda}}$ branches must be investigated to fully
determine whether solutions $[d\mathbf{x}]$ lie within the
particular cones $\Omega^{i}$ in which their stiffnesses
$[\mathbf{H}^{i}]$ apply.
In dense assemblies, only 10--20\% of the contacts are
active (sliding), with $M_{\lambda}\approx 0.1M \text{ to } 0.2M$,
so that the number $M_{\lambda}$ can be quite large.
Although a systematic enumerative procedure was proposed in
\cite{Kuhn:2019b},
with the purpose of finding stiffness pathologies during
the loading of an assembly,
for large assemblies,
searching through all $2^{M_{\lambda}}$ 
branches (i.e., all sliding/non-sliding combinations of
the $M_{\lambda}$ conditions)
is NP-hard, with computational effort growing
exponentially with $M_{\lambda}$.
\par
A further complication is that
each stiffness $[\mathbf{H}^{i}]$ is the sum of mechanical and
geometrical contributions.
The former is attributed to deformation of the particles at
their contacts;
the latter is due to incremental rotations of the contact
forces and to shifting of the assembly's internal geometry.
Such geometric effects are usually asymmetric,
making the full stiffness $[\mathbf{H}^{i}]$ inherently asymmetric
(Section~\ref{sec:GeomH}).
\par
The type of loading (i.e., the \emph{control program}) is embodied
in the $R$ rows of the kinematic constraint matrix $[\mathbf{C}]$
in Eq.~(\ref{eq:form1}c)
(more precisely, the row-space of $[\mathbf{C}]$).
The control program can range from full force-control
(an empty matrix $[\mathbf{C}]$)
to full displacement-control
(a $[\mathbf{C}]$ with $R=6N$ independent rows).
For a particular control program $[\mathbf{C}]$,
the loading
is specified by the \emph{loading parameters}, comprised of the
applied forces $[\dP]$ and the
controlled displacements $[\dC]$.
The displacement constraints of
Eq.~(\ref{eq:form1}c)
reduce the solution space of available movements $[\dX]$ to
a subset of $\mathbb{R}^{6N}$.
Such constraints are associated with walls, platens, the
isochoric restriction, etc., and these constraints
result in complementary reaction forces or pore fluid pressures,
which are added to the applied loads $[d\mathbf{p}]$:
that is,
each row in Eq.~(\ref{eq:form1}c) specifies
a displacement constraint, and each constraint results in
corresponding reaction forces that are added to
the applied loads $[\dP]$
(Section~\ref{sec:FinalForms1}). 
\par
The applied loads $[\dP]$ 
can be controlled by a smaller set of \emph{loading parameters}
$[d\mathbf{q}]$,
and the applied loads can also depend on the particles'
positions $[d\mathbf{x}]$,
so that $[d\mathbf{p}]$ is a function of both $[d\mathbf{q}]$
and $[d\mathbf{x}]$
(Section~\ref{sec:GeomH}).
An example of the former is the chamber pressure in a triaxial cell,
which controls the forces $d\mathbf{p}$ on particles
alongside an enclosing membrane.
Examples of the latter are follower forces,
forces produced by a boundary pressure, or the
self-gravitation of granular asteroids.
When present, the loading parameters $[d\mathbf{q}]$ are included among
the control parameters.
\par
The \emph{load path} is the combination of the control
program and the control parameters:
the combination of $[\mathbf{C}]$, $[\dC]$,
and $[\dP]$ (and/or $[d\mathbf{q}]$).
\subsubsection{Problem~2: linear complementarity problem (LCP)}%
\label{sec:Problem2}
In the field of solid mechanics,
posing a stiffness formulation as
an alternative linear complementarity problem (LCP)
originates with the work of Maier
\cite{Maier:1970a,Maier:1971a,Strang:1977a},
who formulated the discretized
field equations of elasto-plastic continua
as an LCP.
More recent applications to continua include
\cite{Feriani:1996a,Bassi:2012a}.
For a discrete granular system, 
Problem~\ref{eq:form2} is to find 
solutions $[\dLambda]$
associated with the $M_{\lambda}$ contacts that are active:
%
\begin{linenomath}
\begin{subequations}\label{eq:form2}
\begin{align}
  \text{Problem~2: }&\text{find }[\dLambda] \text{ such that}\\
  &\left[\dQbar\right] + \left[\Mbar\right]\left[\dLambda\right] \ge 0\\
  &\left[\dLambda\right] \ge 0\\
  &
   \left[\dLambda\right]^{\text{T}}
   \big(
     \left[\dQbar\right] + \left[\Mbar\right]\left[\dLambda\right]
   \big) = 0
\end{align}
\end{subequations}
\end{linenomath}
where vectors
$[\dQbar],[\dLambda]\in\mathbb{R}^{M_{\lambda}}$,
and matrix
$[\Mbar]\in\mathbb{R}^{M_{\lambda}\times M_{\lambda}}$
(note that $[\dQbar]$ is distinct from the
loading parameters $[d\mathbf{q}]$).
As will be seen,
matrix $[\Mbar]$ and vector $[\dQbar]$ are derived
from the data of Problem~1~---
the elements $[\mathbf{H}^{i}]$, $[d\mathbf{p}]$, $[\mathbf{C}]$,
and $[d\mathbf{c}]$ in  Eq.~(\ref{eq:form1})~---
and once the LCP is solved, the same elements,
along with solution $[\dLambda]$, are then
used to find the associated movements $[\dX]$
(Eq.~\ref{eq:dx23} below).
Matrix $[\Mbar]$ not only includes the stiffness
information of the many matrices $[\mathbf{H}^{i}]$
but also embeds information that distinguishes the individual cones
$\Omega^{i}$ to which the $[\mathbf{H}^{i}]$ apply.
Note that
in expressions~(\ref{eq:form2}b) and~(\ref{eq:form2}c),
the inequality ``$\ge 0$'' applies element-wise to vectors on the left;
and the complementarity condition~(\ref{eq:form2}d) implies that
the individual products
$d\lambda_{i}([\dQbar]+[\Mbar][\dLambda])_{i}=0$
for all $i$
(meaning that if $d\lambda_{i}>0$, then
$([\dQbar]+[\Mbar][\dLambda])_{i}=0$ and vice versa).
%
\par
Problem~2 is abbreviated as \LCP,
to emphasize the essential data,
$[\dQbar]$ and $[\Mbar]$.
The contents of both
$[\dQbar]$ and $[\Mbar]$,
which are derived in Section~\ref{sec:p12final},
depend upon the arrangements of particles,
on the control program $[\mathbf{C}]$,
and on the conditions of the contacts,
in particular, on the $M_{\lambda}$ active contacts
(e.g., contacts that lie tenuously at the friction limit).
Vector $[\dQbar]$ depends additionally upon the
system's incremental loading parameters,
$[\dP]$ and $[\dC]$.
\par
Solutions $[\dLambda]$ contain the incremental progress
of the active contacts
(e.g., their incremental sliding).
If the $i$th increment $d\lambda_{i}$ is greater than zero, then
its mechanism advances (slides, etc.);
whereas, if $d\lambda_{i}$ is zero,
the mechanism does not proceed.
On the other hand,
the $i$th value in Eq.~(\ref{eq:form2}b),
$([\dQbar] + [\Mbar][\dLambda])_{i}$, is positive when
the active mechanism withdraws (e.g., elastic unloading of a contact
on the verge of sliding),
akin to retreating into 
the interior of a yield surface.
The value is zero when the mechanism continues to engage,
similar to remaining on a yield surface.
The $[\dLambda]$ are solved for the \LCP,
and once found,
the incremental particle movements $[\dX]$
can be determined (Eq.~\ref{eq:dx23} below).
\par
Equation~(\ref{eq:form2}c) describes the non-negative orthant of
$\mathbb{R}^{M_{\lambda}}$,
bounded by the $M_{\lambda}$ hyperplanes
$\lambda_{1}=0, \lambda_{2}=0$, etc.
Equation~(\ref{eq:form2}b) reduces this region to a polygon
(possibly unbounded) having
$M_{\lambda}$ additional faces.
Assuming that the region is not diminished to the null set,
the complementarity condition of
Eq.~(\ref{eq:form2}d) further reduces
Problem~2 to finding the polygon's vertices.
%
If vertex solutions exist,
they are located among combinations of $M_{\lambda}$ of the
$2M_{\lambda}$ hyperplanes formed from
$d\lambda_{i}=0$ and $([\dQbar] + [\Mbar][\dLambda])_{j}=0$,
$i,j\in\{1,2,\ldots,M_{\lambda}\}$ with $i\neq j$.
Finding all solutions
(or establishing non-existence) involves a combinatorial
search among the $2^{M_{\lambda}}$ possible vertices,
and such full, enumerative strategies include those
of \cite{Murty:1972a}\S 2.10, \cite{Rohn:2009a},
and the algorithm in \ref{sec:enum}.
%
%
\par
Like Problem~1, a full solution of Problem~2 is NP-hard,
but Problem~\ref{eq:form2} has certain advantages:
\begin{itemize}
\item
Problem~\ref{eq:form2} is distilled to a more elemental
form of reduced dimension, placing
focus on the $M_{\lambda}$ active
movements (e.g., those of contacts that are sliding or are
in a condition of incipient sliding)
rather than on all $6N$ particle movements,
as in Problem~\ref{eq:form1}.
\item
Apart from a brute force, enumerative approach,
efficient methods have been developed for solving
LCPs, motivated by their importance in rigid-body dynamics,
computer game software, economic modeling, and other areas.
Problem~\ref{eq:form2} can also be recast as a quadratic program (QP)
(i.e., Problem~3), for which
other procedures are available.
Such efficient methods, however, are often limited
to special classes of $[\Mbar]$ matrices;
the methods usually result in a single solution,
even when multiple solutions exist;
and the methods are often limited to finding
non-degenerate solutions
(see \cite{Judice:1994a,Cottle:2009a,Erleben:2013a,Enzenhofer:2018b}
for reviews).
\item
A full solution of Problem~1 requires repeatedly solving
Eq.~(\ref{eq:form1}b) for each of the $2^{M_{\lambda}}$ instances
of $[\mathbf{H}^{i}]$, with each instance requiring the inversion
of a $6N\times 6N$ matrix.
Enumerative approaches for solving Problem~2, however, only require the
repeated inversion of smaller
$M_{\lambda}\times M_{\lambda}$ matrices (e.g. \cite{Rohn:2009a,Rohn:2012a}),
and once the $[\dLambda]$ solutions are found,
a $6N\times 6N$ matrix is then inverted
only once to disclose the corresponding vectors $[\dX]$.
\item
Much attention has been given to
questions of existence and
uniqueness for general LCPs, 
leading to guidance on matrix classes that can have
multiple solutions, no solutions at all,
or unstable solutions (Sections~\ref{sec:exist}
and~\ref{sec:sensitivity}).
Helpfully, when the matrix class of $[\Mbar]$ assures a single solution,
one can directly proceed to finding this solution with
efficient, non-enumerative methods.
\item
One pathology, proposed herein,
is similar to Lyapunov instability and can only
be determined with methods of Problem~2,
and the other pathologies
are most directly determined when posed as Problem~2
(Section~\ref{sec:pathologies}).
\end{itemize}
%
%
Note, however, that whereas the $6N\times6N$
matrices $[\mathbf{H}^{i}]$ are usually
block sparse, the $M_{\lambda}\times M_{\lambda}$
matrix $[\Mbar]$ is usually not,
due to its reduced size.
\subsubsection{Problem~3: quadratic program (QP)}%
\label{sec:Problem3}
A quadratic program (QP) is
associated with
the \LCP\ of Problem~\ref{eq:form2}, and
comprises the necessary KKT conditions of the LCP's solutions
\cite{Cottle:1964a,Murty:1972a,Cottle:2009a}:
\begin{linenomath}
\begin{subequations}\label{eq:form3}
\begin{alignat}{2}
\text{Problem~3: }
&\text{Minimize }
  &&z(\dLambda)=
   \left[\dLambda\right]^{\text{T}}
   \big(
     \left[\dQbar\right]
     + \left[\Mbar\right]\left[\dLambda\right]
   \big)\\
&\text{Subject to }\quad
    &&\left[\dQbar\right]
     + \left[\Mbar\right]\left[\dLambda\right] \ge 0\\
  &&&\left[\dLambda\right] \ge 0
\end{alignat}
\end{subequations}
\end{linenomath}
The objective function $z(\dLambda)$ is bounded below by
zero, so that $[\dLambda]$ is a solution of the
\LCP\ of Problem~\ref{eq:form2} if and only if
$z(\dLambda)$ is a global minima with value zero.
On the other hand, a positive global minimum means
that the \LCP\ has no solutions.
Methods for solving Problem~3 usually
face the same restrictions on $[\Mbar]$ mentioned with Problem~2.
\par
Note that a simpler quadratic program,
minimizing
$[\dLambda]^{\text{T}}[\dQbar]
+ \frac{1}{2}[\dLambda]^{\text{T}}[\Mbar][\dLambda]$
subject to $[\dLambda]\ge 0$, is only equivalent
to Problem~\ref{eq:form2} if matrix $[\Mbar]$ is symmetric.
This simpler QP is encountered in the mechanics literature, usually
inspired by its relation to energy functionals.
Because $[\Mbar]$ is usually asymmetric, the more restrictive
QP of Eq.~(\ref{eq:form3}) applies herein.
Also note that the QP for granular systems is often not convex,
due to an $[\Mbar]$ that is not positive semi-definite.
As shown in the examples of
Sections~\ref{sec:LCPillus} and~\ref{sec:examples}, the QP can even
be concave, with a negative definite $[\Mbar]$.
\subsection{Existence, number, and type of LCP solutions}%
\label{sec:exist}
Before attempting to solve a problem,
one should consider questions of existence
and uniqueness.
Although answers are the same for all three Problems,
the questions are most directly addressed
in the context of Problem~2, with which
existence and uniqueness involve criteria on $[\Mbar]$
and $[\dQbar]$.
But even though a number of
criteria have been developed
(e.g. \cite{Judice:1994a,Cottle:2010a}),
no comprehensive
(necessary and sufficient) analytic
means are yet available that can be applied
to arbitrary $[\Mbar]$ and $[\dQbar]$.%
\footnote{Murty \cite{Murty:1988a}\S 3.85 presents a finite
algorithm, attributed to D.~Gale,
that determines whether $[\Mbar]$ is a Q-matrix,
but the method is impractical for all but the smallest
problems.}
Even worse,
we see in later sections that the
matrices $[\Mbar]$ and $[\dQbar]$ of granular systems
often do not fit
known criteria that would allow an unequivocal
\emph{a priori} determination
of existence and uniqueness
(thus requiring an enumerative examination).
Here, we review known results that are most relevant to granular systems.
\par
The set $\mathcal{F}(\dQbar,\Mbar)$ 
of \emph{feasible solutions}
of an \LCP\ is
\begin{equation}\label{eq:feasible}
  \mathcal{F}\big(\dQbar,\Mbar\big)
  =
  \Big\{
  [\dLambda]:\:
  [\dLambda]\ge 0,\:
  [\dQbar] + [\Mbar][\dQbar]\ge 0
  \Big\}
\end{equation}
Set $\mathcal{F}$ is polyhedral in
$\mathbb{R}^{M_{\lambda}}$, and solutions of the
\LCP\ are located at boundary points of $\mathcal{F}$
\cite{Murty:1972a,Pardalos:1988b,Pardalos:1991a}.
An \LCP\ with
non-empty $\mathcal{F}(\dQbar,\Mbar)$ is termed \emph{feasible}.
\par
The \emph{solution set} $\mathcal{S}(\dQbar,\Mbar)$ are
the elements of $\mathcal{F}$ that also satisfy the complementarity condition,
\begin{equation}\label{eq:solution}
  \mathcal{S}\big(\dQbar,\Mbar\big)
  =
  \Big\{
  [\dLambda]:\:
  [\dLambda]\in \mathcal{F}(\dQbar,\Mbar),\:
  [\dQbar]^{\text{T}}\big([\dQbar] + [\Mbar][\dQbar]\big)= 0
  \Big\}
\end{equation}
and, when the set is finite, we denote the number of solutions
of the \LCP\ 
as cardinality $|\mathcal{S}(\dQbar,\Mbar)|$.
A solvable \LCP\ is one with a non-empty solution set:
$\mathcal{S}(\dQbar,\Mbar)\neq\varnothing$.
\par
As with rate-independent materials \cite{Darve:1990a},
the solutions $\mathcal{S}$ of the \LCP\ are
positive-homogeneous of degree~1 in the loading $[\dQbar]$:
\begin{equation}\label{eq:homo}
  [\dLambda] \in \mathcal{S}(\dQbar,\Mbar)
  \;\Rightarrow\;
  \gamma [\dLambda] \in \mathcal{S}(\gamma\dQbar,\Mbar),
  \;
  \forall
  \;
  \gamma \in \mathbb{R}^{+}
\end{equation}
which is readily verified with the criteria of Eq.~(\ref{eq:form2}).
\par
The class of \emph{Q-matrices}
are those $[\Mbar]$ for which the \LCP\ is
solvable (with possibly multiple solutions)
for all vectors $[\dQbar]$;
whereas, the class of \emph{Q\textsubscript{0}-matrices}
are matrices $[\Mbar]$ for which the \LCP\ is
solvable (perhaps with multiple solutions)
whenever the \LCP\ is feasible.
With non-Q\textsubscript{0}-matrices,
a feasible \LCP\ might have no solutions.
\par
\emph{Complementary matrices} will be useful in developing
the results of this section.
A complementary matrix
$[\bar{\mathbf{A}}]$ is formed from an index set
$J\subseteq I$ and its complement $J^{\,\text{C}}$,
where set $I=\{1,2,3,\ldots,M_{\lambda}\}$:
\begin{equation}\label{eq:comp}
  [\bar{\mathbf{A}}_{\bullet,J^{\text{C}}}]=
  [\mathbf{I}_{\bullet,J^{\text{C}}}] \text{ and }
  [\bar{\mathbf{A}}_{\bullet,J}]=[-\Mbar_{\bullet,J}],
  \text{ with }
  J\subseteq I,\; J^{\,\text{C}}= I\!\setminus\! J,\;
  I=\{1,2, \ldots, M_{\lambda}\}
\end{equation}
with matrix $[\Mbar]$ having $2^{M_{\lambda}}$ complementary matrices,
corresponding to the power set of $I$
\cite{Murty:1972a}.
Here, $[\mathbf{I}_{\bullet,J^{\text{C}}}]$
and $[-\bar{\mathbf{M}}_{\bullet,J}]$
are the sub-matrices
of square matrices $[\,\mathbf{I}\,]$ and $[-\Mbar]$,
comprising
all of the rows and the $J^{\text{C}}$ and $J$ columns
of
$[\mathbf{I}]$ and  $[-\bar{\mathbf{M}}]$, respectively.
Complementary matrices will be necessary in applying rule~20,
in distinguishing bifurcation paths, and in
implementing the algorithms of \ref{sec:enum}
and~\ref{sec:setting}.
\par
In regard to the feasibility, existence, and number of solutions,
the following
results are most relevant to granular systems and are,
henceforth, referred to as ``rules'':
\begin{enumerate}
\item
An \emph{S-matrix} is one for which a vector $[d\mathbf{u}]>0$
exists such that $[\Mbar][d\mathbf{u}]>0$.
An \LCP\ is feasible for all $[\dQbar]$
if and only if it is an S-matrix.
Every Q-matrix is an S-matrix (i.e., Q~$\Rightarrow$~S),
which implies that every row of a Q-matrix contains
at least one positive element.
From the definition
of class Q\textsubscript{0}, it follows that
$\text{S}\cap\text{Q}_{0}\Leftrightarrow\text{Q}$.
\item
If $\Mbar$ is positive definite (a \emph{PD-matrix})
then it is a Q-matrix.
Moreover, a PD matrix assures a unique solution of
the \LCP\ for all $[\dQbar]$, or
$\text{PD} \Rightarrow |\mathcal{S}(\dQbar,\Mbar)|=1,\:\forall\,[\dQbar]$.
\par
\rule{1em}{0ex}
Although positive definiteness is efficiently determined
and is emphasized in
the mechanics literature \cite{Maier:1971a,Franchi:1998a},
broader criteria for existence and uniqueness are available.
The $[\Mbar]$ matrices of granular systems are rarely
symmetric, and, as noted in the next rule, a non-symmetric $[\Mbar]$,
even if it is non-PD, can be a Q-matrix with an assured unique solution
of the LCP.
\item
\emph{P-matrices} are those for which all principal
minors are positive.
(A \emph{principal minor} of a matrix
is the determinant of the matrix or the determinant of a submatrix
that remains after removing 
a set of rows and the same set of columns.
A \emph{P\textsubscript{0}-matrix} is one with no non-negative
principal minors.)
P-matrices belong to the class of Q-matrices,
and the \LCP\ has a unique solution for all $[\dQbar]$ if and
only if $[\Mbar]$ is a P-matrix
\cite{Samelson:1958a,Murty:1972a}.
As such,
the class of P-matrices encompasses the smaller class of PD-matrices:
$\text{PD}\Rightarrow\text{P}$.
Moreover, if a matrix is
symmetric then it is PD if and only if it is a P-matrix,
although non-symmetric P-matrices are not
necessarily PD.
From rule~1,
a P-matrix is also an S-matrix: $\mathrm{P}\Rightarrow\mathrm{S}$.
\item
The class of \emph{R-matrices} are also Q-matrices,
but this class is broader than P-matrices.
R-matrices $[\Mbar]$ are those for which the following system
is inconsistent:
\begin{equation}\label{eq:R}
  \begin{aligned}
    &\left([\Mbar][\mathbf{u}]\right)_{i} + t = 0  \quad\text{if } u_{i}>0\\
    &\left([\Mbar][\mathbf{u}]\right)_{i} + t \ge 0 \quad\text{if } u_{i}=0\\
    &\,[\mathbf{u}]\ne0,\;[\mathbf{u}]\ge 0, \quad t\ge 0 
  \end{aligned}
\end{equation}
regardless of non-negative scalar $t$ \cite{Karamardian:1972a,Lee:1997c}.
That is, an R-matrix is one for which 
an $\text{LCP}(t,\Mbar)$ has no solution
$[\mathbf{u}]$
with a positive element $u_{i}$.
R-matrices are Q-matrices, and all P-matrices are R-matrices:
$P\Rightarrow R\Rightarrow Q$.
The related class of R\textsubscript{0}-matrices is described
after item~15 below.
\item
\emph{Z-matrices} $[\Mbar]$ are those with $\bar{M}_{ij}\le 0$
for all $i\neq j$.
Z-matrices belong to the class of Q\textsubscript{0}-matrices, or
$\text{Z} \Rightarrow \text{Q}_{0}$.
From rule~1, $\text{Z}\cap\text{S}\Rightarrow\text{Q}$.
\item
For a matrix $[\Mbar]$ with $\bar{M}_{ij}\ge 0$,
the matrix is a Q-matrix if and only if
$\bar{M}_{ii}> 0$ for all $i$
\cite{Murty:1972a,Cottle:2009a}.
\item
\emph{N-matrices} are those for which
all principal minors are negative
(an \emph{N\textsubscript{0}-matrix} has no non-positive principal
minors).
N-matrices in which not all elements are negative are 
N-matrices \emph{of the first category}, and these matrices belong to
the class Q, with the following cardinalities \cite{Gowda:1993a}:
\begin{equation}\label{eq:Nmatrix1}
 \text{N-matrix and }[\Mbar]\nless 0
 \:\Rightarrow\: |\mathcal{S}(\dQbar,\Mbar)| =
 \begin{cases}
   1 & \text{if }[\dQbar]\ngeq 0 \text{ or }[\dQbar]=0\\
   1\text{ or }2 & \text{if }[\dQbar]\ge 0,\,[\dQbar]\ngtr 0\\
   3 & \text{if }[\dQbar]> 0
 \end{cases}
\end{equation}
where the element-wise inequalities ``$\nless$'', ``$\ngtr$'', and ``$\ngeq$''
mean not all less than, etc.
Kajima and Saigal \cite{Kojima:1979a} and
Mohan and Sridhar \cite{Mohan:1992a} give conditions that distinguish
the cardinalities of 1 and 2 in the second case
of Eq.~(\ref{eq:Nmatrix1}),
and they show that affirmation of the
first and third cases Eq.~\ref{eq:Nmatrix1} is also \emph{sufficient} for
$[\Mbar]$ to be an N-matrix of the first category.
\end{enumerate}
Together, the above rules provide conditions that assure
(i.e., are sufficient for)
existence of solutions of the \LCP, with the following
hierarchies:
$\text{PD}\Rightarrow\text{P}\Rightarrow\text{R}\Rightarrow\text{Q}\Rightarrow\text{S}$,
$\text{Z}\cap\text{S}\Rightarrow\text{Q}$,
$\text{N of first category}\Rightarrow\text{Q}$, and
$\text{rule 6}\Rightarrow\text{Q}$.
Moreover, $[\Mbar]\in\text{P}$ is both necessary and sufficient for
the \LCP\ to have a unique solution for all $[\dQbar]$,
although non-P matrices might have a unique solution for
some $[\dQbar]$.
These results are augmented with the following:
\begin{enumerate}
\setcounter{enumi}{7}
\item
N-matrices in which all elements are negative are 
N-matrices \emph{of the second category}, and these matrices can
be solvable, with the following cardinalities \cite{Gowda:1993a}:
\begin{equation}\label{eq:SecondKind}
 \text{N-matrix and }[\Mbar]< 0
 \:\Rightarrow\: |\mathcal{S}(\dQbar,\Mbar)| =
 \begin{cases}
   0 & \text{if }[\dQbar]\ngeq 0\\
   1 & \text{if }[\dQbar]\ge 0,\,[\dQbar]\ngtr 0\\
   2 & \text{if }[\dQbar]> 0
 \end{cases}
\end{equation}
Affirmation of the third case in Eq.~(\ref{eq:SecondKind})
is sufficient for $[\Mbar]$ to be an N-matrix of the second kind
\cite{Parthasarathy:1990a}.
\item
A matrix $[\Mbar]$ is an \emph{almost N-matrix} if its
determinant is positive but all of its proper principal
minors are negative.
An almost N-matrix of the first category is an almost N-matrix
for which both $[\Mbar]$ and $[\Mbar]^{-1}$ have at least one
positive entry.
If $[\Mbar]$ is also of size 4 or greater, then it is a Q-matrix,
and if all of its elements are negative, then the \LCP\ has
exactly four solutions for every $[\dQbar]$ with
$[\Mbar]^{-1}[\dQbar]<0$, as in \cite{Olech:1991a}.
\end{enumerate}
The above
rules~1--9 apply to particular cases of $[\Mbar]$ and $[\dQbar]$,
but they are not sufficiently general for all cases in matters of
existence and number of solutions.
\par
The next results involve the properties of degeneracy and non-degeneracy.
A \emph{non-degenerate matrix} $[\Mbar]$
is one of which all principal minors
are nonzero; 
whereas, a \emph{degenerate matrix} is one that is not non-degenerate.
A solution $[\dLambda]$ of the \LCP\ is a \emph{degenerate solution} if
both
$d\bar{\lambda}_{i}=0$ and $([\dQbar]+[\Mbar][\dLambda])_{i}=0$
for at least one $i\in I=\{1, 2, \ldots, M_{\lambda}\}$.
A non-degenerate solution is one that is not degenerate.
That is, a \emph{non-degenerate solution} is a solution
for which $[\dQbar]+[\Mbar][\dLambda] + [\dLambda] > 0$.
Note that a non-degenerate matrix can have degenerate solutions.
\par
Degeneracy of a matrix plays a role in
an LCP similar to the singularity of a matrix 
in linear equations,
an equation $[\mathbf{A}][\mathbf{x}]=[\mathbf{b}]$,
such as those of Eq.~(\ref{eq:form1}b) in Problem~1.
Because of the similarity, one expects that
degenerate matrices expose
similar stiffness pathologies with Problem~2
as singularity exposes with Problem~1
(Section~\ref{sec:pathologies}).
Note that degeneracy is more general than singularity:
a singular matrix is degenerate, but not all degenerate matrices
are singular.
\begin{enumerate}
\setcounter{enumi}{9}
\item
The \LCP\ has a finite number (possibly zero) of solutions 
for all $[\dQbar]$ if and only if $[\Mbar]$
is non-degenerate \cite{Murty:1972a}.
Furthermore, if $[\Mbar]$ is non-degenerate,
then every solution is an \emph{isolated solution}
(i.e., locally unique solution).
An isolated solution $[\dLambda]$ is one that has
a neighborhood in $\mathbb{R}^{M_{\lambda}}$
that is devoid of other solutions \cite{Mangasarian:1980a}
(see rule~13).
\item
If $[\Mbar]$ is non-degenerate,
then the number of solutions of \LCP\ has
the same \emph{parity} for all $[\dQbar]$ for which there are no degenerate
solutions,
where parity means either an odd or an even number of solutions
\cite{Murty:1972a}.
As one consequence, if $[\Mbar]$ is non-degenerate, then
a $[\dQbar]$ exists such that there are
an odd number of solutions,
and if these solutions are non-degenerate,
then the matrix is a Q-matrix,
and the \LCP\ will have at least one solution for all $[\dQbar]$.
If the parity is even, then an aberrant condition is assured:
depending upon the input $[\dQbar]$, the \LCP\ will have either
no solutions or multiple solutions.
\par\quad
Broyden \cite{Broyden:1991a} describes a procedure for determining the
parity of non-degenerate matrices having \emph{degenerate} solutions.
For a given $[\Mbar]$ and $[\dQbar]$,
a sum $\omega$ is tallied as follows:
each non-degenerate solution adds 1 to $\omega$,
and each degenerate solution with an \emph{auxiliary matrix}
(described below) that has an odd parity
adds 1 to $\omega$.
The parity of $[\Mbar]$ is the parity of $\omega$.
\par\quad
A solution $[\dLambda^{\ast}]$
is characterized with
three disjoint subsets of the indices $I=\{1,2,\ldots,M^{\lambda}\}$,
with partition $I$:
\begin{equation}\label{eq:sens1}
  \begin{aligned}
  I_{1}^{\ast}&=
   \{i: d\lambda^{\ast}_{i}=0,\;([\dQbar]+[\Mbar][\dLambda^{\ast}])_{i}>0\}\\
  I_{2}^{\ast}&=
   \{i: d\lambda^{\ast}_{i}>0,\;([\dQbar]+[\Mbar][\dLambda^{\ast}])_{i}=0\}\\
  I_{3}^{\ast}&=
   \{i: d\lambda^{\ast}_{i}=0,\;([\dQbar]+[\Mbar][\dLambda^{\ast}])_{i}=0\}
  \end{aligned}
\end{equation}
such that $I_{1}^{\ast}\cup I_{2}^{\ast}\cup I_{3}^{\ast}=I$,
noting that the solution is degenerate when $I_{3}^{\ast}\neq\varnothing$.
(In the context, developed below,
of the set $I$ of active contacts within a granular
assembly,
set $I_{1}^{\ast}$ ceases sliding and elastically unloads;
set $I_{2}^{\ast}$ slides; and
set $I_{3}^{\ast}$ stalls, neither sliding nor unloading.)
By permuting rows and columns of $[\Mbar]$, the \LCP\ can be given
in the following form for the particular solution:
\begin{equation}\label{eq:Broyden}
  \begin{bmatrix}
    \Mbar_{I_{2}^{\ast}I_{2}^{\ast}} &
     \Mbar_{I_{2}^{\ast}I_{3}^{\ast}} &
     \Mbar_{I_{2}^{\ast}I_{1}^{\ast}} \\
    \Mbar_{I_{3}^{\ast}I_{2}^{\ast}} &
     \Mbar_{I_{3}^{\ast}I_{3}^{\ast}} &
     \Mbar_{I_{3}^{\ast}I_{1}^{\ast}} \\
   \Mbar_{I_{1}^{\ast}I_{2}^{\ast}} &
     \Mbar_{I_{1}^{\ast}I_{3}^{\ast}} &
     \Mbar_{I_{1}^{\ast}I_{1}^{\ast}}
  \end{bmatrix}
  \begin{bmatrix}
    \dLambda^{\ast}_{I_{2}^{\ast}} \\
    \mathbf{0}_{I_{3}^{\ast}} \\
    \mathbf{0}_{I_{1}^{\ast}}
  \end{bmatrix}
  +
  \begin{bmatrix}
    \dQbar_{I_{2}^{\ast}} \\
    \dQbar_{I_{3}^{\ast}} \\
    \dQbar_{I_{1}^{\ast}}
  \end{bmatrix}
  =
  \begin{bmatrix}
    \mathbf{0}_{I_{2}^{\ast}} \\
    \mathbf{0}_{I_{3}^{\ast}} \\
    \mathbf{w}_{I_{1}^{\ast}}
  \end{bmatrix}
  ,
  \quad
  [\dLambda^{\ast}_{I_{2}^{\ast}}]>0
  ,
  \;
  [\mathbf{w}_{I_{1}^{\ast}}]>0
\end{equation}
where subscripts give blocks of rows and columns taken
from $[\Mbar]$, $[\dQbar]$, and $[\dLambda^{\ast}]$,
and $[\mathbf{w}]$ is an auxiliary positive vector.
The auxiliary matrix for a degenerate solution 
$[\dLambda^{\ast}]$ is
\begin{equation}
  [\Mbar_{I_{3}^{\ast}I_{3}^{\ast}}]
  -
  [\Mbar_{I_{3}^{\ast}I_{2}^{\ast}}]
  [\Mbar_{I_{2}^{\ast}I_{2}^{\ast}}]^{-1}
  [\Mbar_{I_{2}^{\ast}I_{3}^{\ast}}]
\end{equation}
and the parity of the degenerate solution is the parity
of this auxiliary matrix \cite{Broyden:1991a}.
\item
If $[\Mbar]$ is degenerate,
it is possible to find a vector $[\dQbar]$
for which the  \LCP\ has two distinct solutions,
$[\dLambda_{1}]$ and $[\dLambda_{2}]$, such that any
convex combination of the two vectors is also a solution of the LCP
\cite{Broyden:1991a}.
The solutions form a non-isolated convex
(and possibly unbounded) set in $\mathbb{R}^{M_{\lambda}}$.
For granular systems,
non-isolated solutions result in bifurcations of Types~II and~III,
described in Section~\ref{sec:bif}.
When $[\Mbar]$ is also non-R\textsubscript{0},
the solution set can be both non-isolated and unbounded
(rules~16--19).
Non-degenerate matrices $[\Mbar]$ also
give rise to the stiffness pathology
of path-sensitivity
(Section~\ref{sec:sensitivity}).
\par\quad
These statements for a degenerate $[\Mbar]$ in an LCP
are similar those for a singular matrix $[\mathbf{A}]$ in an equation
$[\mathbf{A}][\mathbf{x}]=[\mathbf{b}]$, which for
certain $[\mathbf{b}]$ can have
non-unique and
non-isolated solutions
and the existence of an open and convex linear solution-space.
\item
The presence of non-isolated solutions results from the
degeneracy of $[\Mbar]$, such that
a non-degenerate solution $[\dLambda^{\ast}]\neq 0$
is isolated if and only if
the matrix $[\Mbar_{I_{2}^{\ast}I_{2}^{\ast}}]$ 
in Eq.~(\ref{eq:Broyden}) is non-singular
\cite{Jansen:1987a}.
\item
When non-isolated solutions exist,
they occur as continuous convex sets $P\subset\mathbb{R}^{M^{\lambda}}$
(line segments, rays, polygonal faces, etc.) within
$d\boldsymbol{\lambda}$-space,
with $P$ possibly being unbounded.
Set $P$ is termed a \emph{convex component} \cite{Jansen:1987a},
such that no convex subset of $\mathcal{S}(\dQbar,\Mbar)$
contains $P$.
The dimension of and the extreme-points of sets $P$
(end-points of line segments,
vertices of polygonal faces, etc.) obey certain rules.
A convex solution set $P$ is characterized by
the index set $C(P)$ that is the union of the
$I_{2}^{\ast}$ indices among solutions $P$,
and by the index set $E(P)$ that is the intersection
of the unions $I_{2}^{\ast}\cup I_{3}^{\ast}$ among
solutions $P$.
The dimension of a non-isolated solution set $P$ is
\cite{Jansen:1987a}
\begin{equation}\label{eq:dimnoiso}
  \text{dim}(P)=|C(P)| - \text{rank}\left([\Mbar_{E(P),C(P)}]\right)
\end{equation}
\item
The extreme-points of a non-isolated solution set $P$
are solutions $[\dLambda^{\ast}]$ such that
$[\Mbar_{I_{2}^{\ast},I_{2}^{\ast}}]$ is non-singular and
$[\dLambda^{\ast}_{I_{2}^{\ast}}]=
-[\Mbar_{I_{2}^{\ast},I_{2}^{\ast}}]^{-1}[\dQbar_{I_{2}^{\ast}}]$,
as in \cite{Jansen:1987a}.
\end{enumerate}
\par
Certain pathologies in
Section~\ref{sec:PI1}
are related to the matrix class R\textsubscript{0}.
An \emph{R\textsubscript{0}-matrix}
(also, pseudo-regular matrix or E$^{\ast}(0)$-matrix \cite{Garcia:1973a})
is one for which the
LCP$(\mathbf{0},\Mbar)$, with $[\dQbar]=[\mathbf{0}]$,
has the unique solution $[\dLambda]=[\mathbf{0}]$, or
$\mathcal{S}(\mathbf{0},\Mbar)=\{\mathbf{0}\}$.
%
\emph{Non-R\textsubscript{0} matrices} $[\Mbar]$ have non-zero
solutions of the LCP$(\mathbf{0},\Mbar)$,
such that $\mathcal{S}(\mathbf{0},\Mbar)\neq[\mathbf{0}]$.
In the context of Problem~2, neutral equilibrium results from
non-R\textsubscript{0} matrices $[\Mbar]$;
whereas, with Problem~1, neutral equilibrium
arises when a matrix $[\mathbf{H}^{i}]$ is singular
and the null-space has a non-null intersection with
$\Omega^{i}$ (Section~\ref{sec:NE}).
\par
Because the solution set $\mathcal{S}(\dQbar,\Mbar)$
of an \LCP\ is a positive-homogeneous degree-1
multi-valued function of the loading $[\dQbar]$,
a non-R\textsubscript{0} matrix $[\Mbar]$
has an unbounded solution set
for the case of zero loading,
$[\dQbar]=[\mathbf{0}]$,
since any non-zero solution $[\dLambda]$ of the LCP$(\mathbf{0},\Mbar)$
also has
the unbounded solutions $\gamma[\dLambda]$,
for all $\gamma\in\mathbb{R}^{+}$
(Eq.~\ref{eq:homo}).
\begin{enumerate}
\setcounter{enumi}{15}
\item
The solution set $\mathcal{S}(\dQbar,\Mbar)$ is bounded
for all $[\dQbar]$ if and only if $[\Mbar]$
is R\textsubscript{0} \cite{Jansen:1987a}.
That is, unbounded solutions can only occur (and will occur) with
non-R\textsubscript{0} matrices, a result that impacts the
availability of certain bifurcations
(Section~\ref{sec:bif}).
\end{enumerate}
%
%
%
As a final consequence,
a non-R\textsubscript{0} matrix $[\Mbar]$ is
non-R and, therefore, non-Q, so that solutions will not exist for
some $[\dQbar]$.
These consequences of non-R\textsubscript{0} matrices
in an LCP resemble
those of a singular matrix $[\mathbf{A}]$
in a linear equation
$[\mathbf{A}][\mathbf{x}]=[\mathbf{b}]$:
unbounded solutions for certain $[\mathbf{b}]$
and the lack of solutions for others.
\par
Although the definition of class R\textsubscript{0} is
akin to that of non-singular matrices,
determining R\textsubscript{0} status is more complex,
since a singular $[\Mbar]$ can be either
R\textsubscript{0} or non-R\textsubscript{0},
a non-R\textsubscript{0} matrix can be non-singular, and
a degenerate matrix can be either
R\textsubscript{0} or non-R\textsubscript{0}.
The following conditions apply to
R\textsubscript{0} and non-R\textsubscript{0} status:
\begin{enumerate}
\setcounter{enumi}{16}
\item
As a sufficient (but not necessary) condition,
all R-matrices are also R\textsubscript{0}-matrices:
$\text{R}\Rightarrow\text{R}_{0}$ \cite{Lee:1997c}.
\item
As a sufficient (but not necessary) condition,
non-degenerate matrices belong to R\textsubscript{0}
\cite{Dutta:2022a}.
\item
As a sufficient (but not necessary) condition,
matrices that are both non-singular and
\emph{column competent} belong to R\textsubscript{0}
\cite{Dutta:2022a}.
A matrix $[\Mbar]$ is column competent if the products
$d\lambda_{i}([\Mbar][\dLambda])_{i}=0$ for all
$i\in\{1,2,\ldots,M_{\lambda}\}$ implies
that $[\Mbar][\dLambda]=[\mathbf{0}]$.
(Note that elements of
$[\dLambda ]$ can be negative in this definition.)
Xu \cite{Xu:1999a} has shown that a matrix is
column competent if and only if for all principal minors,
the determinants $|\Mbar_{J,J}|$,
that are zero (thus rendering the matrix degenerate),
the set $J$ of columns $\Mbar_{\bullet,J}$ are linearly dependent,
where index set $J$ is a subset of $I$: $J\subseteq I$.
\item
Mohan \cite{Mohan:1993a} obtained the following necessary and sufficient
condition:
a matrix $[\Mbar]$
is non-R\textsubscript{0} if and only if
any of its complementary matrices
is \emph{strongly degenerate}.
A complementary matrix $[\bar{\mathbf{A}}]$ (see Eq.~\ref{eq:comp})
is strongly degenerate if
there exists a $[\dLambda]\ge 0$ with $[\dLambda]\neq [\mathbf{0}]$
and $[\bar{\mathbf{A}}][\dLambda]=[\mathbf{0}]$.
\end{enumerate}
Applying each of rules~17--20 has a difficulty of order
$2^{M_{\lambda}}$.
\subsection{Illustration of LCP principles}\label{sec:LCPillus}
\begin{figure}
  \centering
  \includegraphics{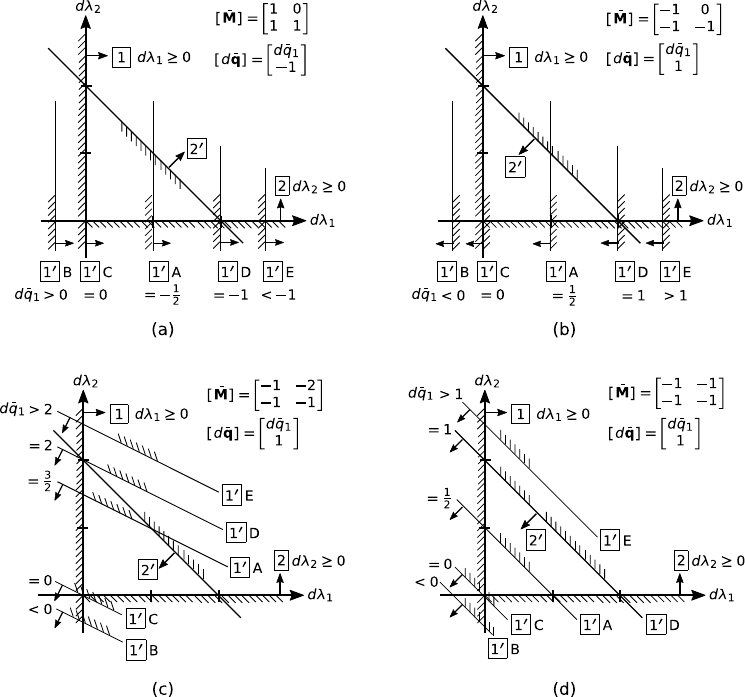}
  \caption{Examples of four LCPs.
           Results are summarized in Table~\ref{table:ExampleLCPs}.
           The half-planes \fbox{1} and \fbox{2} correspond to
           $d\lambda_{1}\ge 0$ and $d\lambda_{2}\ge 0$. 
           The regions \fbox{1$^{\prime}$} and \fbox{2$^{\prime}$}
           are $\Mbar_{1,\bullet}\cdot\dLambda+d\bar{q}_{1}\ge0$
           and $\Mbar_{2,\bullet}\cdot\dLambda+d\bar{q}_{2}\ge0$.
           \label{fig:ExampleLCPs}}
\end{figure}

Figure~\ref{fig:ExampleLCPs} and Table~\ref{table:ExampleLCPs} illustrate
these principles. 
\begin{table}
  \caption{Summary of the two-dimensional LCP in Fig.~\ref{fig:ExampleLCPs}.
           \label{table:ExampleLCPs}}
  \centering
  \small
  \begin{threeparttable}
  \begin{tabular}{cccccr@{, $d\bar{q}_{1}\:$}lcccc}
    \toprule
         & \multicolumn{4}{c}{$[\Mbar]$} & \multicolumn{2}{c}{} & 
      \multicolumn{2}{c}{Solutions, $|\mathcal{S}(\dQbar,\Mbar)|$} & & \\
    \cmidrule(r){2-5}
    \cmidrule(r){8-9}
    Fig. & Deg.\tnote{1} & P\tnote{1} & N\tnote{1} & 
      Parity & \multicolumn{2}{c}{Case, $d\bar{q}_{1}$} & 
      Non-deg.\tnote{2} & Deg.\tnote{2} &
      Non-isol.\tnote{3} & 
      Sens.\tnote{3}\\
    \midrule
    \ref{fig:ExampleLCPs}a & No & Yes & No & Odd & 
       A & $=-0.5$ & 1 & 0 &  0 & 0 \\
       \multicolumn{5}{c}{\multirow{4}{*}
                            {
                            $\displaystyle
                             [\Mbar]=
                             \begin{bmatrix} 1 & 0\\ 1 & 1\end{bmatrix}
                             \text{, }
                             [\dQbar]=
                             \begin{bmatrix} d\bar{q}_{1}\\ -1\end{bmatrix}
                            $
                            }
                           } 
                            & B & $>0$ & 1 & 0 & 0 & 0 \\
       \multicolumn{5}{c}{} & C & $=0$ & 0 & 1\tnote{o} & 0 & 0  \\
       \multicolumn{5}{c}{} & D & $=-1$ & 0 & 1\tnote{o} & 0 & 0  \\
       \multicolumn{5}{c}{} & E & $<-1$ & 1 & 0 & 0 & 0 \\
    \midrule
    \ref{fig:ExampleLCPs}b & No & No & No & Even & 
       A & $=0.5$ & 
       4 & 0 & 0 & 0 \\
       \multicolumn{5}{c}{\multirow{4}{*}
                            {
                            $\displaystyle
                             [\Mbar]=
                             \begin{bmatrix} -1 & 0\\ -1 & -1\end{bmatrix}
                             \text{, }
                             [\dQbar]=
                             \begin{bmatrix} d\bar{q}_{1}\\ 1\end{bmatrix}
                            $
                            }
                           } 
                            & B & $<0$ & 0 & 0 & -- & -- \\
       \multicolumn{5}{c}{} & C & $=0$ & 0 & 2\tnote{e} & 0 & 2  \\
       \multicolumn{5}{c}{} & D & $=1$ & 2 & 1\tnote{e} & 0 & 1  \\
       \multicolumn{5}{c}{} & E & $>1$ & 2 & 0 & 0 & 0 \\
    \midrule
    \ref{fig:ExampleLCPs}c & No & No & Yes\tnote{$\ast$} & Even & 
       A & $=1.5$ & 2 & 0 & 0 & 0 \\
       \multicolumn{5}{c}{\multirow{4}{*}
                            {
                            $\displaystyle
                             [\Mbar]=
                             \begin{bmatrix} -1 & -2\\ -1 & -1\end{bmatrix}
                             \text{, }
                             [\dQbar]=
                             \begin{bmatrix} d\bar{q}_{1}\\ 1\end{bmatrix}
                            $
                            }
                           } 
                            & B & $<0$ & 0 & 0 & -- & -- \\
       \multicolumn{5}{c}{} & C & $=0$ & 0 & 1\tnote{e} & 0 & 1  \\
       \multicolumn{5}{c}{} & D & $=2$ & 1 & 1\tnote{o} & 0 & 0  \\
       \multicolumn{5}{c}{} & E & $>2$ & 2 & 0 & 0 & 0 \\
    \midrule
    \ref{fig:ExampleLCPs}d & Yes & No & No & --\tnote{\dag} & 
       A & $=0.5$ & 2 & 0 & 0 & 0 \\
       \multicolumn{5}{c}{\multirow{4}{*}
                            {
                            $\displaystyle
                             [\Mbar]=
                             \begin{bmatrix} -1 & -1\\ -1 & -1\end{bmatrix}
                             \text{, }
                             [\dQbar]=
                             \begin{bmatrix} d\bar{q}_{1}\\ 1\end{bmatrix}
                            $
                            }
                           } 
                            & B & $<0$ & 0 & 0 & -- & -- \\
       \multicolumn{5}{c}{} & C & $=0$ & 0 & 1 & 0 & 1  \\
       \multicolumn{5}{c}{} & D & $=1$ & 1 & 2\tnote{\ddag} & 
         1\tnote{\#} & 1\tnote{\#}  \\
       \multicolumn{5}{c}{} & E & $>1$ & 2 & 0 & 0 & 0 \\
   \bottomrule
  \end{tabular}
  \begin{tablenotes}
    \item[1] Whether $[\Mbar]$ is degenerate (Deg.),
             a P-matrix (P), and/or an N-matrix (N).
    \item[2] Numbers of non-degenerate (Non-deg.) 
             and degenerate (Deg.) solutions.
             The total number of solutions is the sum
             of these two numbers.
    \item[3] Numbers of non-isolated (Non-isol.)
    and path-sensitive (Sens.) solutions.
    \item[$\ast$] N-matrix of the second kind.
    \item[e] Even-valued degenerate solutions.
    \item[o] Odd-valued degenerate solutions.
    \item[\dag] Parity not applicable to non-degenerate matrices
      $[\Mbar]$.
    \item[\ddag] The solutions (0,1) and (1,0) are degenerate.
    \item[\#] Points on the closed line segment from (0,1) to (1,0) are
              non-isolated.  Points on the open line segment
              are path-sensitive.
  \end{tablenotes}
  \end{threeparttable}
\end{table}
The example is a two-dimensional problem with unknowns
$d\lambda_{1}$ and $d\lambda_{2}$.
For granular systems,
matrix $[\Mbar]$ derives from the arrangements of
the particles and their current contact forces;
whereas, the vector $[\dQbar]$ depends upon the direction
of the incremental loading
(Sections~\ref{sec:formulate1} and~\ref{sec:formulate2}).
This two-dimensional example
is the analog of a two-contact granular system presented
in Section~\ref{sec:three}.
\par
Four matrices $[\Mbar]$ are considered, each with five vectors
$[\dQbar]$ that differ in the single element $d\bar{q}_{1}$.
The results in Table~\ref{table:ExampleLCPs} include
whether $[\Mbar]$ is degenerate, a P-matrix or an N-matrix, and
its parity.
For each $[\Mbar]$ and each of
five values of $d\bar{q}_{1}$, cases~A through E,
the table also gives the number of solutions
(both non-degenerate and degenerate), and the number of these
solutions that are not isolated and that exhibit path sensitivity.
The latter is a type of stiffness pathology
described in Section~\ref{sec:sensitivity}.
\par
The four $[\Mbar]$ include a P-matrix (\ref{fig:ExampleLCPs}a),
an N-matrix (\ref{fig:ExampleLCPs}c),
a matrix that is neither a P- nor M-matrix (\ref{fig:ExampleLCPs}b),
and a matrix that is degenerate (\ref{fig:ExampleLCPs}d).
Although all four matrices are in class R\textsubscript{0},
only the first matrix is an R-matrix.
For each $[\Mbar]$ and $[\dQbar]$,
the feasible region $\mathcal{F}(\dQbar,\Mbar)$
is formed from four half-spaces:
$d\lambda_{i}\ge 0$ and $([\dQbar]+[\Mbar][\dLambda])_{j}\ge 0$,
with $i,j\in\{1,2\}$. 
Solutions are possible at the $2^2=4$ intersections
of these half-spaces,
lying at the vertices of $\mathcal{F}$:
the intersections,
$d\lambda_{1}=0$ and $d\lambda_{2}=0$;
$d\lambda_{1}=0$ and $([\dQbar]+[\Mbar][\dLambda])_{2}=0$;
$([\dQbar]+[\Mbar][\dLambda])_{1}=0$ and $d\lambda_{2}=0$; and
$([\dQbar]+[\Mbar][\dLambda])_{1}=0$ and $([\dQbar]+[\Mbar][\dLambda])_{2}=0$.
\par
In three of the twenty instances,
the feasible space $\mathcal{F}$ is null, because of
inconsistent half-spaces (cases~B of~\ref{fig:ExampleLCPs}b, 
\ref{fig:ExampleLCPs}c, and~\ref{fig:ExampleLCPs}d).
The possibility of a null $\mathcal{F}$ is exposed by
the matrix $[\Mbar]$ not being an S-matrix (rule~1).
However, the three $[\Mbar]$ matrices are also Z-matrices,
so they are also Q\textsubscript{0}-matrices, and one more
solution must exist whenever $[\dQbar]$ renders the \LCP\ feasible
(rule~5).
\par
Among the twenty instances,
the numbers of solutions range from 0 to 4,
both isolated and non-isolated:
\begin{itemize}
\item
With Fig.~\ref{fig:ExampleLCPs}a,
$[\Mbar]$ is both a P-matrix and PD
(and, in either case, is a Q-matrix),
and a single solution is found for
all $[\dQbar]$.
Note that the $[\Mbar]$ is also deemed a Q-matrix,
because it meets the conditions of rule~6.
In two cases, 
the solution is degenerate, with each solution having
odd parity (rule~11).
\item
In Fig.~\ref{fig:ExampleLCPs}b,
matrix $[\Mbar]$ is non-degenerate but is neither
a P-matrix nor an N-matrix
(although it is an almost N-matrix, rule~9).
The matrix has even parity, and the five cases include instances with
0, 2, 3, and 4 solutions (for the case having 3 solutions, the single
non-degenerate solution has even parity, yielding a total
even parity, rule~11).
\item
With Fig.~\ref{fig:ExampleLCPs}c,
$[\Mbar]$ is an N-matrix of the second kind,
and the cases with 0, 1, and 2 solutions are in
agreement with rule~8. 
\item
With the three Figs.~\ref{fig:ExampleLCPs}a,
\ref{fig:ExampleLCPs}b, and \ref{fig:ExampleLCPs}c
$[\Mbar]$ is non-degenerate,
so that each solution (when a solution exists) is isolated 
for all $[\dQbar]$ (rule~10).
Because $[\Mbar]$ is non-degenerate,
it belongs to class R\textsubscript{0} (rule~18).
\item
Case~D of Fig.~\ref{fig:ExampleLCPs}d
is, perhaps, the most interesting.
Because the matrix is degenerate and 
the two half-spaces, $([\dQbar]+[\Mbar][\dLambda])_{1}\ge 0$
and $([\dQbar]+[\Mbar][\dLambda])_{2}\ge 0$, coincide,
a continuous set of non-isolated solutions exists
as a line segment in $\dLambda$-space:
the closed segment between the extreme points, $(0,1)$ and $(1,0)$.
This result is an example of rules~12--14
(the line segment has dimension 1,
which is the difference of $|C(P)|=|\{1,2\}|=2$ and
$\mathtt{rank}([\Mbar])=1$),
and the result
is a counter-example
to rule~10 for non-degenerate matrices.
The two extreme points of the line segment follow from
rule~15.
Note that the existence of the non-isolated solutions is not
because $[\Mbar]$ is singular, but because the matrix is degenerate
(as a counter-example,
the non-singular but degenerate matrix $[\Mbar]=[0, -1; -1, -1]$
has non-isolated solutions when $\dQbar=[0, 1]^{\text{T}}$).
\item
With the three Figs.~\ref{fig:ExampleLCPs}b--d,
several solutions are path-sensitive,
because small changes in $[\dQbar]$ can cause abrupt changes
(or the emergence or loss) of solutions
(Section~\ref{sec:sensitivity}).
\end{itemize}
\section{\normalsize Formulating Problem~1}\label{sec:formulate1}
Regardless of form, the three Problems in Section~\ref{sec:forms}
seek incremental particle
movements within an assembly,
in response to increments of the applied
external forces and of the imposed displacement constraints.
This section develops Problem~1, which also serves as the basis
for Problems~2 and~3.
Herein, we assume that
the movements during time increment $dt$ 
are quasi-static, so that inertial and viscous
effects are neglected, and that equilibrium is maintained in both the
reference ($t$) and displaced ($t+dt$) configurations.
\subsection{Incremental equilibrium of particles}\label{sec:equilp}
\begin{figure}
  \centering
  \includegraphics{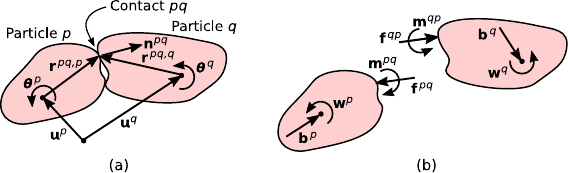}
  \caption{Two contacting particles, $p$ and $q$:
           (a)~geometry; and (b)~forces.
           \label{fig:ContactVectors}}
\end{figure}

An assembly has $N$ particles, possibly non-convex,
that touch at $M$ contacts,
with $M_{\lambda}$ of the contacts,
usually a minority of $M$, being at the friction limit.
At time $t$, the condition of a particle $p$ is given by four $3\times 1$
vectors:
$\mathbf{u}^{p}$ is the location of a material reference
point attached to the particle;
$\boldsymbol{\theta}^{p}$ is the vector of its orientation cosines;
and
$\mathbf{b}^{p}$ and $\mathbf{w}^{p}$ are the external
force and moment that act upon $p$ at its reference point
(Fig.~\ref{fig:ContactVectors}).
The condition of a contact of $p$ with particle $q$ is given by
the \emph{contact vector} $\mathbf{r}^{pq,p}$, directed
from $p$'s reference point to contact $pq$ with $q$;
$\mathbf{n}^{pq}$ is the outward unit \emph{normal vector} to the
surface of $p$ at $pq$;
and $\mathbf{f}^{pq}$ and $\mathbf{m}^{pq}$
are the contact force and moment exerted upon $p$
by $q$ at $pq$.
Particles $p$ and $q$ can share multiple contacts, with each
contact having separate force, moment, and contact and normal vectors.
The particles' surfaces are assumed smooth
(twice continuously differentiable)
at their contacts.
\par
To aid in constructing the geometric part
of the assembly's stiffness,
each contact is assigned two variants, $pq$ and $qp$,
the latter referring to the same contact but with respect to particle $q$.
As such, the assembly is treated as having $2M$ contacts.
The two variants are related:
with, $\mathbf{r}^{pq,p}=\mathbf{r}^{qp,p}$,
$\mathbf{n}^{pq}=-\mathbf{n}^{qp}$,
$\mathbf{f}^{pq}=-\mathbf{f}^{qp}$, etc.
\par
Hereafter, focus is upon incremental ``$d$'' changes in
the quantities: $d\mathbf{u}^{p}$, $d\mathbf{b}^{p}$,
$d\mathbf{f}^{pq}$, etc.
We gather the 
incremental translations and 
rotations of the $N$ particles
into the stacked column vector $[d\mathbf{u}/d\boldsymbol{\theta}]$,
the incremental changes in the applied external forces and moments
into the stacked vector $[d\mathbf{b}/d\mathbf{w}]$,
and the incremental contact forces and moments into the stacked
vector $[d\mathbf{f}/d\mathbf{m}]$, with the following notation:
\begin{equation}\label{eq:xpf}
  \left[\dX\right]_{6N \times 1} =
  \left[
    \begin{array}{@{\extracolsep{\fill}}c@{\extracolsep{\fill}}}
      d\mathbf{u} \\\hdashline[1pt/1pt]
      d\boldsymbol{\theta}
    \end{array}
  \right]
  \;\text{, }\quad
  \left[\dP\right]_{6N \times 1} =
  \left[
    \begin{array}{@{\extracolsep{\fill}}c@{\extracolsep{\fill}}}
      d\mathbf{b} \\\hdashline[1pt/1pt]
      d\mathbf{w}
    \end{array}
  \right]
  \;\text{, and}\quad
  \left[\,\dFF\,\right]_{2(6M) \times 1} =
  \left[
    \begin{array}{@{\extracolsep{\fill}}c@{\extracolsep{\fill}}}
      d\mathbf{f} \\\hdashline[1pt/1pt]
      d\mathbf{m}
    \end{array}
  \right]
\end{equation}
\par
Because the system is in equilibrium at $t$,
the contact and external forces on each particle $p$ are balanced:
\begin{equation}\label{eq:Equilibrium0}
  \left.
  \begin{aligned}
  &-\sum_{q}
    \mathbf{f}^{pq} =
  \mathbf{b}^{p}, \quad
  \\
  &-\sum_{q}
    \left(
      \mathbf{m}^{pq} +
      \mathbf{r}^{pq,p}\times \mathbf{f}^{pq}
    \right) =
  \mathbf{w}^{p}
  \end{aligned}
  \;
  \right\}
  \;
  \rightsquigarrow\;
  \left[\mathbf{A}\right]_{6N \times 2(6M)}
  \left[\,\boldsymbol{\mathfrak{f}}\,\right]
  =
  \left[\mathbf{p}\right]
\end{equation}
where each summation includes all of $p$'s contacts $pq$ with
adjacent particles $q$.
The summations for all $N$ particles are collected (here represented by
the arrow ``$\rightsquigarrow$'') with the statics
matrix $[\mathbf{A}]$.
Note that the $qp$ variants, $\mathbf{f}^{qp}$ and $\mathbf{m}^{qp}$,
are summed in the equilibrium of particle $q$, to compute
$\mathbf{b}^{q}$ and $\mathbf{w}^{q}$.
\par
As stated, the primary concern
is the \emph{incremental} form of Eq.~(\ref{eq:Equilibrium0}),
in which a mixed control of forces and movements,
$[\dP]$ and $[\dC]$, are imposed on the system
(the $[\dC]$ being present in Eq.~(\ref{eq:form1}c)
as movement constraints).
This form
requires an accounting of changes in
$\mathbf{f}^{pq}$, $\mathbf{m}^{pq}$, and $\mathbf{r}^{pq,p}$
during $dt$.
The sources of these changes include
deformations of $p$ and of its neighbors
in the vicinity of contacts $pq$;
rotations of vectors due to mutual (rigid) rotations of particles pairs;
rotations of vectors due to rolling or twirling at the contacts;
and shifting of the contact points.
The incremental form of Eq.~(\ref{eq:Equilibrium0}) is
\begin{equation} \label{eq:db1}
  \begin{gathered}
  -\sum_{q}
  \left(
  \underbrace{\mathfrak{d}\mathbf{f}^{pq}}_{\text{m}}
  +
  \underbrace{\delta\hat{\mathbf{f}}^{pq}}_{\text{g-2}}
  +
  \underbrace{d\boldsymbol{\theta}^{p} \times \mathbf{f}^{pq}}_{\text{g-3}}
  \right)
  =
  d\mathbf{b}^{p}
  \\ 
  -\sum_{q}
  \left(\rule{0ex}{2.5ex}
  \underbrace{\mathfrak{d}\mathbf{m}^{pq}
              + \mathbf{r}^{pq,p}\times \mathfrak{d}\mathbf{f}^{pq}}_{\text{m}}
  +
  \underbrace{\delta\mathbf{r}^{pq,p} \times \mathbf{f}^{pq}}_{\text{g-1}}
  +
  \underbrace{\delta\hat{\mathbf{m}}^{pq}
              +\; \mathbf{r}^{pq,p}\times \delta\hat{\mathbf{f}}^{pq}}_{\text{g-2}}
  +
  \underbrace{d\boldsymbol{\theta}^{p} \times \mathbf{m}^{pq}}_{\text{g-3}}
  \right)
  =
  d\mathbf{w}^{p}
  \end{gathered}
\end{equation}
for the incremental equilibrium of $p$ \cite{Kuhn:2019b}.
The equations for all $N$ particles
are gathered into a stiffness matrix $[\mathbf{H}]$ of the
entire assembly (see Eq.~\ref{eq:form1}b), 
which will have both mechanical and geometric parts, as described below.
The underbraces in Eq.~(\ref{eq:db1}) refer to the contributions
to contact $pq$ from the mechanical ``m''
and from various geometric ``g'' parts,
all contributing to $[\mathbf{H}]$.
The incremental quantities are as follows:
\begin{itemize}
\item
Contact force and moment increments,
$\mathfrak{d}\mathbf{f}^{pq}$ and 
$\mathfrak{d}\mathbf{m}^{pq}$,
are due to indentation, slip, and other
deformations of the two particles at $pq$.
(Note that the ``$d$'' vectors,
$d\mathbf{f}^{pq}$ and $d\mathbf{m}^{pq}$ in Eq.~\ref{eq:xpf},
also result
from rigid rotations of the particle pair, which can occur
even in the absence of any deformation of the contact.
Also, note that $\mathfrak{d}\mathbf{f}^{pq}=-\mathfrak{d}\mathbf{f}^{qp}$ and 
$\mathfrak{d}\mathbf{m}^{pq}=-\mathfrak{d}\mathbf{m}^{qp}$.)
\item
The corotated
increment $\delta\mathbf{r}^{pq,p}$ is the change in
the radial vector $\mathbf{r}^{pq,p}$, as seen by an
observer attached to a (possibly) rotating particle $p$.
The increment $d\mathbf{r}^{pq,p}$ is one
seen by an observer attached to the global frame, and it
is related to $\delta\mathbf{r}^{pq,p}$ as follows:
$d\mathbf{r}^{pq,p}=
\delta\mathbf{r}^{pq,p} + d\boldsymbol{\theta}^{p}\times\mathbf{r}^{pq,p}$,
where $d\boldsymbol{\theta}^{p}$ is $p$'s rotation seen
by the global observer.
Unlike $d\mathbf{r}^{pq,p}$,
increment $\delta\mathbf{r}^{pq,p}$ is objective.
\item
The corotated increments
$\delta\hat{\mathbf{f}}^{pq}$ and $\delta\hat{\mathbf{m}}^{pq}$
are changes in the contact force and moment
that are produced by rotations of the contact normal $d\mathbf{n}^{pq}$
(for example, by rolling)
or by twisting of two particles about normal $\mathbf{n}^{pq}$,
rather than by deformations of the particles.
\end{itemize}
These increments are described in more detail below and
are illustrated with examples.
\par
The increments $\mathfrak{d}\mathbf{f}^{pq}$ and 
$\mathfrak{d}\mathbf{m}^{pq}$
depend on the material and frictional properties of the particles,
and they are produced by the relative movement and rotation
of $p$ and $q$ at their contact $pq$.
As such, these increments
are functions of the contact deformation vector
$[\delta\mathbf{x}^{\text{def, }pq}]$,
which is composed of the two sub-vectors,
$[\delta\mathbf{u}^{\text{def, }pq}]$ and
$[\delta\boldsymbol{\theta}^{\text{def, }pq}]$:
\begin{equation}\label{eq:dudef}
  \begin{aligned}
  \left[\delta\mathbf{x}^{\text{def, }pq}\right]_{6\times 1} &=
  \def\arraystretch{1.20}
  \left[
    \begin{array}{@{\extracolsep{\fill}}c@{\extracolsep{\fill}}}
      \Dudefpq \\\hdashline[1pt/1pt]
      \Dthetadefpq
    \end{array}
  \right]
  =
  \def\arraystretch{1.20}
  \left[
  \begin{array}{@{\extracolsep{\fill}}c@{\extracolsep{\fill}}}
    d\mathbf{u}^{q} - d\mathbf{u}^{p} +
    \left(
      d\boldsymbol{\theta}^{q}\times\mathbf{r}^{pq,q} -
      d\boldsymbol{\theta}^{p}\times\mathbf{r}^{pq,p}
    \right) \\\hdashline[1pt/1pt]
    d\boldsymbol{\theta}^{q} - d\boldsymbol{\theta}^{p}
  \end{array}
  \right]
  \\
  &\;\rightsquigarrow\;
%
  \left[\delta\mathbf{x}^{\text{def}}\right]
  =
  \def\arraystretch{1.20}
  \left[
    \begin{array}{@{\extracolsep{\fill}}c@{\extracolsep{\fill}}}
      \delta\mathbf{u}^{\text{def}} \\\hdashline[1pt/1pt]
      \delta\boldsymbol{\theta}^{\text{def}}
    \end{array}
  \right]
  =
  \left[ \mathbf{B} \right]_{2(6M)\times 6N}
  \def\arraystretch{1.00}
  \left[
    \dX
  \right]
  \end{aligned}
\end{equation}
where $\mathbf{r}^{pq,p}$ and $\mathbf{r}^{pq,q}$
are the radial vectors that point
from the two particles' reference points to their shared contact $pq$
(Fig.~\ref{fig:ContactVectors}).
Note that $\delta\mathbf{u}^{\text{def},pq}=-\delta\mathbf{u}^{\text{def},qp}$
and
$\delta\boldsymbol{\theta}^{\text{def},pq}=-\delta\boldsymbol{\theta}^{\text{def},qp}$.
%
The relationships between the
contacts' deformation vectors and the particles' movements are
gathered for all $M$ contacts 
(i.e., arrow $\rightsquigarrow$),
where $[\mathbf{B}]$ is the kinematic matrix
(or the rigidity matrix, as in \cite{Agnolin:2007a,Agnolin:2007c}),
and the assembly's vector 
$[\delta\mathbf{x}^{\text{def}}]=[\delta\mathbf{u}^{\text{def}}/\delta\boldsymbol{\theta}^{\text{def}}]$
is of size $2(6M)\times 6N$,
with factor~2
since it contains both
$pq$ and $qp$ variants.
As a condition of equilibrium,
static and kinematic matrices, $[\mathbf{A}]$ and $[\mathbf{B}]$,
are dual, with $[\mathbf{A}]=[\mathbf{B}]^{\text{T}}$.
\par
The incremental change of the radial vector
$\delta\mathbf{r}^{pq,p}$ in Eq.~(\ref{eq:db1})
is produced by deformation at the contact and by rolling
of the particles $p$ and $q$.
The increment is the
sum of normal ``n'' and tangential ``t'' parts:
\begin{linenomath}
\begin{align}\label{eq:delrpq}
  \delta\mathbf{r}^{pq,p} &=
  \delta\mathbf{r}^{\text{n},pq,p} +
  \delta\mathbf{r}^{\text{t},pq,p} \\
  \delta\mathbf{r}^{\text{n},pq,p} &=
  \frac{1}{2}\left(
    \Dudefpq
    \cdot
    \mathbf{n}^{pq}
  \right)
  \mathbf{n}^{pq} \\
  \label{eq:drKPKQ}
  \delta\mathbf{r}^{\text{t},pq,p} &=
  -\left(
    \mathbf{K}^{pq,p} + \mathbf{K}^{pq,q}
  \right)^{\dagger} \cdot
  \left[
    \Dthetadefpq \times \mathbf{n}^{pq}
    - \mathbf{K}^{pq,q}\cdot
      \left(
        \Dudefpq
        - \left(
          \Dudefpq \cdot
          \mathbf{n}^{pq}
        \right) \mathbf{n}^{pq}
      \right)
  \right]
\end{align}
\end{linenomath}
where $\mathbf{n}^{pq}$ is the contact normal vector
directed outward from $p$ (Fig.~\ref{fig:ContactVectors}),
and
$\mathbf{K}^{pq,p}$ and $\mathbf{K}^{pq,q}$ are the curvature
tensors (matrices)
of the two particles' surfaces at their shared contact point $pq$
(see \cite{Montana:1988a,Kuhn:2004b}).
For the $pq$ and $qp$ variants of a contact,
$\mathbf{n}^{pq}=-\mathbf{n}^{qp}$,
$\mathbf{K}^{pq,p}=\mathbf{K}^{qp,p}$, and
$\mathbf{K}^{pq,q}=\mathbf{K}^{qp,q}$.
The curvature matrices are singular, as they are surjective
mappings from the three-dimensional space of contact movements onto
the two-dimensional contact tangent plane,
and a generalized inverse, such as
the Moore-Penrose ``$\,\dagger\,$'' inverse must be used
in Eq.~(\ref{eq:drKPKQ}) \cite{Kuhn:2004b}.
\par
The force and moment increments,
$\delta\hat{\mathbf{f}}^{pq}$ and $\delta\hat{\mathbf{m}}^{pq}$,
in Eq.~(\ref{eq:db1})
result from rotations of the contact normal $\mathbf{n}^{pq}$
(the increment $\delta\mathbf{n}^{pq}$,
as seen by an observer attached to $p$) and from the
twisting of $p$ and $q$ about the normal.
\begin{linenomath}
\begin{align} \label{eq:dfhat}
  \delta\hat{\mathbf{f}}^{pq} &=
    \mathbf{f}^{pq} \times
      \left( \delta\mathbf{n}^{pq} \times \mathbf{n}^{pq} \right)
    - \frac{1}{2}\left(
                 \delta\boldsymbol{\theta}^{\text{def},pq}
                 \cdot \mathbf{n}^{pq}\right) \mathbf{f}^{pq}
                 \times \mathbf{n}^{pq}
  \\ \label{eq:dmhat}
  \delta\hat{\mathbf{m}}^{pq} &=
    \mathbf{m}^{pq} \times
      \left( \delta\mathbf{n}^{pq} \times \mathbf{n}^{pq} \right)
    - \frac{1}{2}\left(
                 \delta\boldsymbol{\theta}^{\text{def},pq}
                 \cdot \mathbf{n}^{pq}\right) \mathbf{m}^{pq}
                 \times \mathbf{n}^{pq}
\end{align}
\end{linenomath}
These force
increments 
are independent of the corresponding increments
$\mathfrak{d}\mathbf{f}^{pq}$ and 
$\mathfrak{d}\mathbf{m}^{pq}$
that are produced by material deformations within the contact $pq$.
With the first terms on the right of
Eqs~(\ref{eq:dfhat}) and~(\ref{eq:dmhat}),
rolling produces a rotation of the contact normal
$\delta\mathbf{n}^{pq}$ (as seen by
an observer attached to $p$) that results
from the particles' curvatures \cite{Kuhn:2004b},
\begin{equation} \label{eq:deltan}
  \delta\mathbf{n}^{pq} =
  - \mathbf{K}^{pq,p} \cdot
    \delta\mathbf{r}^{\text{t},pq,p}
\end{equation}
where $\delta\mathbf{r}^{\text{t},pq,p}$ is defined
in Eq.~(\ref{eq:drKPKQ}).
Note that when particles $p$ and $q$ shift,
the contact normals can rotate, with
$d\mathbf{n}^{pq}=-d\mathbf{n}^{qp}$
(when referred to the global frame),
but the rotations $\delta\mathbf{n}^{pq}$ and $-\delta\mathbf{n}^{qp}$
can differ, as these are observed in their
local frames.
The second terms on the right of Eqs.~(\ref{eq:dfhat})--(\ref{eq:dmhat})
are rotations of force and moment due to relative twisting
of the particles about $\mathbf{n}^{pq}$.
\par
\begin{figure}
  \centering
  \includegraphics{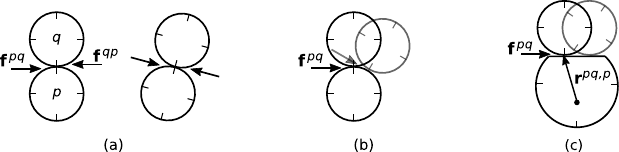}
  \caption{Examples of contact force increments:
           (a)~effect of a rigid rotation of the pair;
           (b) and~(c)~effects of rolling and contact curvatures.
           \label{fig:examples}}
\end{figure}

Figure~\ref{fig:examples} illustrates geometric
effects on various increments.
%
In Fig.~\ref{fig:examples}a,
the two particles rotate as a rigid pair,
with the original contact force $\mathbf{f}^{pq}$ on $p$
rotating with the pair.
(Note that other forces, not shown in the figure, are
required to counteract $\mathbf{f}^{pq}$.)
No deformation occurs at the contact,
so that $\mathfrak{d}\mathbf{f}^{pq}=\mathbf{0}$,
and there is no rolling at the contact,
so that $\delta\hat{\mathbf{f}}^{pq}=\mathbf{0}$
(i.e., an observer attached to $p$ sees no rotation
of $\mathbf{f}^{pq}$).
However, a downward change $d\mathbf{f}^{pq}$
is viewed by a global observer, due to rotation
$d\boldsymbol{\theta}^{p}$.
This global change results from the
term $d\boldsymbol{\theta}^{p} \times \mathbf{f}^{pq}$
in Eq.~(\ref{eq:db1}\textsubscript{1}),
so that an upward external force $d\mathbf{b}^{p}$ on $p$
(or a change of other contact forces) is required to maintain
equilibrium.
\par
Figs.~\ref{fig:examples}b and~\ref{fig:examples}c
demonstrate the role of the particles' curvatures
on their incremental equilibrium.
In Fig.~\ref{fig:examples}b,
contact force $\mathbf{f}^{pq}$ on disk $p$
is horizontal at time $t$.
Disk $p$ remains stationary while
disk $q$ rolls across $p$ with rotation $d\theta^{q}$ but with no change
in the deformation-based force
(i.e.,
$\Dudefpq=\mathfrak{d}\mathbf{f}^{pq}=\mathbf{0}$).
If the two disks have the same radius $R$,
then the radial vector $\mathbf{r}^{pq,p}$ is
shifted to the right by the increment
$\delta\mathbf{r}^{pq,p} = \frac{1}{2}R|d\theta^{q}|$,
as found with Eq.~(\ref{eq:delrpq}).
The upward normal $\mathbf{n}^{pq}$ shifts to the right by
$\delta\mathbf{n}^{pq}=\frac{1}{2}|d\theta^{q}|$.
When viewed by $p$,
the force $\mathbf{f}^{pq}$ is changed by a downward
increment
$\delta\hat{\mathbf{f}}^{pq}=\frac{1}{2}|\mathbf{f}^{pq}||d\theta^{q}|$,
so that an upward incremental external force $d\mathbf{b}^{p}$
is required to maintain equilibrium.
No increment of external moment, $d\mathbf{w}^{p}$, ensues,
since the cross products in~(\ref{eq:db1}\textsubscript{2}) are zero.
\par
Fig.~\ref{fig:examples}c
is similar to Fig.~\ref{fig:examples}b,
except that particle $q$ rolls across a \emph{flat} surface of
particle $p$.
As before,
disk $p$ remains stationary, and $q$ rolls through an angle
$d\theta^{q}$ in a manner that produces no change
in the deformation-based force
(i.e.,
$\Dudefpq=\mathfrak{d}\mathbf{f}^{pq}=\mathbf{0}$).
If, as before, the two radial distances are $R$, then
a rightward shift in $\mathbf{r}^{pq,p}$, with
$\delta\mathbf{r}^{pq,p} = \frac{1}{2}R|d\theta^{q}|$,
is observed by $p$.
However,
the normal vector and the $p$-viewed contact force are unchanged:
$\delta\mathbf{n}^{pq}=\delta\hat{\mathbf{f}}^{pq}=\mathbf{0}$.
The body force is also unchanged,
$d\mathbf{b}^{p}=\mathbf{0}$,
but an anti-clockwise external moment 
$d\mathbf{w}=\frac{1}{2}R|\mathbf{f}^{pq}||d\theta^{q}|$
is required by the
$\delta\mathbf{r}^{pq,p} \times \mathbf{f}^{pq}$
term in~(\ref{eq:db1}\textsubscript{2}).
\subsection{Geometric stiffness of an assembly}\label{sec:GeomH}
Equations~(\ref{eq:db1}) are the conditions of
equilibrium at $t+dt$ 
for a particle $p$ that is in equilibrium at $t$
(Eq.~\ref{eq:Equilibrium0})
but which undergoes rolling and rotation as well as material
changes in its contact forces.
The equations not only include terms
with $\mathfrak{d}$ and $\delta$ increments
of force and moment, but they also include terms with the full
forces and moments at time $t$~--- the terms
$\mathbf{f}^{pq}$ and $\mathbf{m}^{pq}$~--- and
these terms have a geometric basis
that is independent of the particles'
material properties but are dependent on the particles'
sizes and shapes.
Four geometric stiffnesses contribute to the
full assembly stiffness $[\mathbf{H}]$.
Three geometric stiffnesses, ``{g-1}'', ``g-2'', and ``g-3'',
have an internal origin, arising from the
shifting and rotation of contact forces~---
the internal analog of external follower forces.
The three stiffnesses
depend upon (and are proportional to)
the current forces and upon the particles' sizes and
surface curvatures.
A fourth ``g-4''
stiffness accounts for
any position-dependent external forces.
\par
The ``g-1'' term in Eq.~(\ref{eq:db1}\textsubscript{1}) is included in the
incremental moment equilibrium of particle $p$, arising from
changes in the radial vectors $\mathbf{r}^{pq,p}$.
Increment $\delta\mathbf{r}^{pq,p}$ depends on
the movements $d\mathbf{u}^{p}$, $d\mathbf{u}^{q}$,
$d\boldsymbol{\theta}^{p}$, and $d\boldsymbol{\theta}^{q}$
(Eqs.~\ref{eq:dudef}--\ref{eq:drKPKQ}).
The terms for all $2M$ contacts (both $pq$ and $qp$ variants)
are gathered into assembly stiffness
$[\mathbf{H}^{\text{g-1}}]$:
\begin{equation}\label{eq:Hg1}
  \left.\begin{array}{l}
  \mathbf{0}\\
  -\delta\mathbf{r}^{pq,p} \times \mathbf{f}^{pq}
  \end{array}\right\}_{6\times 1}
  \;\rightsquigarrow\;
  \left[\mathbf{H}^{\text{g-1}}\right]_{6N\times 6N}
  \left[
    \begin{array}{@{\extracolsep{\fill}}c@{\extracolsep{\fill}}}
      d\mathbf{u} \\\hdashline[1pt/1pt]
      d\boldsymbol{\theta}
    \end{array}
  \right]
\end{equation}
where 
the zero vector 
represents a
nil contribution 
to the force equilibrium of $p$
(a g-1 term is absent in Eq.~\ref{eq:db1}\textsubscript{1}).
The cross product on the left of
Eq.~(\ref{eq:Hg1}) leads to an asymmetric
stiffness $[\mathbf{H}^{\text{g-1}}]$.
Stiffness $[\mathbf{H}^{\text{g-1}}]$ can also be viewed
as an alteration of the statics matrix $[\mathbf{A}]$,
with $H_{ij}^{\text{g-1}}=(\partial A_{ik}/\partial dx_{j})\mathfrak{f}_{k}$
(see Eq.~\ref{eq:Equilibrium0}).
This type of geometric stiffness is commonly included in continuum and
finite element models (e.g., \cite{Bazant:1991a}),
but rarely in discrete models.
The other geometric stiffnesses, $[\mathbf{H}^{\text{g-2}}]$
and $[\mathbf{H}^{\text{g-3}}]$, described below,
are peculiar to granular materials and
result from the rotations of discrete rigid grains.
\par
The ``g-2'' terms in Eq.~(\ref{eq:db1})
account for changes,
$\delta\hat{\mathbf{f}}^{pq}$ and $\delta\hat{\mathbf{m}}^{pq}$,
in the contact forces 
that are produced by rotations of the contact normal $d\mathbf{n}^{pq}$
and the twirling of $\mathbf{f}^{pq}$ and $\mathbf{m}^{pq}$
about $\mathbf{n}^{pq}$.
After substituting
Eqs.~(\ref{eq:dfhat})--(\ref{eq:deltan}),
the ``g-2'' terms in Eq.~(\ref{eq:db1})
depend linearly on
the movements $d\mathbf{u}^{p}$, $d\mathbf{u}^{q}$,
$d\boldsymbol{\theta}^{p}$, and $d\boldsymbol{\theta}^{q}$,
and these terms are gathered into a contact stiffness matrix
$[\mathbf{E}^{\text{g-2}}]$ and a corresponding geometric
stiffness matrix $[\mathbf{H}^{\text{g-2}}]$ of the assembly:
\begin{equation}\label{eq:Hg2}
  \left.
    \begin{aligned}
      &\,\delta\hat{\mathbf{f}}^{pq}\\
      &
      \delta\hat{\mathbf{m}}^{pq}
    \end{aligned}
  \right\}_{6\times 1}
  \rightsquigarrow
   \left[ \mathbf{E}^{\text{g-2}}\right]_{2(6M)\times 6N}
  \left[
    \begin{array}{@{\extracolsep{\fill}}c@{\extracolsep{\fill}}}
      d\mathbf{u} \\\hdashline[1pt/1pt]
      d\boldsymbol{\theta}
    \end{array}
  \right]
  \quad\text{and}\quad
  \left[\mathbf{H}^{\text{g-2}}\right]_{6N\times 6N}
  =
  \left[\mathbf{A}\right]
  \left[\mathbf{E}^{\text{g-2}}\right]
\end{equation}
where $[\mathbf{A}]$ is the statics matrix of
Eq.~(\ref{eq:Equilibrium0}) and the transpose of $[\mathbf{B}]$
in Eq.~(\ref{eq:dudef}).
\par
The ``g-3'' terms in Eq.~(\ref{eq:db1}) also depend linearly on
the movements $d\mathbf{u}^{p}$, $d\mathbf{u}^{q}$,
$d\boldsymbol{\theta}^{p}$, and $d\boldsymbol{\theta}^{q}$,
and these terms are gathered into the 
third geometric stiffness matrix $[\mathbf{H}^{\text{g-3}}]$:
\begin{equation}\label{eq:Hg3}
  \left.
    \begin{aligned}
      &-d\boldsymbol{\theta}^{p} \times \mathbf{f}^{pq}\\
      &-d\boldsymbol{\theta}^{p} \times \mathbf{m}^{pq}
    \end{aligned}
  \right\}_{6\times 1}
  \rightsquigarrow
  \left[\mathbf{H}^{\text{g-3}}\right]_{6N\times 6N}
  \left[
    \begin{array}{@{\extracolsep{\fill}}c@{\extracolsep{\fill}}}
      d\mathbf{u} \\\hdashline[1pt/1pt]
      d\boldsymbol{\theta}
    \end{array}
  \right]
\end{equation}
%
\par
A final geometric stiffness $[\mathbf{H}^{\text{g-4}}]$ applies
when the
external forces, $[\mathbf{b}]$ and $[\mathbf{w}]$, depend on
the particles' positions and orientations,
$[\mathbf{u}]$ and $[\boldsymbol{\theta}]$,
and on possible loading parameters $[d\mathbf{q}]$.
%
For such follower-type forces and loading parameters,
$[\mathbf{p}]= [\mathbf{p}(\mathbf{x}, \mathbf{q})]$
with increment
%
\begin{equation}\label{eq:dbG}
  \left[d\mathbf{p}\right]
  =
  \left[
    \begin{array}{@{\extracolsep{\fill}}c@{\extracolsep{\fill}}}
      d\mathbf{b} \\\hdashline[1pt/1pt]
      d\mathbf{w}
    \end{array}
  \right]
  =
  \left[
    \partial \mathbf{p}/\partial \mathbf{x}
  \right]_{6N\times 6N}
  \left[ d\mathbf{x}\right]
  +
  \left[
    \partial \mathbf{p}/\partial \mathbf{q}
  \right]_{6N\times L}
  \left[ d\mathbf{q}\right]_{L\times 1}
\end{equation}
where the final term
gives the loading increments $[d\mathbf{p}]$ produced by
increments of the $L$ loading parameters, $d\mathbf{q}$
(e.g., increments in the chamber pressure of a triaxial apparatus,
increments of applied dead platen loads, etc.).
%
%
In the first term on the right of Eq.~(\ref{eq:dbG}),
the matrix $[\partial \mathbf{p}/\partial\mathbf{x}]$
gives increments in the position-dependent forces and
constitutes the fourth geometric stiffness,
%
\begin{equation}\label{eq:Hg4}
  \left[\mathbf{H}^{\text{g-4}}\right]_{6N\times 6N}
  =
  - \left[
      \partial \mathbf{p}/\partial \mathbf{x}
    \right]_{6N\times 6N}
\end{equation}
where the change in sign results from shifting the contribution of 
$[\partial\mathbf{p}/\partial\mathbf{x}][\dX]$
to $[\dP]$ from the right side
of Eq.~(\ref{eq:form1}b). 
Note that
when all external loads are independent of the particles'
positions (as with non-follower dead loads),
the problem is simplified, with
$[\mathbf{H}^{\text{g-4}}]=[\mathbf{0}]$.
Henceforth, the force increment $[\dP]$ in Eq.~(\ref{eq:form1})
is understood as the applied loading increment
$[\partial\mathbf{p}/\partial\mathbf{q}][d\mathbf{q}]$
in Eq.~(\ref{eq:dbG}).
%
\subsection{Stiffness of a single contact}\label{sec:linfrict}
The mechanical part of an assembly's stiffness,
$[\mathbf{H}^{\text{m}}]$, results from the
``$\mathfrak{d}$'' increments of contact force
that originate from the
contacts' resistance to deformation and sliding.
This section presents the
simplest model: the standard
linear-frictional model of a spring in series with a frictional slider.
Two equivalent forms of stiffness
are presented:
one intended for Problem~1; the other used for the
LCP of Problems~2 and the QP of Problem~3.
More complex contact models are discussed at the end
of this section.
%
%
\par
In the linear-frictional model,
the contact is elastic
until the friction limit (i.e., frictional yield surface)
is reached.
Elastic stiffness is characterized
with positive normal and tangential stiffnesses,
$k$ and $\alpha k$,
and friction coefficient $\mu$.
Upon reaching the friction limit,
the \emph{possibility} of slip is active, and
behavior is incrementally nonlinear
with two branches:
an elastic (no-slip) branch
and a sliding (slip) branch.
The branch that is subsequently taken is
determined by the direction of the incremental contact
deformation $\Dudefpq$, of Eq.~(\ref{eq:dudef}).
Contacts that are not yet active have a single, elastic branch.
\par
The frictional yield envelope is  
\begin{equation}\label{eq:yield}
  G^{pq}_{\text{s}} =
  \mathbf{h}^{pq}_{\text{s}}\cdot\mathbf{f}^{pq}
  +
  \mu\mathbf{n}^{pq}\cdot\mathbf{f}^{pq}
  \le 0
\end{equation}
where subscript ``s'' corresponds to sliding.
The equation simply asserts that the tangent force
$|\mathbf{h}^{pq}_{\text{s}}\cdot\mathbf{f}^{pq}|$
cannot exceed $\mu$ times
the normal force $|\mathbf{n}^{pq}\cdot\mathbf{f}^{pq}|$
(note that $\mathbf{n}^{pq}\cdot\mathbf{f}^{pq}$ is negative for compressive
force).
An active contact with
incipient sliding occurs when equality is reached,
$G_{\text{s}}^{pq}=0$.
Unit vector $\mathbf{h}^{pq}_{\text{s}}$ lies within the contact's
tangent plane, and is the projection of the current force's direction
onto the plane:
\begin{equation}\label{eq:Hdirection}
\mathbf{h}^{pq}_{\text{s}} = 
            \frac{  \mathbf{f}^{pq} - (\mathbf{n}^{pq} \cdot\mathbf{f}^{pq} )
                       \mathbf{n}^{pq}   }
            {| \mathbf{f}^{pq} - (\mathbf{n}^{pq} \cdot\mathbf{f}^{pq} )
            \mathbf{n}^{pq} |}
\end{equation}
That is, $\mathbf{h}^{pq}_{\text{s}}$ is the direction
of incipient slip.
\par
With Problem~1,
the incremental force~--- the $\mathfrak{d}\mathbf{f}^{pq}$
in Eq.~(\ref{eq:db1})~--- derives from contact
stiffness $\mathbf{F}^{pq}$:
\begin{align}
\label{eq:dfpq}
&\mathfrak{d}\mathbf{f}^{pq} =
 \mathbf{F}^{pq}\cdot\Dudefpq
 \\
\label{eq:ContactF}
&\mathbf{F}^{pq} = \begin{cases}
 \Felastpq
  & \text{if } G^{pq}_{\text{s}}<0 \text{ or } 
    \mathbf{g}_{\text{s}}^{pq}\cdot\Dxdefpq\leq 0 \\
  \mathbf{F}^{\text{sliding},\,pq} =
  \Felastpq -
  \mathbf{h}^{pq}_{\text{s}}\otimes\mathbf{g}^{pq}_{\text{s}}
  & \text{if } G^{pq}_{\text{s}}=0 
    \text{ and } \mathbf{g}_{\text{s}}^{pq}\cdot\Dxdefpq>0
 \end{cases}
 \\
 \label{eq:Felasticpq}
 &\Felastpq =
  k\left[ \alpha\mathbf{I} + (1-\alpha)\mathbf{n}^{pq}\otimes\mathbf{n}^{pq}  \right]
 \\
 \label{eq:gpq}
 &\mathbf{g}^{pq}_{\text{s}}=
 \left[
  \begin{array}{@{\extracolsep{0ex}}c;{1pt/1pt}c@{\extracolsep{\fill}}}
          k\left( \alpha\mathbf{h}^{pq}_{\text{s}} + \mu\mathbf{n}^{pq} \right)
           & \mathbf{0}
  \end{array}
  \right]
\end{align}
where the two stiffness branches
(elastic and sliding) appear in Eq.~(\ref{eq:ContactF}),
and $\mathbf{n}^{pq}\otimes\mathbf{n}^{pq}$ is the
dyadic product $n^{pq}_{i} n^{pq}_{j}$.
Vector $\mathbf{g}^{pq}_{\text{s}}$
is a $1\times 6$
vector that combines $1\times 3$ translation and rotation parts.
Likewise,
the $\Dxdefpq$ vector is a $6\times 1$ stack
of movements $\Dudefpq$ and $\Dthetadefpq$,
\begin{equation}
  \left[\Dxdefpq\right]_{6\times 1}
  =
  \left[
    \begin{array}{@{\extracolsep{\fill}}c@{\extracolsep{\fill}}}
      \Dudefpq \\\hdashline[1pt/1pt]
      \Dthetadefpq
    \end{array}
  \right]
\end{equation}
similar to Eq.~(\ref{eq:xpf}\textsubscript{1}).
The contact force is unaffected by relative rotation
$\Dthetadefpq$, hence the zero sub-vector in $\mathbf{g}^{pq}_{\text{s}}$
of Eq.~(\ref{eq:gpq}).
Vector $\mathbf{g}^{pq}_{\text{s}}$ is directed outward from the
yield surface, $G^{pq}_{\text{s}}=0$,
such that $\mathbf{g}^{pq}_{\text{s}}\cdot\Dudefpq>0$
for the slip branch of Eq.~(\ref{eq:ContactF}\textsubscript{2}).
Two branches are available when the contact is
active (with $G^{pq}_{\text{s}}=0$), 
but for elastic
contacts with $G^{pq}_{\text{s}}<0$, only the single stiffness
of Eq.~(\ref{eq:ContactF}\textsubscript{1}) applies.
\par
With Problems~2 and~3, an equivalent formulation is used,
one with an embedded complementarity condition.
The auxiliary scalar $\Lambdas$ is introduced,
representing the incremental slip distance,
and the incremental force $\mathfrak{d}\mathbf{f}^{pq}$
is given by
\begin{equation}\label{eq:dfdlambda}
  \mathfrak{d}\mathbf{f}^{pq} =
  \Felastpq
  \cdot
  \left(
    \Dudefpq
    -
    \Lambdas\mathbf{h}^{pq}_{\text{s}}
  \right)
\end{equation}
where $\mathbf{h}^{pq}_{\text{s}}$ is the incipient slip direction,
and $\Felastpq$ is given by Eq.~(\ref{eq:Felasticpq}).
If the contact has not reached the friction limit,
then the slip increment $\Lambdaf$ is zero and
Eq.~(\ref{eq:ContactF}\textsubscript{1}) applies.
On the other hand, if the contact is active, with
$G^{pq}_{\text{s}}=0$,
then $\Lambdaf$ is positive, but only when the
consistency condition, $dG^{pq}_{\text{s}}=0$, is met:
\begin{equation}\label{eq:dGs}
   dG^{pq}_{\text{s}} =
   \left(\mathbf{h}^{pq}_{\text{s}}+\mu\mathbf{n}^{pq}\right)
   \cdot
   \mathfrak{d}\mathbf{f}^{pq} =
   \mathbf{g}^{pq}_{\text{s}}\!
   \cdot\delta\mathbf{x}^{\text{def, }pq}
   - \mathbf{g}^{pq}_{\text{s}}\!\cdot\mathbf{h}^{pq}_{\text{s}}\Lambdas
   = 0 
\end{equation}
in which Eqs.~(\ref{eq:ContactF}\textsubscript{1})
and~(\ref{eq:dfdlambda}) have been substituted.
%
These conditions are summarized in the following
expressions, intended for Problems~2 and~3:
\begin{equation}\label{eq:dfForm2}
 \mathfrak{d}\mathbf{f}^{pq} =
  \left\{
    \begin{alignedat}{2}
      &\:\Felastpq\cdot\Dudefpq
       &&\text{ if }G^{pq}_{\text{s}}<0
      \\
      &\left.
        \begin{aligned}
          & \Felastpq\cdot
            \left(\Dudefpq - \Lambdas\mathbf{h}^{pq}_{\text{s}}\right)\\
          & \quad
            \Lambdas \ge 0\\
          & \quad
            d\bar{G}^{pq}_{\text{s}}=-dG^{pq}_{\text{s}}=
            -
            \left(
              \mathbf{g}^{pq}_{\text{s}}\!
              \cdot\delta\mathbf{x}^{\text{def, }pq}
              - \mathbf{g}^{pq}_{\text{s}}\!\cdot\mathbf{h}^{pq}_{\text{s}}\Lambdas
            \right)
            \ge 0\\
          & \quad
            \Lambdas d\bar{G}^{pq}_{\text{s}} = 0
        \end{aligned}
      \right\}
      &&
      \text{ if }G^{pq}_{\text{s}}=0
    \end{alignedat}
  \right.
\end{equation}
which includes the complementarity condition,
$\Lambdas d\bar{G}^{pq}_{\text{s}} = 0$, for an active
contact (a contact with $G^{pq}_{\text{s}}=0$).
Note that the ``bar'' increment $d\bar{G}^{pq}_{\text{s}}=-dG^{pq}_{\text{s}}$
is introduced, so that the ``$\ge$'' inequality
is consistent with the standard forms of
Eqs.~(\ref{eq:form2}) and~(\ref{eq:form3}).
\par
Beyond the linear-frictional contact considered above,
more complex (perhaps, more realistic)
contact models have been used in granular modeling,
such as the Hertz, simplified Mindlin, Mindlin--Deresiewicz,
and other models
\cite{Mindlin:1953a,Jager:2005a,Cole:2016a}. 
Unlike the simpler model,
the incremental stiffness $k$ in these
non-linear models depends upon the current force $\mathbf{f}$,
and such models require the stiffness and current force
to be updated at each increment of an LCP simulation.
A similar updating is used in DEM modeling.
Even the micro-slip model of Mindlin--Deresiewicz, 
in which slip occurs when the normal forces
is less than the sliding force
$\mu|\mathbf{n}\cdot\mathbf{f}|$,
can be modeled in an incremental manner,
with the complementarity condition supplied
by the J\"{a}ger algorithm \cite{Jager:2005a,Kuhn:2011a}.
\subsection{Combined stiffness of an assembly}\label{sec:assemstiff}
An assembly's combined stiffness
is developed in this section,
and final statement of Problem~1 is derived in
Section~\ref{sec:FinalForms1},
incorporating the displacement boundary conditions
of Eq.~(\ref{eq:form1}c).
Multiple mechanisms
can pertain at contacts
(for example, sliding, rolling, and twisting friction types),
and $M_{\lambda}$ is the total number of active contacts
(or of active mechanisms among the contacts,
when multiple mechanisms apply).
In counting $M_{\lambda}$, the $pq$ and $qp$ variants
are merged as a single contact.
%
\par
Each assembly stiffness $[\mathbf{H}^{i}]$ 
in Eq.~(\ref{eq:form1}d) pertains
within its $i$th cone of displacements,
$\Omega^{i}\in\mathbb{R}^{6N}$,
with this cone corresponding to the $i$th combination of
sliding/non-sliding branches of the $M_{\lambda}$
active contacts.
With two branches per active contact,
a total of $B=2^{M_{\lambda}}$ combinations are possible:
$i\in\{1,2,\ldots,B\}$.
For the $i$th combination of branches,
the mechanical ``m'' stiffness $\mathbf{F}^{pq}$
of each $pq$ contact
is assigned from the two stiffnesses in Eq.~(\ref{eq:ContactF}),
and the stiffnesses of all $M$ contacts are then assembled
for the entire assembly,
\begin{equation}
  \left[
    \begin{array}{@{\extracolsep{\fill}}c@{\extracolsep{\fill}}}
      \mathfrak{d}\mathbf{f}^{pq} \\[0.1ex]\hdashline[1pt/1pt]
      \mathfrak{d}\mathbf{m}^{pq}
    \end{array}
  \right]_{6\times 1}
  =
  \left[
    \begin{array}{@{\extracolsep{\fill}}c@{\extracolsep{\fill}}}
      \mathbf{F}^{pq,i}_{\;3\times 6} \\[0.1ex]\hdashline[1pt/1pt]
      \mathbf{M}^{pq,i}_{\;3\times 6}
    \end{array}
  \right]_{6\times 6}
  \left[
    \begin{array}{@{\extracolsep{\fill}}c@{\extracolsep{\fill}}}
      \Dudefpq \\[0.1ex]\hdashline[1pt/1pt]
      \Dthetadefpq
    \end{array}
  \right]_{6\times 1}
  \rightsquigarrow\;
  \left[
    \begin{array}{@{\extracolsep{\fill}}c@{\extracolsep{\fill}}}
      \mathfrak{d}\mathbf{f} \\[0.1ex]\hdashline[1pt/1pt]
      \mathfrak{d}\mathbf{m}
    \end{array}
  \right]_{2(6M)\times 1}
  =
  \def\arraystretch{1.00}
  \left[
    \begin{array}{@{\extracolsep{\fill}}c@{\extracolsep{\fill}}}
      \mathbf{F}^{i} \\\hdashline[1pt/1pt]
      \mathbf{M}^{i}
    \end{array}
  \right]_{2(6M)\times 2(6M)}
  \def\arraystretch{1.20}
  \left[
    \begin{array}{@{\extracolsep{\fill}}c@{\extracolsep{\fill}}}
      \delta\mathbf{u}^{\text{def}} \\\hdashline[1pt/1pt]
      \delta\boldsymbol{\theta}^{\text{def}}
    \end{array}
  \right]
\end{equation}
in which the collected material-derived (mechanical)
forces and moments,
$[\mathfrak{d}\mathbf{f} / \mathfrak{d}\mathbf{m}]$,
depend upon the collected deformation motions,
$[\delta\mathbf{u}^{\text{def}} / \delta\boldsymbol{\theta}^{\text{def}}]$.
Substituting the kinematics matrix $[\mathbf{B}]$
of Eq.~(\ref{eq:dudef}),
\begin{equation}\label{eq:dfrakB}
  \def\arraystretch{1.00}
  \left[
    \begin{array}{@{\extracolsep{\fill}}c@{\extracolsep{\fill}}}
      \mathfrak{d}\mathbf{f} \\\hdashline[1pt/1pt]
      \mathfrak{d}\mathbf{m}
    \end{array}
  \right]_{2(6M)\times 1}
  =
  \def\arraystretch{1.00}
  \left[
    \begin{array}{@{\extracolsep{\fill}}c@{\extracolsep{\fill}}}
      \mathbf{F}^{i} \\\hdashline[1pt/1pt]
      \mathbf{M}^{i}
    \end{array}
  \right]
  \def\arraystretch{1.00}
  \left[
    \begin{array}{@{\extracolsep{\fill}}c@{\extracolsep{\fill}}}
      \delta\mathbf{u}^{\text{def}} \\\hdashline[1pt/1pt]
      \delta\boldsymbol{\theta}^{\text{def}}
    \end{array}
  \right]
  =
  \def\arraystretch{1.00}
  \left[
    \begin{array}{@{\extracolsep{\fill}}c@{\extracolsep{\fill}}}
      \mathbf{F}^{i} \\\hdashline[1pt/1pt]
      \mathbf{M}^{i}
    \end{array}
  \right]
  \left[\mathbf{B}\right]
  \def\arraystretch{1.00}
  \left[
    \begin{array}{@{\extracolsep{\fill}}c@{\extracolsep{\fill}}}
      d\mathbf{u} \\\hdashline[1pt/1pt]
      d\boldsymbol{\theta}
    \end{array}
  \right]
\end{equation}
\par
Finally, the static matrix $[\mathbf{A}]$ of Eq.~(\ref{eq:Equilibrium0})
accomplishes the summations of Eq.~(\ref{eq:db1}),
combining
the mechanical contact
forces to find their contribution to the external forces $[\dP]$,
so that the mechanical stiffness $[\mathbf{H}^{\text{m},i}]$
of the $i$th branch is
\begin{equation}\label{eq:Hmprod}
  \left[ \mathbf{H}^{\text{m},i} \right]_{6N\times 6N}
  =
  \left[ \mathbf{A} \right]_{6N\times 2(6M)}
  \def\arraystretch{1.00}
  \left[
    \begin{array}{@{\extracolsep{\fill}}c@{\extracolsep{\fill}}}
      \mathbf{F}^{i} \\\hdashline[1pt/1pt]
      \mathbf{M}^{i}
    \end{array}
  \right]_{2(6M)\times 2(6M)}
  \left[ \mathbf{B} \right]_{2(6M)\times 6N}
\end{equation}
\par
In Eq.~(\ref{eq:db1}), the increment of 
external force and moment on each particle $p$
is the sum of mechanical and geometric contributions
of the contacts.
Likewise, the assembly stiffness
$[\mathbf{H}^{i}]$ is the sum
of the mechanical stiffness 
and four geometric stiffnesses:
\begin{linenomath}
\begin{align}\label{eq:Himg}
  &\left[\mathbf{H}^{i}\right]=
         \left[\mathbf{H}^{\text{m},i}\right]
         + \left[\mathbf{H}^{\text{g}}\right]\\
  \label{eq:Hg}
  &\left[ \mathbf{H}^{\text{g}} \right]
    =  \left[ \mathbf{H}^{\text{g-1}} \right]
      + \left[ \mathbf{H}^{\text{g-2}} \right]
      + \left[ \mathbf{H}^{\text{g-3}} \right]
      + \left[ \mathbf{H}^{\text{g-4}} \right]
\end{align}
\end{linenomath}
\par
The stiffnesses $[\mathbf{H}^{i}]$ are rarely symmetric.
When even a single active contact continues to slide,
its stiffness is non-symmetric
(see Eq.~\ref{eq:ContactF}\textsubscript{2}),
and the assembly's mechanical stiffness $[\mathbf{H}^{\text{m},i}]$
becomes non-symmetric.
Moreover,
all four geometric stiffness matrices
will usually be non-symmetric,
also leading to a non-symmetric $[\mathbf{H}^{i}]$.
\par
Solution of Problem~1
requires a systematic means of interrogating
all $2^{M_{\lambda}}$ stiffness branches $[\mathbf{H}^{i}]$ and
searching for solutions $[\dX]$ that lie within the
cone $\Omega^{i}$ 
for which stiffness $[\mathbf{H}^{i}]$ is intended
(see Eq.~\ref{eq:form1}c).
Noting that the sliding/non-sliding condition of a contact $pq$ is determined
in Eq.~(\ref{eq:ContactF}) by the sign of
$\mathbf{g}^{pq}_{\text{s}}\cdot\Dxdefpq$,
we collect the conditions of
all $M_{\lambda}$ active contacts and compare these conditions
with those of a particular region $\Omega^{i}$.
In this approach,
the vectors $\mathbf{g}^{pq}$ of
Eq.~(\ref{eq:gpq}) are gathered into the $M_{\lambda}$
rows of a matrix $[\mathbf{G}]$,
with one row for each active contact $pq$,
\begin{equation}
\mathbf{g}^{pq}_{\text{s}}
\rightsquigarrow
\left[\mathbf{G}_{\text{s}}\right]_{M_{\lambda}\times 2(6M)}
\end{equation}
%
%
%
%
A Boolean comparison vector $[\mathbf{s}^{i}]_{M_{\lambda}\times 1}$
of 0's and 1's
is created for each of the $2^{M_{\lambda}}$ branches:
with 1's for those
active contacts whose movements during $dt$ are assumed to produce
frictional slip in $\Omega^{i}$,
and 0's for active contacts whose movements are assumed to be elastic.
As such, a region $\Omega^{i}$ is defined as
%
\begin{equation}\label{eq:condition}
\Omega^{i}=
\left\{[d\mathbf{x}]\in\mathbb{R}^{6N}\!:\;
  \right(
  \left[\mathbf{G}_{\text{s}}\right]
  \left[\mathbf{B}\right]
  \left[\dX\right]
  > 0\left)
  =\left[\mathbf{s}^{i}\right]\right\}
\end{equation}
where the quantity in parentheses is a Boolean
vector of 0's (false) and 1's (true).
In this manner, a potential solution of Eq.~(\ref{eq:form1b}b), $[\dX]$,
can be compared with the $2^{M_{\lambda}}$ vectors
$[\mathbf{s}^{i}]$ for a match with the $\Omega^{i}$ cones.
The union of cones $\Omega^{i}$ is the full
range $\mathbb{R}^{6N}$ of vectors $[\dX]$
(Eq.~\ref{eq:form1}d).
\subsection{Displacement constraints and final form of
Problem~1}\label{sec:FinalForms1}
The stiffness relation of Eq.~(\ref{eq:form1}b)
is augmented with
$R$ equality constraints on the particles' displacements and rotations,
in Eq.~(\ref{eq:form1}c).
These constraints
are stipulated with the control program
matrix $[\mathbf{C}]_{R\times 6N}$
and the increment vector $[d\mathbf{c}]_{R\times 1}$, which can represent
kinematic
restrictions imposed by platens, walls, the isochoric restriction,
etc.
The constraints reduce the solution space of movements
$[d\mathbf{x}]$ to a subspace of $\mathbb{R}^{6N}$, and
this smaller sub-space can preclude stiffness pathologies
that might otherwise
be available in the full space $\mathbb{R}^{6N}$.
Challamel and colleagues \cite{Challamel:2009a,Lerbet:2020a}
have shown, however, that adding constraints
(e.g., adding rows to $[\mathbf{C}]$) can, paradoxically,
destabilize an otherwise stable linear system.
The $R$ constraints give rise to $R$ reaction forces
that are necessary to keep the loads $[\dP]$
within the reduced range of the stiffness matrix $[\mathbf{H}]$.
With Problems~2 and~3,
a reckoning of the displacement constraints,
presented in Section~\ref{sec:p12final}, leads
to final expressions for the
$[\Mbar]$ and $[\dQbar]$ in Eqs.~(\ref{eq:form2})
and~(\ref{eq:form3}).
%
%
\par
As in Eq.~(\ref{eq:form1}c), a set of $R$ linear
non-homogeneous conditions are placed upon
the particles' displacements and rotations, $[d\mathbf{x}]$,
\begin{equation}\label{eq:nonhomogeneousC}
  \left[\mathbf{C}\right]_{R\times 6N}
  \left[d\mathbf{x}\right]
  =
  \left[d\mathbf{c}\right]_{R\times 1}
\end{equation}
in which matrix $[\mathbf{C}]$ is assumed to have full row rank.
Because $[\dX]$ is restricted by both
Eqs.~(\ref{eq:form1}b) and~(\ref{eq:nonhomogeneousC}),
certain applied loads $[\dP]$ will lie outside the column space of
$[\mathbf{H}]$ in Eq.~(\ref{eq:form1}a).
As such,
a vector of reaction forces $[d\mathbf{y}]$ is added to
$[\dP]$ to bring the combined force within the range
of $[\mathbf{H}]$, while
conforming with the displacement constraints:
\begin{equation}\label{eq:Hdxdpdy}
  \left[\mathbf{H}\right]
  \left[d\mathbf{x}\right] =
  \left[\dP\right] +
  \left[d\mathbf{y}\right]
\end{equation}
In this section, generic matrix $[\mathbf{H}]$ represents
any of the $B=2^{M_{\lambda}}$ stiffnesses
$[\mathbf{H}^{i}]$, with each
applying to its particular domain,
$[\dX]\in\Omega^{i}$,
as defined in Eq.~(\ref{eq:condition}).
\par
By itself, matrix $[\mathbf{C}]$ has far fewer rows than
columns, but
the general solution of Eq.~(\ref{eq:nonhomogeneousC})
is
\begin{equation}\label{eq:xcdc}
  \left[d\mathbf{x}\right] =
  \left[\mathbf{C}\right]^{(1)}
  \left[d\mathbf{c}\right] +
  \left(
    \left[\,\mathbf{I}\,\right]
    -
    \left[\mathbf{C}\right]^{(1)}
    \left[\mathbf{C}\right]
  \right)
  [d\mathbf{v}],\quad\forall\,
  [d\mathbf{v}]\in\mathbb{R}^{6N}
\end{equation}
where $[\mathbf{C}]^{(1)}$ is any $\{1\}$-inverse of
$[\mathbf{C}]$,
and $[d\mathbf{v}]$ is an arbitrary $6N\times 1$ vector
(see \cite{BenIsrael2003a}\S 2.1).
The basis of this statement is as follows.
When $[\mathbf{C}]$ is multiplied by the first part
on the right of Eq.~(\ref{eq:xcdc}),
as in Eq.~(\ref{eq:nonhomogeneousC}),
the result is $[d\mathbf{c}]$,
since a $\{1\}$-inverse has the property
$[\mathbf{C}][\mathbf{C}]^{(1)}[\mathbf{C}]=[\mathbf{C}]$.
Because of the same property,
multiplying $[\mathbf{C}]$ by the second part on the right
gives the zero vector.
\par
Note that a product $[\mathbf{C}]^{(1)}[\mathbf{C}]$ is idempotent and
is a projector onto the range $\mathcal{R}$ of $[\mathbf{C}]^{(1)}$
along the
direction of null space $\mathcal{N}$ of $[\mathbf{C}]$,
so that any vector $[d\mathbf{z}]$ has the additive decomposition,
$[d\mathbf{z}]=[\dX'] + [d\mathbf{y}']$, that is associated
with a $\{1\}$-inverse,
%
%
%
\begin{equation}\label{eq:decomp}
  \begin{alignedat}{2}
    &[d\mathbf{x}'] =
     \left(
      [\,\mathbf{I}\,] - [\mathbf{C}]^{(1)}[\mathbf{C}]
     \right)
     [d\mathbf{z}],
     &&
     [d\mathbf{x}']\in 
     \mathcal{N}\left([\mathbf{C}]\right)
    \\
    &[d\mathbf{y}'] =
     \left(
      [\mathbf{C}]^{(1)}[\mathbf{C}]
     \right)
     [d\mathbf{z}],
     &\quad&
     [d\mathbf{y}']\in 
       \mathcal{R}\left([\mathbf{C}]^{(1)}\right)
  \end{alignedat}
\end{equation}
Because it is readily computed, the Moore--Penrose inverse
$[\mathbf{C}]^{\dagger}$ is chosen as the $\{1\}$-inverse,
and because $[\mathbf{C}]$ is assumed to have full row rank,
$[\mathbf{C}]^{\dagger}=
 [\mathbf{C}]^{\text{T}}([\mathbf{C}][\mathbf{C}]^{\text{T}})^{-1}$.
With the Moore-Penrose inverse $[\mathbf{C}]^{\dagger}$,
the decomposition of Eq.~(\ref{eq:decomp}) is orthogonal,
formed from the projections
of $[\dZ]$ onto $\mathcal{L}=\mathcal{N}([\mathbf{C}])$ and its
orthogonal complement
$\mathcal{L}^{\perp}=\mathcal{R}([\mathbf{C}]^{\dagger})$,
\begin{linenomath}
\begin{align}\label{eq:CMP}
  &\left[\mathbf{C}\right]^{\dagger} =
      \left[\mathbf{C}\right]^{\text{T}}
      \left(
        \left[\mathbf{C}\right]
        \left[\mathbf{C}\right]^{\text{T}}
      \right)^{-1}
  \\
  \label{eq:PL}
  &\begin{alignedat}{2}
    &\left[\dX'\right] = 
     \left[\mathbf{P}_{\mathcal{L}}\right]
     [d\mathbf{z}]
     \in\mathcal{L}=\mathcal{N}([\mathbf{C}]),
    &\quad&\left[\mathbf{P}_{\mathcal{L}}\right] =
      \left[\,\mathbf{I}\,\right] -
      \left[\mathbf{C}\right]^{\dagger}
      \left[\mathbf{C}\right]
    \\
    &\left[d\mathbf{y}'\right] =
     \left[\mathbf{P}_{\mathcal{L}^\perp}\right]
     [d\mathbf{z}]
     \in\mathcal{L}^{\perp}=\mathcal{R}([\mathbf{C}]^{\dagger}),
    &&\left[\mathbf{P}_{\mathcal{L}^\perp}\right] =
      \left[\mathbf{C}\right]^{\dagger}
      \left[\mathbf{C}\right]
  \end{alignedat}
\end{align}
\end{linenomath}
Recognizing the second term of Eq.~(\ref{eq:xcdc}) as the
$[\dX']$ in Eqs.~(\ref{eq:decomp}) and~(\ref{eq:PL}), the Eq.~(\ref{eq:xcdc})
is written as
\begin{equation}
  \left[d\mathbf{x}\right] =
  \left[\mathbf{C}\right]^{(1)}
  \left[d\mathbf{c}\right] +
  \left[\dX'\right]
  ,\quad
  \left[\dX'\right]\in\mathcal{N}(\left[\mathbf{C}\right])
\end{equation}
Substituting $[\dX]$ into Eq.~(\ref{eq:Hdxdpdy}) and using
the Moore-Penrose inverse as the $\{1\}$-inverse,
the Eq.~(\ref{eq:Hdxdpdy}) is written as
the constrained linear equation
\begin{equation}
  \left[\mathbf{H}\right]
  \left[\dX'\right]
  -
  \left[d\mathbf{y}\right]
  =
  \left[\dP\right]
  -
  \left[\mathbf{H}\right]
  \left[\mathbf{C}\right]^{\dagger}
  \left[d\mathbf{c}\right]
  ,\quad
  \left[\dX'\right]\in\mathcal{N}(\left[\mathbf{C}\right])
\end{equation}
As shown in \cite{Bott:1953a} and in \cite{BenIsrael2003a}\S 2.10,
the consistency of this equation is equivalent
to the consistency of
\begin{equation}
  \big(
  \left[\mathbf{H}\right]
  \left[\mathbf{P}_{\mathcal{L}}\right]
  +
  \left[\mathbf{P}_{\mathcal{L}^{\perp}}\right]
  \big)
  \left[d\mathbf{z}\right]
  =
  \left[\dP\right]
  -
  \left[\mathbf{H}\right]
  \left[\mathbf{C}\right]^{\dagger}
  \left[d\mathbf{c}\right]
\end{equation}
and consistency is assured when the matrix
$([\mathbf{H}][\mathbf{P}_{\mathcal{L}}]+[\mathbf{P}_{\mathcal{L}^{\perp}}])$
is non-singular.
If the matrix is non-singular, the displacements
$[\dX]$ and reaction forces $[d\mathbf{y}]$ are generated
with the Bott--Duffin inverse of $[\mathbf{H}]$ constrained to
$\mathcal{L}$, written as
$[\mathbf{H}]^{(-1)}_{(\mathcal{L})}$:
\begin{linenomath}
\begin{align}
  &\left[\mathbf{H}\right]^{(-1)}_{(\mathcal{L})}
   =
   \left[\mathbf{P_{\mathcal{L}}}\right]
    \left(
      \left[\mathbf{H}\right]
      \left[\mathbf{P}_{\mathcal{L}}\right]
      +
      \left[\mathbf{P}_{\mathcal{L}^{\perp}}\right]
    \right)^{-1}
  \\
  &\left[\dX'\right]=
    \left[\mathbf{P}_{\mathcal{L}}\right]
    \left[d\mathbf{z}\right]
    =
    \left[\mathbf{H}\right]^{(-1)}_{(\mathcal{L})}
    \left(
      \left[\dP\right]
      -
      \left[\mathbf{H}\right]
      \left[\mathbf{C}\right]^{\dagger}
      \left[d\mathbf{c}\right]
    \right)
    \\
    \label{eq:dxconstrained1}
  &\left[\dX\right]=
    \left[\mathbf{C}\right]^{\dagger}
    \left[d\mathbf{c}\right]
    +
    \left[\mathbf{H}\right]^{(-1)}_{(\mathcal{L})}
    \left(
      \left[\dP\right]
      -
      \left[\mathbf{H}\right]
      \left[\mathbf{C}\right]^{\dagger}
      \left[d\mathbf{c}\right]
    \right)
    \\
    \label{eq:dyconstrained1}
   &\left[d\mathbf{y}\right]
    =
    -\left(
      \left[\dP\right]
      -
      \left[\mathbf{H}\right]
      \left[\mathbf{C}\right]
      \left[d\mathbf{c}\right]
      -
      \left[\mathbf{H}\right]
      \left[\dX'\right]
    \right)
    =
    -\left[\dP\right]
    +
    \left[\mathbf{H}\right]
    \left[d\mathbf{x}\right]
\end{align}
\end{linenomath}
as in \cite{Bott:1953a,Yonglin:1990a,BenIsrael2003a}.
Note that in these equations,
the Moore--Penrose inverse
$[\mathbf{C}]^{\dagger}$ can be replaced with any
$\{1\}$-inverse, $[\mathbf{C}]^{(1)}$.
\par
To the author's knowledge,
the case of a matrix
$([\mathbf{H}][\mathbf{P}_{\mathcal{L}}]+[\mathbf{P}_{\mathcal{L}^{\perp}}])$
that is \emph{singular}
has not been discussed in the mechanics literature
(even though this case leads to Type~II bifurcations,
discussed in Section~\ref{sec:bif}, and the case
occurs in all examples in Section~\ref{sec:examples}).
Whether the matrix is singular or non-singular,
Chen~\cite{Yonglin:1990a} has shown that the solution(s)
of a constrained system of equations, such as
Eq.~(\ref{eq:Hdxdpdy})
subject to the constraint of Eq.~(\ref{eq:nonhomogeneousC}),
can be expressed with a \emph{generalized} Bott--Duffin
inverse.
When
$([\mathbf{H}][\mathbf{P}_{\mathcal{L}}]+[\mathbf{P}_{\mathcal{L}^{\perp}}])$
is non-singular, the solution $[\dX]$ is unique and given
by Eq.~(\ref{eq:dxconstrained1});
otherwise, the solution is non-unique.
The generalized Bott--Duffin inverse proposed by
Chen \cite{Yonglin:1990a},
$[\mathbf{H}]_{(\mathcal{L})}^{(\dagger)}$,
is the product of the projection matrix
$[\mathbf{P}_{\mathcal{L}}]$ and the
Moore--Penrose pseudo-inverse of
$([\mathbf{H}][\mathbf{P}_{\mathcal{L}}]+[\mathbf{P}_{\mathcal{L}^{\perp}}])$,
\begin{equation}\label{eq:GBD}
  \left[\mathbf{H}\right]_{(\mathcal{L})}^{(\dagger)}
  =
  \left[\mathbf{P}_{\mathcal{L}}\right]
  \left(
  \left[\mathbf{H}\right]
  \left[\mathbf{P}_{\mathcal{L}}\right]
  +
  \left[\mathbf{P}_{\mathcal{L}^{\perp}}\right]
  \right)^{\dagger}
\end{equation}
in which ``$(\dagger)$'' designates the Moore--Penrose 
generalized $\{1,2,3,4\}$-inverse.
Unlike the inverse $[\mathbf{C}]^{\dagger}$,
the inverse $[\mathbf{H}]_{(\mathcal{L})}^{(\dagger)}$ 
is computed from the singular value
decomposition of sum
$\left(
  [\mathbf{H}]
  [\mathbf{P}_{\mathcal{L}}]
  +
  [\mathbf{P}_{\mathcal{L}^{\perp}}]
  \right)$
when the sum lacks full rank.
%
With the generalized Bott--Duffin inverse,
the solution of Eqs.~(\ref{eq:nonhomogeneousC})--(\ref{eq:Hdxdpdy}) is
\begin{equation}\label{eq:dxdyGBD}
  \begin{aligned}
    &\left[\dX\right]=
     \left[\mathbf{C}\right]^{\dagger}
     \left[d\mathbf{c}\right]
     +
     \left[\mathbf{H}\right]^{(\dagger)}_{(\mathcal{L})}
     \left(
       \left[\dP\right]
       -
       \left[\mathbf{H}\right]
       \left[\mathbf{C}\right]^{\dagger}
       \left[d\mathbf{c}\right]
     \right)
     +
     \left[\mathbf{P}_{\mathcal{L}}\right]
     \left[\mathbf{P}_{\mathcal{N}
([\mathbf{H}][\mathbf{P}_{\mathcal{L}}]+[\mathbf{P}_{\mathcal{L}^{\perp}}])}\right]
     \left[d\mathbf{v}\right]
     \\
     &\left[d\mathbf{y}\right]
     =
     -\left[\dP\right]
     +
     \left[\mathbf{H}\right]
     \left[d\mathbf{x}\right]
  \end{aligned}
\end{equation}
where $[d\mathbf{v}]$ is an arbitrary vector, and
$[\mathbf{P}_{\mathcal{N}
([\mathbf{H}][\mathbf{P}_{\mathcal{L}}]+[\mathbf{P}_{\mathcal{L}^{\perp}}])}]$
is the projection matrix onto the null space of
$([\mathbf{H}][\mathbf{P}_{\mathcal{L}}]+[\mathbf{P}_{\mathcal{L}^{\perp}}])$
(see \cite{Yonglin:1990a}).
If
$([\mathbf{H}][\mathbf{P}_{\mathcal{L}}]+[\mathbf{P}_{\mathcal{L}^{\perp}}])$
is non-singular, then
$[\mathbf{H}]_{(\mathcal{L})}^{(\dagger)} = [\mathbf{H}]_{(\mathcal{L})}^{(-1)}$;
the null space is simply the zero vector $[\mathbf{0}]$;
the final term in
Eq.~(\ref{eq:dxdyGBD}\textsubscript{1}) can be ignored;
and the solution in Eq.~(\ref{eq:dxdyGBD}) coincides with that of
Eqs.~(\ref{eq:dxconstrained1})--(\ref{eq:dyconstrained1}).
%
%
%
\par
For a granular system that is subject to displacement constraints,
the matrix
$([\mathbf{H}][\mathbf{H}_{\mathcal{L}}]+[\mathbf{H}_{\mathcal{L}^{\perp}}])$
plays the central role in finding $[\dX]$
and in resolving questions of
uniqueness and stability.
Recognizing this centrality,
the Problem~1 of Eq.~({\ref{eq:form1}}) is restated in the
following equivalent form:
\begin{linenomath}
\begin{subequations}\label{eq:form1b}
\begin{align}
  \text{Problem~1: }
  & \text{find } \left[d\mathbf{z}\right] \text{such that}
  \\
  &
    \left(
      \left[\mathbf{H}\right]
      \left[\mathbf{P}_{\mathcal{L}}\right]
      +
      \left[\mathbf{P}_{\mathcal{L}^{\perp}}\right]
    \right)
    \left[d\mathbf{z}\right]
    =
    \left[d\mathbf{p}\right]
    -
    \left[\mathbf{H}\right]
    \left[\mathbf{C}\right]^{\dagger}
    \left[d\mathbf{c}\right]
  \\
  & \left[\mathbf{H}\right] \in
    \left\{
      \left[\mathbf{H}^{i}\right]
      ,\;
      i = 1, 2, \ldots , B
    \right\}
  ,\quad
  \left[d\mathbf{x}\right]
  =
  \left[\mathbf{C}\right]^{\dagger}
  \left[d\mathbf{c}\right] +
  \left[\mathbf{P}_{\mathcal{L}}\right]
  \left[d\mathbf{z}\right]
    \in\Omega^{i} \subset \mathbb{R}^{6N}
  \\
  & \bigcup_{i=1}^{B}\Omega^{i}
    =
   \mathbb{R}^{6N}
   ,\quad
   \Omega^{i}\cap\Omega^{j} = \varnothing, i\neq j
\end{align}
\end{subequations}
\end{linenomath}
where
the term $[\mathbf{P}_{\mathcal{L}}][\dZ]$ in Eq.~(\ref{eq:form1b}c)
is meant to encompass the null-space
$\mathcal{N}([\mathbf{H}^{i}][\mathbf{P}_{\mathcal{L}}] +
[\mathbf{P}_{\mathcal{L}^{\perp}}])$
when the matrix
$([\mathbf{H}^{i}][\mathbf{H}_{\mathcal{L}}]+[\mathbf{H}_{\mathcal{L}^{\perp}}])$
is singular.
\par
Regardless of the form,
one is reminded by either Eq.~(\ref{eq:form1}c) or~(\ref{eq:form1b}c)
that the operator $[\mathbf{H}]$ is incrementally nonlinear, and
each branch $[\mathbf{H}^{i}]$ is only relevant for $[\dX]$
that lie within its domain $\Omega^{i}$.
Any apparent solution of
Eq.~(\ref{eq:form1}b) or~(\ref{eq:form1b}b)
must satisfy this condition
(see Eq.~\ref{eq:condition}).
\section{\normalsize Formulating Problems~2 and~3}\label{sec:formulate2}
Problems~2 and~3 require construction of
matrix $[\Mbar]$ and vector $[\dQbar]$.
These components are now assembled from the stiffness elements
and from the constraint projections that were derived
in the previous section.
\subsection{Assembly stiffness for Problems~2 and~3}
As an intermediate step in constructing
$[\Mbar]$ and $[\dQbar]$,
consider the following mixed linear complementarity problem:
\begin{linenomath}
\begin{subequations}\label{eq:form2b}
\begin{align}
  \renewcommand*{\arraystretch}{1.2}
%
  \left[
      \begin{array}{@{\extracolsep{0ex}}c;{1pt/1pt}c@{\extracolsep{\fill}}}
        \mathbf{H}_{\mathbf{xx}}
        & \mathbf{H}_{\mathbf{x}\boldsymbol{\lambda}}
        \\\hdashline[1pt/1pt]
        \mathbf{H}_{\boldsymbol{\lambda}\mathbf{x}}
        & \mathbf{H}_{\boldsymbol{\lambda}\boldsymbol{\lambda}}
        \\\hdashline[1pt/1pt]
        \mathbf{C} & \mathbf{0}
        \\\hdashline[1pt/1pt]
        \mathbf{0}
        & \mathbf{I}
      \end{array}
  \right]
  \left[
    \begin{array}{l@{\extracolsep{\fill}}}
        \dX\,
        \\\hdashline[1pt/1pt]
        \dLambda\,
    \end{array}
  \right]
  &
  \begin{array}{c}
    =
    \\
    =
    \\
    =
    \\
    \ge
  \end{array}
  \left[
    \begin{array}{c@{\extracolsep{\fill}}}
     \dP
     \\\hdashline[1pt/1pt]
     d\bar{\mathbf{G}}\ge\mathbf{0}
     \\\hdashline[1pt/1pt]
     d\mathbf{c}
     \\\hdashline[1pt/1pt]
     \mathbf{0}
    \end{array}
  \right]
  \\
  \left[\dLambda\right]^{\text{T}}
  \Big[
      \begin{array}{@{\extracolsep{0ex}}c;{1pt/1pt}c@{\extracolsep{\fill}}}
        \mathbf{H}_{\boldsymbol{\lambda}\mathbf{x}}
        & \mathbf{H}_{\boldsymbol{\lambda}\boldsymbol{\lambda}}
      \end{array}
  \Big]
  \left[
    \begin{array}{l@{\extracolsep{\fill}}}
        \dX\,
        \\\hdashline[1pt/1pt]
        \dLambda\,
    \end{array}
  \right]
  &\, = \; 0
\end{align}
\end{subequations}
\end{linenomath}
The first row in Eq.~(\ref{eq:form2b}a) assures equilibrium
in the displaced condition and is the counterpart of
Eq.~(\ref{eq:form1}b) of Problem~1.
The second row collects the consistency conditions
of the $M_{\lambda}$ active (yielding) contacts,
as in the expression for $d\bar{G}^{pq}_{\text{s}}$
in Eq.~(\ref{eq:dfForm2}).
The third and fourth rows of Eq.~(\ref{eq:form2b}a)
are recognized as the displacement constraints of
Eq.~(\ref{eq:form1}c) and the $\dLambda$-orthant
of Eq.~(\ref{eq:form2}c), respectively, and
in which vector $[\dLambda]\in\mathbb{R}^{M_{\lambda}}$
collects the $M_{\lambda}$ 
scalars $d\lambda^{pq}$
of active contacts.
\par
As before, vectors
$[\dX],[\dP]\in\mathbb{R}^{6N}$ and $[\dC]\in\mathbb{R}^{R}$.
But instead of 
the material-derived (mechanical) contact forces
$[\mathfrak{d}\mathbf{f} / \mathfrak{d}\mathbf{m}]$
being given by Eq.~(\ref{eq:ContactF}),
as used in Problem~1,
the forces are now linear functions of
the contact deformations
$[\delta\mathbf{u}^{\text{def}} / \delta\boldsymbol{\theta}^{\text{def}}]$
and of the
active variables $[\dLambda]$,
as in Eq.~(\ref{eq:dfForm2}).
The dependence of force on the contact deformations is embodied
in sub-matrix $[\mathbf{H}_{\mathbf{xx}}]$;
whereas dependence on the active $\lambda$ variables is manifested
in $[\mathbf{H}_{\mathbf{x}\boldsymbol{\lambda}}]$.
\par
Similar to stiffness $[\mathbf{H}]$ in Eqs.~(\ref{eq:form1}b)
and~(\ref{eq:Himg}),
sub-matrix $[\mathbf{H}_{\mathbf{xx}}]$ has both
mechanical and geometric parts.
The mechanical part originates from
the elastic contact stiffnesses in
Eq.~(\ref{eq:dfForm2}), 
and these individual stiffnesses
are assembled for the assembly,
\begin{equation}\label{eq:Felastdu}
  \left[
    \begin{array}{@{\extracolsep{\fill}}c@{\extracolsep{\fill}}}
      \Felastpq_{3\times 6} \\[0.1ex]\hdashline[1pt/1pt]
      \rule{0ex}{2.4ex}\Melastpq_{3\times 6}
    \end{array}
  \right]_{6\times 6}
  \left[
    \begin{array}{@{\extracolsep{\fill}}c@{\extracolsep{\fill}}}
      \Dudefpq \\[0.1ex]\hdashline[1pt/1pt]
      \rule{0ex}{2.4ex}\Dthetadefpq
    \end{array}
  \right]_{6\times 1}
  \rightsquigarrow\;
%
%
  \def\arraystretch{1.00}
  \left[
    \begin{array}{@{\extracolsep{\fill}}c@{\extracolsep{\fill}}}
      \mathbf{F}^{\text{elastic}} \\\hdashline[1pt/1pt]
      \rule{0ex}{2.4ex}\mathbf{M}^{\text{elastic}}
    \end{array}
  \right]_{2(6M)\times 2(6M)}
  \def\arraystretch{1.20}
  \left[
    \begin{array}{@{\extracolsep{\fill}}c@{\extracolsep{\fill}}}
      \delta\mathbf{u}^{\text{def}} \\\hdashline[1pt/1pt]
      \rule{0ex}{2.4ex}\delta\boldsymbol{\theta}^{\text{def}}
    \end{array}
  \right]
\end{equation}
%
The product
$[\mathbf{B}][\dX]$
is substituted for
$[\delta\mathbf{u}^{\text{def}}/\delta\boldsymbol{\theta}^{\text{def}}]=
[\delta\mathbf{x}^{\text{def}}]$,
and the contact forces are totaled with the statics matrix $[\mathbf{A}]$
to find
the mechanical, elastic $\left[ \mathbf{H}_{\mathbf{xx}} \right]$ matrix,
%
\begin{equation}\label{eq:Hxxel}
  \left[\mathbf{H}^{\text{m, elastic}}_{\mathbf{xx}}\right]
  =
  \left[ \mathbf{A} \right]_{6N\times 2(6M)}
  \def\arraystretch{1.00}
  \left[
    \begin{array}{@{\extracolsep{\fill}}c@{\extracolsep{\fill}}}
      \mathbf{F}^{\text{elastic}} \\\hdashline[1pt/1pt]
      \rule{0ex}{2.4ex}\mathbf{M}^{\text{elastic}}
    \end{array}
  \right]_{2(6M)\times 2(6M)}
  \left[ \mathbf{B} \right]_{2(6M)\times 6N}
\end{equation}
(compare with Eq.~\ref{eq:Hmprod}),
and this elastic stiffness is added to 
the geometric stiffness $[\mathbf{H}^{\text{g}}]$
to arrive at the full sub-matrix $[\mathbf{H}_{\mathbf{xx}}]$,
\begin{equation}\label{eq:Hxx1}
  \left[ \mathbf{H}_{\mathbf{xx}} \right]_{6N\times 6N}
  =
  \left[
    \mathbf{H}^{\text{m,elastic}}_{\mathbf{xx}}
  \right]
  +
  \left[\mathbf{H}^{\text{g}}\right]
\end{equation}
where the geometric stiffness $[\mathbf{H}^{\text{g}}]$ is given in
Eqs.~(\ref{eq:Hg1})--(\ref{eq:Hg4}) and~(\ref{eq:Hg}).
\par
Matrix $[\mathbf{H}_{\mathbf{x}\boldsymbol{\lambda}}]$
derives from the dependence of contact forces,
$[\mathfrak{d}\mathbf{f} / \mathfrak{d}\mathbf{m}]$,
on the active variables $[\dLambda]$,
and the $6N\times M_{\lambda}$ matrix
has one column for each active contact mechanism
(i.e., contacts with $G_{\text{s}}^{pq}=0$,
Section~\ref{sec:linfrict}).
Referring to Eq.~(\ref{eq:dfdlambda}), 
the contact force is 
reduced by the product
$(\Felastpq\cdot\mathbf{h}^{pq}_{\text{s}})\Lambdas$.
The products for individual contacts
are gathered into a matrix product, 
$[\mathbf{L}][d\boldsymbol{\lambda}]$,
with one row in $[\mathbf{L}]$
for each variant, $pq$ and $qp$,
of the $M$ contacts and their six force components:
\begin{equation}\label{eq:LdLambda}
  -\left(\Felastpq\cdot\mathbf{h}^{pq}_{\text{s}}\right)\Lambdas
  \;\rightsquigarrow\;
  \left[
    \mathbf{L}
  \right]_{2(6M)\times M_{\lambda}}
  \left[
   \dLambda
  \right]_{M_{\lambda}\times 1}
\end{equation}
The only non-zero rows in $[\mathbf{L}]$ are
those that correspond to 
active contacts, with $G^{pq}_{\text{s}}=0$; whereas, the
rows in $[\mathbf{L}]$ that correspond to
\emph{inactive} contacts, unaffected
by the $d\lambda^{pq}$, are filled with zeros.
To sum the contact forces and compute particle forces,
the statics matrix is multiplied
by $[\mathbf{L}]$, yielding
matrix $[\mathbf{H}_{\mathbf{x}\boldsymbol{\lambda}}]$:
\begin{equation}\label{eq:gatherSl}
  \left[\mathbf{H}_{\mathbf{x}\boldsymbol{\lambda}}\right]
  =
  \left[\mathbf{A}\right]
  \left[\mathbf{L}\right]
\end{equation}
\par
The sub-matrices
$[\mathbf{H}_{\boldsymbol{\lambda}\mathbf{x}}]$
and $[\mathbf{H}_{\boldsymbol{\lambda}\boldsymbol{\lambda}}]$
in Eq.~(\ref{eq:LdLambda}b)
bring about incremental changes $d\bar{G}^{pq}_{\text{s}}$
that are consistent with
the yield conditions of the $M_{\lambda}$
active contacts.
Considering the increments $d\bar{G}^{pq}_{\text{s}}$ in
Eq.~(\ref{eq:dGs}), 
matrix
$[\mathbf{H}_{\boldsymbol{\lambda}\mathbf{x}}]$ is assembled
from vectors $\mathbf{g}^{pq}_{\text{s}}$ of individual active
contacts, which are multiplied by
the contact movements $\Dxdefpq$;
whereas,
$[\mathbf{H}_{\boldsymbol{\lambda}\boldsymbol{\lambda}}]$
is assembled from the contacts' scalar products
$\mathbf{g}^{pq}_{\text{s}}\cdot\mathbf{h}^{pq}_{\text{s}}$,
which are multiplied by the increments
$d\lambda^{pq}_{s}$.
As such, matrices
$[\mathbf{H}_{\boldsymbol{\lambda}\mathbf{x}}]$ and
$[\mathbf{H}_{\boldsymbol{\lambda}\boldsymbol{\lambda}}]$
are constructed as
\begin{linenomath}
\begin{align}
  &-\mathbf{g}^{pq}_{\text{s}} \cdot \delta\mathbf{x}^{\text{def, }pq}
  \;\rightsquigarrow\;
  \left[\bar{\mathbf{G}}\right]_{M_{\lambda}\times 2(6M)}
  \left[\delta\mathbf{x}^{\text{def}}\right]
  \\
  &[\mathbf{H}_{\boldsymbol{\lambda}\mathbf{x}}]
  =
  \left[\bar{\mathbf{G}}\right]\left[\mathbf{B}\right]
  \\
  \label{eq:gdotl}
  &\mathbf{g}^{pq}_{\text{s}}\cdot\mathbf{h}^{pq}_{\text{s}}
  \;\rightsquigarrow\;
  \left[\mathbf{H}_{\boldsymbol{\lambda}\boldsymbol{\lambda}}\right]
\end{align}
\end{linenomath}
where the changes in sign, 
$\mathbf{g}^{pq}_{\text{s}}\rightarrow -\mathbf{g}^{pq}_{\text{s}}$
and $[\bar{\mathbf{G}}]=[-\mathbf{G}]$,
are consistent with the standard form
of Eq.~(\ref{eq:form2}b).
Note that the intermediate matrix
$[\bar{\mathbf{G}}]$ is multiplied by the kinematics
matrix $[\mathbf{B}]$, which acts upon the particle
movements $[\dX]$ to yield the contact
movements $[\delta\mathbf{x
}^{\text{def}}]$.
Also note that matrix $[\mathbf{H}_{\boldsymbol{\lambda}\boldsymbol{\lambda}}]$
is diagonal.
\subsection{Final form of Problems~2 and~3}\label{sec:p12final}
%
Matrix $[\Mbar]$ and vector $[\dQbar]$
embody all of the essential information
of Problems~2 and~3:
the current contact forces and orientations,
the contacts' stiffnesses,
the applied loads, the contacts' curvatures,
and the controlled displacements.
The intermediate Eq.~(\ref{eq:form2b}) is used
in deriving $[\Mbar]$ and $[\dQbar]$,
by noting the
two equalities present in the equation's
first and third rows: 
\begin{equation}\label{eq:hxxdx}
  \begin{aligned}
    &\left[\mathbf{H}_{\mathbf{xx}}\right]
    \left[\dX\right]
    +
    \left[\mathbf{H}_{\mathbf{x}\boldsymbol{\lambda}}\right]
    \left[d\boldsymbol{\lambda}\right]
    =
    \left[\dP\right]
    +
    \left[d\mathbf{y}\right]
    \\
    &\left[\mathbf{C}\right]
    \left[\dX\right]
    =
    \left[d\mathbf{c}\right]
  \end{aligned}
\end{equation}
where we now include the reaction forces $[d\mathbf{y}]$
that are needed to bring the total external force
$[\dP]+[\dY]$ into
the subspace of $\mathbb{R}^{6N}$ that is
permitted by the displacement constraints
of Eq.~(\ref{eq:hxxdx}\textsubscript{2}).
The two inequalities of Eq.~(\ref{eq:form2b}a),
given in its second and fourth rows, are considered later.
\par
To begin, the two
Eqs.~(\ref{eq:hxxdx}) are treated in a similar manner
as in Section~\ref{sec:FinalForms1},
noting that matrix $[\mathbf{H}_{\mathbf{xx}}]$ takes the place
of $[\mathbf{H}]$,
and the first equality now includes the
product $[\mathbf{H}_{\mathbf{x}\boldsymbol{\lambda}}][\dLambda]$.
The displacement solution $[\dX]$ and reaction forces $[d\mathbf{y}]$ are
now found with the Bott--Duffin inverse of
$[\mathbf{H}_{\mathbf{xx}}]$ constrained to $\mathcal{L}$:
\begin{linenomath}
\begin{align}
  &\left[\mathbf{H}_{\mathbf{xx}}\right]^{(-1)}_{(\mathcal{L})}
   =
   \left[\mathbf{P_{\mathcal{L}}}\right]
    \left(
      \left[\mathbf{H}_{\mathbf{xx}}\right]
      \left[\mathbf{P}_{\mathcal{L}}\right]
      +
      \left[\mathbf{P}_{\mathcal{L}^{\perp}}\right]
    \right)^{-1}
  \\
  \label{eq:dx23}
  &\left[\dX\right]=
    \left[\mathbf{C}\right]^{\dagger}
    \left[d\mathbf{c}\right]
    +
    \left[\mathbf{H}_{\mathbf{xx}}\right]^{(-1)}_{(\mathcal{L})}
    \left(
      \left[\dP\right]
      -
      \left[\mathbf{H}_{\mathbf{x}\boldsymbol{\lambda}}\right]
      \left[\dLambda\right]
      -
      \left[\mathbf{H}_{\mathbf{xx}}\right]
      \left[\mathbf{C}\right]^{\dagger}
      \left[d\mathbf{c}\right]
    \right)
    \\
   \label{eq:dy23}
   &\left[d\mathbf{y}\right]
    =
    -\left[\dP\right]
    +
    \left[\mathbf{H}_{\mathbf{xx}}\right]
    \left[d\mathbf{x}\right]
    +
    \left[\mathbf{H}_{\mathbf{x}\boldsymbol{\lambda}}\right]
    \left[d\boldsymbol{\lambda}\right]
\end{align}
\end{linenomath}
Substituting this expression for $[\dX]$ into the
inequality in the second row of Eq.~(\ref{eq:form2b}a),
\begin{linenomath}
\begin{multline}
    \left[\mathbf{H}_{\boldsymbol{\lambda}\mathbf{x}}\right]
    \left[\mathbf{H}_{\mathbf{xx}}\right]^{(-1)}_{(\mathcal{L})}
    \left[\dP\right]
    +
    \left[\mathbf{H}_{\boldsymbol{\lambda}\mathbf{x}}\right]
    \left(
      \left[\,\mathbf{I}\,\right]
      -
      \left[\mathbf{H}_{\mathbf{xx}}\right]^{(-1)}_{(\mathcal{L})}
      \left[\mathbf{H}_{\mathbf{xx}}\right]
    \right)
    \left[\mathbf{C}\right]^{\dagger}
    \left[d\mathbf{c}\right]
    \\
    +
    \left(
      \left[\mathbf{H}_{\boldsymbol{\lambda}\boldsymbol{\lambda}}\right]
      -
      \left[\mathbf{H}_{\boldsymbol{\lambda}\mathbf{x}}\right]
      \left[\mathbf{H}_{\mathbf{xx}}\right]^{(-1)}_{(\mathcal{L})}
      \left[\mathbf{H}_{\mathbf{x}\boldsymbol{\lambda}}\right]
    \right)
    \left[\dLambda\right]
    \;\ge\; 0
\end{multline}
\end{linenomath}
In this inequality, one recognizes
matrix $[\Mbar]$ and vector $[\dQbar]$ of the
\LCP\ of Eq.~(\ref{eq:form2}),
\begin{linenomath}
\begin{align}
  \label{eq:Mbar}
  &\left[\Mbar\right]
   =
   \left[\mathbf{H}_{\boldsymbol{\lambda}\boldsymbol{\lambda}}\right]
   -
   \left[\mathbf{H}_{\boldsymbol{\lambda}\mathbf{x}}\right]
   \left[\mathbf{H}_{\mathbf{xx}}\right]^{(-1)}_{(\mathcal{L})}
   \left[\mathbf{H}_{\mathbf{x}\boldsymbol{\lambda}}\right]
  \\
  \label{eq:dQbar}
  &\left[\dQbar\right]
   =
   \left[\mathbf{H}_{\boldsymbol{\lambda}\mathbf{x}}\right]
    \left[\mathbf{H}_{\mathbf{xx}}\right]^{(-1)}_{(\mathcal{L})}
    \left[\dP\right]
    +
    \left[\mathbf{H}_{\boldsymbol{\lambda}\mathbf{x}}\right]
    \left(
      \left[\,\mathbf{I}\,\right]
      -
      \left[\mathbf{H}_{\mathbf{xx}}\right]^{(-1)}_{(\mathcal{L})}
      \left[\mathbf{H}_{\mathbf{xx}}\right]
    \right)
    \left[\mathbf{C}\right]^{\dagger}
    \left[d\mathbf{c}\right]
\end{align}
\end{linenomath}
as required in Problems~2 and~3.
Once solved,
$[\dLambda]$ is substituted in
Eqs.~(\ref{eq:dx23}) and~(\ref{eq:dy23}) to
find the movements $[\dX]$ and reaction forces $[\dY]$.
\par
Vector $[\dQbar]$ is seen to
depend on increments of both the applied loading $[\dP]$ and
imposed displacements $[\dC]$; whereas,
matrix $[\Mbar]$ is derived from the four
``$\mathbf{H}$'' submatrices in
Eq.~(\ref{eq:form2b}a),
as modified by the displacement constraints.
The $[\mathbf{H}_{\boldsymbol{\lambda}\boldsymbol{\lambda}}]$
in Eq.~(\ref{eq:Mbar})
is diagonal,
but the matrix $[\Mbar]$ is rarely symmetric:
even if matrix
$[\mathbf{H}^{\text{m,elastic}}_{\mathbf{xx}}]$
in Eq.~(\ref{eq:Hxx1}) is symmetric,
the geometric contribution
$[\mathbf{H}^{\text{g}}]$ is usually non-symmetric,
and the product
$[\mathbf{H}^{\text{m,elastic}}_{\mathbf{xx}}][\mathbf{P}_{\mathcal{L}}]$
is non-symmetric even if both matrices are symmetric, thus
making
$[\mathbf{H}_{\mathbf{xx}}]^{(-1)}_{(\mathcal{L})}$
non-symmetric.
Furthermore, friction does not obey the normality condition of
plasticity, so usually
$[\mathbf{H}_{\boldsymbol{\lambda}\mathbf{x}}]^{\text{T}}\neq
[\mathbf{H}_{\mathbf{x}\boldsymbol{\lambda}}]$,
another source of asymmetry of $[\Mbar]$
(compare $\mathbf{F}^{\text{elastic, }pq}\cdot\mathbf{h}^{pq}$
and $\mathbf{g}^{pq}$ in Eqs.~\ref{eq:dfdlambda} and~\ref{eq:dfForm2}).
\section{\normalsize Stiffness pathologies}\label{sec:pathologies}
A \emph{stiffness pathology}
is a condition allowing aberrant transitions,
perhaps to an extent considered failure,
in the particles' displacements $[\dX]$,
either during a prescribed loading path 
or, with some pathologies, during a suspension of loading.
The various pathologies are summarized in Table~\ref{table:summary1}
and are described below.
Besides granular materials,
these pathologies apply to other systems  and materials,
such as those with internal friction, plasticity,
and other irreversible mechanisms
that lead to complementarity conditions.
\par
A pathology can be one of \emph{path} or of \emph{state}
\cite{Petryk:1991a,Petryk:1993a}:
a pathology of path
depends upon the control program $[\mathbf{C}]$ 
and on the control parameters,
$[\dP]$ and $[\dC]$;
whereas, a pathology of state disregards further loading
and is dependent only on the control program.
That is,
a pathology of path depends both upon the stiffness $[\mathbf{H}^{i}]$,
the control program $[\mathbf{C}]$,
and the control parameters $[\dP]$ and $[\dC]$; whereas,
a pathology of state depends only on the $[\mathbf{H}^{i}]$
and $[\mathbf{C}]$.
With Problems~2 and~3,
the control program is conveyed by $[\Mbar]$, and
the control parameters are conveyed by $[\dQbar]$,
which derives from parameters $[\dP]$ and $[\dC]$
(Eqs.~\ref{eq:Mbar}--\ref{eq:dQbar}).
Because loading increments are assumed to be infinitesimal
and the \LCP\ is positive-homogeneous in $[\dQbar]$,
it is the direction of the loading parameters that is relevant.
%
\par
Although stiffness pathologies
can be ascertained, in principle, with any of the three
Problems, Table~\ref{table:summary1}
gives sensible guidance.
Problems~1 and~2 are most appropriate for most pathologies, since
conventional methods for solving the QP of Problem~3 involve
minimization techniques that find a single solution, even
when multiple solutions exist.
The pathology of path-sensitivity, proposed herein,
is only defined in the context of the LCP of Problem~2.
%
%
\begin{table}
  \centering
  \small
  \caption{Summary of eight stiffness pathologies.
           \label{table:summary1}}
  \begin{threeparttable}
  \begin{tabular}{lccp{8cm}}
    \toprule
    Pathology & Type & Problems & Description and criteria
    \\
    \midrule
    Bifurcation, Type~I & Path & 1, 2 & 
      Multiple isolated solutions of
      Eqs.~(\ref{eq:form1}), (\ref{eq:form2}), or (\ref{eq:form3}).
      Each Type~I solution has a single direction an magnitude.
    \\
    Bifurcation, Type~II & Path & 1, 2 & 
      Non-isolated bounded solutions of
      Eqs.~(\ref{eq:form1}), (\ref{eq:form2}), or (\ref{eq:form3}).
      Each Type~II solution is a bounded range of solutions.
    \\
    Bifurcation, Type~III & Path & 1, 2 & 
      Non-isolated unbounded solutions with non-R\textsubscript{0}
      $[\Mbar]$.
      Each Type~II is an unbounded range of solutions.
    \\
    Path instability & Path & 1, 2 &
        A form of instability in which
        multiple equilibrium solutions exist,
        but in which there are path-unstable
        solutions $[\dX^{\alpha}]$ having a
        path entropy $\mathcal{I}_{2}$
        less than the maximum
        (Eqs.~\ref{eq:I2}--\ref{eq:PI2}).
    \\
    Path sensitivity & Path & 2 &
        Path-sensitive solutions are those that
        can change (or be eliminated) with slight perturbations
        of the system.
        Solution $[\dLambda^{\ast}]$ and matrix $[\Mbar]$
        for which ``$\nRightarrow$'' applies in Eq.~(\ref{eq:sens2}).
    \\
    Path incongruity & Path & 1, 2, 3 &
      Absence of a solution for the given control program
      and control parameters (i.e., 
      absence of solutions for the given $[\dQbar]$).
    \\
    Neutral equil. & State & 1, 2 &
        A condition of non-uniqueness in which unbounded
        solutions are available with neutral loading
        under the given control program.
        Occurs when
        matrix $[\mathbf{H}^{i}]$ satisfies Eq.~(\ref{eq:NE}),
        or  matrix $[\Mbar]$ is
        non-R\textsubscript{0}.
    \\
    Unsustainability & State & 1, 2 &
        A potential for spontaneous movement of a system
        in equilibrium under a given control program.
        Occurs when
        matrix $[\mathbf{H}^{i}]$ satisfies Eq.~(\ref{eq:Ins1})
        or conditions of Eq.~(\ref{eq:instabProb2a}).
    \\
    \bottomrule
  \end{tabular}
%
  \end{threeparttable}
\end{table}
\subsection{Bifurcation: Types~I, II and~III}\label{sec:bif}
Rather than accepting bifurcation as an abrupt change
in behavior,
bifurcation is defined herein as the
existence of multiple equilibrium solutions, $[\dX]$ or $[\dLambda]$,
that emanate from the current equilibrium state for a given
loading path~---
for a given control program $[\mathbf{C}]$ together with
given control parameters, $[\dP]$ and $[\dC]$ \cite{Chambon:2004a}.
Bifurcation, which is a loss of uniqueness, involves
a choice among multiple \emph{equilibrium paths},
and, as such, is a pathology of path rather than of state.
Although bifurcated paths can be found by solving any of
the three Problems,
applying the rules of Section~\ref{sec:exist} to an LCP 
most directly determines whether a system admits multiple solutions.
As a separate matter, a bifurcated solution can be either stable or unstable,
which is addressed in Section~\ref{sec:PI1}.
\par
Bifurcation paths are of three types:
(I) isolated, (II) continuous but bounded,
and (III) continuous and unbounded. 
Distinctions among the three types are most easily
identified in the context of the \LCP\ of Problem~2.
For example, Type~1 solutions are simply ones that are isolated,
which correspond to a non-singular
submatrix $[\Mbar_{I_{2}^{\ast},I_{2}^{\ast}}]$
(rules~12--13 \cite{Murty:1972a}, or alternatively,
to a non-singular complementary matrix $[\Abar]$, as
in the algorithm of \ref{sec:enum}).
In the examples of Fig.~\ref{fig:ExampleLCPs} and Table~\ref{table:ExampleLCPs},
several cases are given with multiple bifurcation solutions,
and except for Case~D of Fig.~\ref{fig:ExampleLCPs}d,
the bifurcation paths are isolated.
\par
With Problem~1,
a Type~I isolated bifurcation path occurs when the matrix
$([\mathbf{H}^{i}][\mathbf{P}_{\mathcal{L}}]+[\mathbf{P}_{\mathcal{L}^{\perp}}])$
of an $i$th branch
is \emph{non-singular} and the solution $[\dX]$ lies within
the branch's domain $\Omega^{i}$.
When multiple isolated paths are present,
they occur as solutions of Problem~1
for \emph{different} stiffness
branches $[\mathbf{H}^{i}]$.
\cite{Shanley:1947a,Hill:1960a}).
%
\par
In the context of Problem~2,
a Type~II, continuous-bounded bifurcation is a non-isolated
solution and occurs as a
continuous compact set of solutions when the
submatrix $[\Mbar_{I_{2}^{\ast},I_{2}^{\ast}}]$
is singular (rules~12--15 of Section~\ref{sec:exist}).
As such, a Type~II bifurcation is only possible when
the matrix $[\Mbar]$ is degenerate
(rule~10),
allowing solutions along a compact edge or face of
lower dimension in
$d\boldsymbol{\lambda}$-space.
The dimension of the solution set is given by
rule~14, and the solution set is bounded as the convex hull of
its extreme-points, described in rule~15.
A continuous-bounded solution is illustrated in Case~D of
Fig.~\ref{fig:ExampleLCPs}d.
In this case, the index set $J=I$
in Eq.~(\ref{eq:comp})
has the complementary
matrix $[\bar{\mathbf{A}}]=-[\Mbar]$, which is singular, allowing
for non-isolated solutions along the continuous line
segment (of dimension~1) between $(0,1)$ and $(1,0)$.
(Note that this case also has an isolated, 
Type~I bifurcation solution at $(0,0)$.)
\par
In the context of Problem~1,
a Type~II continuous-bounded bifurcation occurs when the $i$th matrix
$([\mathbf{H}^{i}][\mathbf{P}_{\mathcal{L}}]+[\mathbf{P}_{\mathcal{L}^{\perp}}])$
is \emph{singular} and a continuous set of solutions
$[\dX]$
is available
within the branch's domain $\Omega^{i}$,
such that $[\dX]\subset\Omega^{i}$.
The continuous set of a Type~II
solution is confined to $\Omega^{i}$ and is
bounded within $\Omega^{i}$.
This condition corresponds to a base solution
(the first two terms on the right of Eq.~\ref{eq:dxdyGBD}\textsubscript{1}),
which is found with the generalized Bott--Duffin inverse,
plus the continuous $[d\mathbf{v}]$ part.
\par
A Type~III continuous-unbounded bifurcation occurs if and only if
matrix $[\Mbar]$ is non-R\textsubscript{0},
since the solution set $\mathcal{S}(\dQbar,\Mbar)$ becomes
unbounded
(rule~16).
The same rules of Type~II bifurcations also apply (rules~12--15),
but the set of extreme points in $\dLambda$-space
of a Type~III bifurcation are insufficient to encompass
the unbounded $\mathcal{S}(\dQbar,\Mbar)$.
%
\par
Of the three types of bifurcation,
only Type~III has an analog with
linear equations, $[\mathbf{A}][\mathbf{x}]=[\mathbf{b}]$:
when $[\mathbf{A}]$ is singular
(the analog of a degenerate $[\Mbar]$),
an unbounded solution set applies for certain
$[\mathbf{b}]$ (the analog of $[\dQbar]$).
However,
analogs of Types~I and~II bifurcations are not found with linear
equations, as these types only arise in the context
of incrementally non-linear problems,
which give rise to the inequality constraints of Problem~2.
\subsection{Path instability}\label{sec:PI1}
%
When multiple equilibrium (bifurcation)
solutions (paths) exist that
satisfy the control program and control parameters,
one or more solutions
can follow favored, stable paths,
with the remaining solutions being unstable
and tending to redirect toward a stable path.
An example is the Shanley column with an elasto-plastic hinge,
for which three equilibrium branches exist
\cite{Shanley:1947a,Bazant:1988b}.
An elasto-plastic column that is loaded
to the tangent limit can remain straight with
continued loading $d\mathbf{p}$
(i.e., the fundamental deformation $d\mathbf{x}^{0}$),
or the column can buckle to the left or to the right.
The fundamental (straight) deformation exhibits path instability,
as the column is inclined to buckle by the criterion
given below. 
\par
Hill \cite{Hill:1961a} and
Ba\v{z}ant \cite{Bazant:1988b,Bazant:1989b}
addressed the issue
in a continuum setting
by deriving the second-order
increments of internal entropy.
For a granular material,
this increment,
$\mathcal{I}_{2}([\dX^{\alpha}])$,
along the $\alpha$th solution path $[\dX^{\alpha}]$
is half
the difference of the second-order work done
by the increments of reaction forces $[\dY^{\alpha}]$
and of applied forces $[\dP]$
for the solution:
%
\begin{equation}\label{eq:I2}
  \mathcal{I}_{2}\left(\left[\dX^{\alpha}\right]\right)
  =
  {\textstyle\frac{1}{2}}
  \left(
    \left[\dY^{\alpha}\right]^{\text{T}}
    \left[\dX^{\alpha}\right]
    -
    \left[\dP\right]^{\text{T}}
    \left[\dX^{\alpha}\right]
  \right)
\end{equation}
The vectors $[\dX^{\alpha}]$, $[\dY^{\alpha}]$, 
and $[\dP^{\alpha}]$ are determined for
each solution of Problem~1,
or from the $[\dLambda^{\alpha}]$ solutions of Problem~2
(after applying Eqs.~\ref{eq:dxdyGBD}).
%
Ba\v{z}ant hypothesized that a system approaches
equilibrium along the branch that minimizes $\mathcal{I}_{2}$.
Petryk \cite{Petryk:1991a,Petryk:1993a}, also using energy arguments,
found that a necessary condition
for a 0th solution $[d\mathbf{x}^{0}]\in A$
to be stable is
\begin{equation}\label{eq:PI3}
  \mathcal{I}_{2}(d\mathbf{x}^{0}) \le
  \mathcal{I}_{2}(d\mathbf{x}^{\alpha}),\quad
  \forall
  \left[d\mathbf{x}^{\alpha}\right]\in A
\end{equation}
for the set of bifurcation paths (i.e. solutions),
$A=\{[d\mathbf{x}^{0}],[d\mathbf{x}^{1}],[d\mathbf{x}^{2}],\ldots\}$.
Contrarily,
a sufficient condition that a
path $[d\mathbf{x}^{0}]\in A$
exhibits \emph{path instability} is
\begin{equation}\label{eq:PI2}
  \exists\,[d\mathbf{x}^{\alpha}]\in A,\quad
  \mathcal{I}_{2}(d\mathbf{x}^{\alpha}) <
  \mathcal{I}_{2}(d\mathbf{x}^{0})
\end{equation}
such that the 0th path does not minimize $\mathcal{I}_{2}$.
That is, the stable path offers the least resistance to
(i.e., the greatest compliance with) the
loading.
\subsection{Path-sensitivity}\label{sec:sensitivity}
Although not yet applied in mechanics,
a condition of \emph{solution stability}
has received attention in the LCP mathematics literature,
where solution stability refers to insensitivity of a solution
of an \LCP\ to perturbations of the underlying data,
$[\dQbar]$ and $[\Mbar]$.
To distinguish \emph{solution instability} from the
path instability of Section~\ref{sec:PI1}
(and to avoid obliging one notion of stability over others),
the term \emph{path-sensitivity} is used herein,
and its absence 
is called \emph{path-insensitivity}.
Path-sensitivity (or path-insensitivity)
can only be determined in relation to
an LCP of Problem~2. 
%
\par
Ha \cite{Ha:1985a} defined path-insensitivity as follows:
a solution
$\dLambda^{\ast}\in\mathcal{S}(\dQbar^{\ast},\Mbar^{\ast})$
of the $\text{LCP}(\dQbar^{\ast},\Mbar^{\ast})$
is insensitive if
there are neighborhoods $\mathcal{V}\subset\mathbb{R}^{M_{\lambda}}$
of the solution $\dLambda^{\ast}$
and
$\mathcal{U}\subset
\mathbb{R}^{M_{\lambda}}\times\mathbb{R}^{M_{\lambda}\times M_{\lambda}}$
of the data $(\dQbar^{\ast},\Mbar^{\ast})$
such that (i) for all $(\dQbar,\Mbar)\in\mathcal{U}$,
the solution set
$\mathcal{S}_{\mathcal{V}}(\dQbar,\Mbar)\in\mathcal{S}(\dQbar,\Mbar)\cap\mathcal{V}$
is non-empty,
and (ii)
$\text{sup}\{|\mathbf{y}-\dLambda^{\ast}|:\mathbf{y}\in\mathcal{S}_{\mathcal{V}}\}\rightarrow 0$
as
$(\dQbar,\Mbar) \rightarrow (\dQbar^{\ast},\Mbar^{\ast})$.
This path-insensitivity,
meaning that infinitesimal changes in $[\Mbar]$ and $[\dQbar]$
produce infinitesimal but measurable changes in the solution $[\dLambda]$,
is described as Lipschitz-like continuity
by Huyen and Yao \cite{Huyen:2019a}.
Path-sensitivity of static systems
is also similar to Lyapunov stability of dynamic systems,
insofar as path-insensitivity applies to the incremental change
$[\dQbar]$ rather than a full trajectory of future
movement.
\par
Moreover, the
definition of path-insensitivity
(called solution stability by Ha \cite{Ha:1985a})
encompasses a broad range of perturbations
(or a perturbation class \cite{Nicot:2012c})
for a quasi-static system
described by $[\Mbar]$ and $[\dQbar]$.
Perturbations of $[\Mbar]$ include slight changes in
a granular system's
particle positions, the particles' curvatures at their contacts,
the current contact forces, the contacts' stiffnesses,
the contacts' friction coefficients,
the regime of the control program, etc.;
whereas, perturbations of $[\dQbar]$ include all of these plus
slight changes
in the applied forces $[\dP]$ or controlled displacements
$[\dC]$.
%
\par
Huyen and Nguyen \cite{Huyen:2016a}
proved that a solution $[\dLambda^{\ast}]$
satisfies this Lipschitz-like continuity
if and only if a certain regularity condition is satisfied.
If this regularity condition is
satisfied for a solution $[\dLambda^{\ast}]$,
the solution
is path-insensitive; otherwise, the solution is path-sensitive.
%
%
%
Determining regularity is based upon the 
three index subsets defined in Eq.~(\ref{eq:sens1}), and
regularity and path-insensitivity mean that a vector
$[\mathbf{u}']_{M_{\lambda}\times 1}$ that meets certain conditions
implies that $[\mathbf{u}']$ is zero \cite{Huyen:2019a}:
\begin{equation}\label{eq:sens2}
  \left.
    \begin{aligned}
      &\text{For each }i\in I_{1}^{\ast},\;u'_{i}=0\\
      &\text{For each }i\in I_{2}^{\ast},\;
        ([\Mbar]^{\text{T}}[\mathbf{u}'])_{i}=0\\
      &\text{For each }i\in I_{3}^{\ast},\;u'_{i}=0,
       \text{ or }
       ([\Mbar]^{\text{T}}[\mathbf{u}'])_{i}=0,
       \text{ or }
       (
       u'_{i}\le 0
       \text{ and }
       ([\Mbar]^{\text{T}}[\mathbf{u}'])_{i}\ge 0
       )
    \end{aligned}
  \right\}
  \Rightarrow
  [\mathbf{u}']=[\mathbf{0}]
\end{equation}
meaning that regularity requires that
the implication ``$\Rightarrow$'' follows from the conditions on
$I_{1}^{\ast}$ and $I_{2}^{\ast}$ combined with any one
(individually)
of the three conditions on $I_{3}^{\ast}$.
The solution $[\dLambda^{\ast}]$ is path-insensitive if and only if
it satisfies this regularity condition; otherwise (i.e., when a negated
implication ``$\nRightarrow$'' applies),
$[\dLambda^{\ast}]$
is path-sensitive.
As such, regularity depends on the matrix $[\Mbar]$
and on the partition of the index set
$I=I_{1}^{\ast}\cup I_{2}^{\ast}\cup I_{3}^{\ast}=\{1,2,\ldots,M_{\lambda}\}$
of a particular solution
$[\dLambda^{\ast}]$
for the control parameters $[\dQbar]$.
Note that the Huyen and Nguyen criterion \cite{Huyen:2016a}
is both sufficient
and necessary for path-insensitivity.
Also
note that this form of instability makes no appeal to
notions of energy or second-order work.
\par
Huyen and Pang \cite{Huyen:2019a} noted the following
special cases:
if $I_{1}^{\ast}=I$, the regularity condition is automatically satisfied
by $[\dLambda^{\ast}]$;
if $I_{2}^{\ast}=I$, the regularity condition is satisfied if and only if
$[\Mbar]$ is nonsingular;
and if $I_{3}^{\ast}=I$, the regularity condition is satisfied if and only if
$[\Mbar]$ is a P-matrix.
%
%
\par
An algorithm for determining whether a solution $[\dLambda]$
is path-sensitive (non-regular) is presented in \ref{sec:sensalg}.
\par
Phung and Tin \cite{Phung:2015a} show that matrices $[\Mbar]$ for
which the \LCP\ is Lipschitz-like
continuous at all $[\dQbar]$ belong to a restricted
matrix class:
the class of $k$-matrices, for which the \LCP\ has
the same number $k$ of solutions for all $[\dQbar]$
that lie within the interior of the space of $[\dQbar]$
for which the LCP is solvable.
Unless $[\Mbar]$ is a $k$-matrix,
the granular system can suffer path-sensitivity for some
loadings $[\dQbar]$.
(On the other hand, 
P-matrices are 1-matrices, so are path-insensitive for all $[\dQbar]$.)
\par
As an example, Case~C of Fig.~\ref{fig:ExampleLCPs}c
with $d\bar{q}_{1}=0$ has the single
solution $[\dLambda]=[0, 0]^{\text{T}}$, for which the index subsets are
$I_{1}^{\ast}=\{2\}$, $I_{2}^{\ast}=\varnothing$, and $I_{3}^{\ast}=\{1\}$.
Following Eq.~(\ref{eq:sens2}) and
combining the assumption of $u'_{2}=0$ with either of the assumptions
$u'_{1}=0$ or $-u'_{1}-u'_{2}=0$ implies that both $u'_{1}$ and $u'_{2}$
are zero;
however, combining $u'_{2}=0$ with the assumption of
$u'_{1}\le 0$ and $-u'_{1}-u'_{2}\ge 0$ allows $u'_{1}\le 0$,
which does not imply that $[\mathbf{u}']=[\mathbf{0}]$.
Therefore, the implication is not affirmed,
and this solution is path-sensitive, a stiffness pathology.
This conclusion is also supported by Fig.~\ref{fig:ExampleLCPs}c,
which shows that the solution of $[\dLambda]=[0, 0]^{\text{T}}$
applies when $d\bar{q}_{1}= 0$, but with the slightest reduction
of $d\bar{q}_{1}$, the solution vanishes,
and no solutions exist when $d\bar{q}_{1}<0$.
\par
On the other hand, Case~D of Fig.~\ref{fig:ExampleLCPs}c
with $d\bar{q}_{1}=2$
has the solution $[\dLambda]=[0, 1]^{\text{T}}$, and the index subsets are
$I_{1}^{\ast}=\varnothing$, $I_{2}^{\ast}=\{2\}$, and $I_{3}^{\ast}=\{1\}$.
Combining the assumption $-2u'_{1}-u'_{2}=0$
(as in Eq.~\ref{eq:sens2}\textsubscript{2})
with any one of the three assumptions
$u'_{1}=0$ or $-u'_{1}-u'_{2}=0$ or
$u'_{1}\le 0$ and $-u'_{1}-u'_{2}\ge 0$
implies that $[\mathbf{u}']=[\mathbf{0}]$.
This solution is not path-sensitive. 
Referring to Fig.~\ref{fig:ExampleLCPs}c,
small changes in $[\dQbar]$ (for example, slightly
shifting the lines \fbox{$1'$} and/or \fbox{$2'$} up or down)
produce small changes in the
solution $[\dLambda]=[0, 1]^{\text{T}}$,
thus confirming continuity (regularity) of the solution.
\subsection{Path-incongruity}\label{sec:incongruity}
Path-incongruity is the absence of
solutions for the intended loading path~---
the intended control program $[\mathbf{C}]$ in combination
with the control parameters $[\dP]$ and $[\dC]$~---
such that the system is
incongruous  with (inconsistent with)
this path.
A possible lack of solutions can be determined
with any of the three Problems,
but with the LCP of Problem~2, several established rules give
conditions
that are sufficient to \emph{preclude} path-incongruity:
if matrix $[\Mbar]$ is positive definite (PD, rule~2)
or a P-matrix (rule~3), a unique
solution is assured for all $[\dQbar]$;
a solution, possibly not unique, is assured for all $[\dQbar]$,
if the matrix is an R-matrix (rule~4), is entirely non-negative
with positive diagonal elements (rule~6),
is both a Z-matrix and S-matrix (rules~1 and~5),
or is an N-matrices of the first category
(rule~7);
and
at least one solution exists for N-matrices of the second category,
provided that $[\dQbar]$ is non-negative (rule~8).
\par
Although one often assumes that a response exists when
a system is loaded, a response might not be available.
For example, with the simple linear-frictional spring--slider of
Eqs.~(\ref{eq:dfpq})--(\ref{eq:gpq}),
a force solution $\mathfrak{d}\mathbf{f}^{pq}$ 
exists for any prescribed displacement $\Dxdefpq$;
however,
displacements solutions might not exist when force
$\mathfrak{d}\mathbf{f}^{pq}$ is prescribed:
the relation
$\mathfrak{d}\mathbf{f}^{pq}=\mathbf{F}^{pq}\cdot\Dudefpq$
is not invertible, as
certain forces are disallowed by the yield condition of
Eq.~(\ref{eq:yield}).
Likewise, a granular assembly can be incongruous with certain
loading parameters,
if $[\dP]$ and $[\dC]$ are inconsistent
with the yield conditions of the system's contacts:
for example, when peak strength is reached, but a greater static loading
is attempted,
causing the system to enter a dynamic domain (collapse). 
Matrices $[\Mbar]$ that do not meet the conditions mentioned above
can engender incongruous loading parameters 
that are inconsistent with available particle motions.
%
\subsection{Neutral equilibrium}\label{sec:NE}
Like bifurcation,
neutral equilibrium is a condition of non-uniqueness,
but one with neutral equilibrium
unbounded solutions that are available with neutral loading.
In a continuum setting,
neutral equilibrium is associated with attaining
the peak stress under a particular control program.
Herein, neutral equilibrium is defined as
the existence of
adjacent equilibrium configurations $[\dX]$,
infinitely close to the current state,
that can be reached with neutral loading, $[\dP]=[\dC]=[\mathbf{0}]$;
that are consistent with the displacement constraints
of the given control program,
$[\dX]\in\mathcal{N}([\mathbf{C}])$; and that are within
the domain $\Omega^{i}$ of a
stiffness branch  $[\mathbf{H}^{i}]$.
Called a \emph{loss of control} \cite{Nova:1994a} or a
\emph{divergence instability},
neutral equilibrium is 
a form of incipient bifurcation, but it 
is a condition of the current state and
control program, rather than of the
subsequent loading path, $[\dP]$ and $[\dC]$.
Neutral equilibrium can also lead to
a \emph{loss of control}
during subsequent loading:
when adjacent equilibrium configurations can be reached
with neutral loading,
a subsequent intended loading might be outside the system's range
(a pathology discussed in Section~\ref{sec:incongruity}).
\par
In the context of Problem~1,
neutral equilibrium occurs when
the determinant of matrix
$([\mathbf{H}^{i}][\mathbf{P}_{\mathcal{L}}]+[\mathbf{P}_{\mathcal{L}^{\perp}}])$
is singular, provided that
the matrix's null space is consistent with the displacement
constraints and lies (wholly or partially)
within domain $\Omega^{i}$:
\begin{equation}\label{eq:NE}
  \text{det}
  \left(
    [\mathbf{H}^{i}][\mathbf{P}_{\mathcal{L}}]+[\mathbf{P}_{\mathcal{L}^{\perp}}]
  \right)
  =0
  \quad
  \text{and}
  \quad
  \mathcal{N}
  \left(
    [\mathbf{H}^{i}][\mathbf{P}_{\mathcal{L}}]+[\mathbf{P}_{\mathcal{L}^{\perp}}]
  \right)
  \cap
  \mathcal{N}([\mathbf{C}])
  \cap
  \Omega^{i}
  \neq
  [\mathbf{0}]
\end{equation}
This condition was discussed in relation to the generalized
Bott--Duffin inverse in
Eqs.~(\ref{eq:GBD})--(\ref{eq:dxdyGBD}),
and applies to Eq.~(\ref{eq:dxdyGBD})
with a neutral loading, $[\dC]=[\dX]=[\mathbf{0}]$.
If the determinant of matrix
$([\mathbf{H}^{i}][\mathbf{P}_{\mathcal{L}}]+[\mathbf{P}_{\mathcal{L}^{\perp}}])$ is zero,
a subsequent loading can lie outside
the range of this matrix, making
the loading unavailable to the system and its control regime.
\par
Although the Eq.~(\ref{eq:NE}) 
provides criteria for
neutral equilibrium,
its implementation can be problematic,
particularly when the dimension of the null-space
$\mathcal{N}([\mathbf{H}^{i}][\mathbf{P}_{\mathcal{L}}]+[\mathbf{P}_{\mathcal{L}^{\perp}}])$ is greater than 1.
When the dimension is only~1, the problem is straightforward.
One simply determines whether the single basis vector $[d\mathbf{x}]$
of the null space, or its reversal $[-d\mathbf{x}]$, belongs
to the region $\Omega^{i}$, by testing the
condition in Eq.~(\ref{eq:condition}),
with both
$\left[\rule{0ex}{2ex}\boldsymbol{\mathfrak{s}}\left([d\mathbf{x}]\right)=1\right]$
and
$\left[\rule{0ex}{2ex}\boldsymbol{\mathfrak{s}}\left([-d\mathbf{x}]\right)=1\right]$.
(Note that $\Omega^{i}$ is a pointed cone,
so both $[d\mathbf{x}]$ and $[-d\mathbf{x}]$ must be tested.)
The situation is more complex when the null space is of dimension
greater than 1.
In this case, all of the basis vectors might lie outside of
$\Omega^{i}$, even though a linear combination of the vectors
can lie inside of $\Omega^{i}$. %
In this case, one must investigate a range of linear combinations of the
the basis vectors of
$\mathcal{N}([\mathbf{H}^{i}][\mathbf{P}_{\mathcal{L}}]+[\mathbf{P}_{\mathcal{L}^{\perp}}])$ to determine whether any combinations
lie within $\Omega^{i}$.
This complexity is avoided by pursuing Problem~2.
\par
With the LCP of Problem~2, the question of neutral equilibrium is
straightforward:  neutral equilibrium occurs when
matrix $[\Mbar]$ is non-R\textsubscript{0},
which can be determined with rules 16--19 of Section~\ref{sec:exist}.
Because a non-R\textsubscript{0} matrix also assures an unbounded
solution set of the \LCP, 
the pathologies of neutral equilibrium and
Type~III unbounded bifurcation share the same criterion
(rule~16 and Section~\ref{sec:bif}).
The distinction is that unbounded bifurcation is a
pathology of path and is associated with the
control parameters that yield the particular direction of the
unbounded movements;
whereas, neutral equilibrium merely establishes the
potential of unbounded movement for the current state.
%
%
\subsection{Unsustainability}\label{sec:instequil}
Instability of equilibrium,
as originated for continua by Hill \cite{Hill:1958a},
occurs when a departure from equilibrium
is energetically available 
when infinitesimal movement in certain directions
can produce finite velocities within the
system without an increment of the loading parameters 
\cite{Hill:1958a,Bazant:1991a,Petryk:1991a,Nicot:2012a,Kuhn:2019b}.
Nicot et al. \cite{Nicot:2007a} show that this condition
implies that the current state is unsustainable,
in that nearby states can be reached without a change
in the control parameters, by way of a dynamic process.
For discrete granular systems, we refer to this pathology as
\emph{unsustainability} to distinguish it from
path instability, as defined in Section~\ref{sec:PI1}.
Unsustainability is an inherently dynamic condition,
but in the context of quasi-static loading,
the pathology only signals a \emph{latent potential}
for spontaneous movement that could
be realized in particular directions
under the given control program.
In this sense, unsustainability is similar to
neutral equilibrium:
it is a latent
condition of the current state,
rather than an actualized condition of the intended loading direction.
Although unsustainability can enable collapse,
the example of Section~\ref{section:disks49} shows that
the unsustainable condition does not necessarily lead to collapse
and failure, but can precipitate internal adjustments
that are resolved within a granular system,
even as it is hardening and even when
a quasi-static analysis obviates dynamic (inertial) effects
(an interpretation is given in Section~\ref{sec:expaths}).
\par
Unsustainability is defined as follows:
if the loading rate $[\dot{\mathbf{p}}]$ is suspended
(i.e., with dead loads $[\mathbf{p}]$ help constant)
while also maintaining the
control program $[\mathbf{C}]$
but not advancing the constrained displacements
(i.e., $[\dC]=[\mathbf{0}]$),
an increase in kinetic energy is 
favored when the second-order internal work, 
$\ddot{W}_{\text{int}}=
[\dot{\mathbf{x}}]^{\text{T}}[\mathbf{H}^{i}][\dot{\mathbf{x}}]$,
is negative in a movement direction
$[\dot{\mathbf{x}}]$ that is consistent with $[\mathbf{H}^{i}]$,
or $[\dot{\mathbf{x}}]\in\Omega^{i}$
(\ref{sec:setting}).
Because movements must lie within the cone
$\Omega^{i}$ of
directional stiffness $[\mathbf{H}^{i}]$
and must be consistent with the displacement constraints,
unsustainability, in the context of Problem~1,
is defined herein as
\begin{equation}\label{eq:Ins1}
  \exists
  \left[\dot{\mathbf{z}}\right]
  :\;\;
  \left[\mathbf{P}_{\mathcal{L}}\right]
  \left[\dot{\mathbf{z}}\right]
  \in\Omega^{i}
  ,\;\;
  \ddot{W}_{\text{int}}=
  \left[\dot{\mathbf{z}}\right]^{\text{T}}
  \left(
  \left[\mathbf{P}_{\mathcal{L}}\right]^{\text{T}}
  \left[\mathbf{H}^{i}\right]
  \left[\mathbf{P}_{\mathcal{L}}\right]
  \right)
  \left[\dot{\mathbf{z}}\right]
  < 0
\end{equation}
where the product
$[\dot{\mathbf{x}}]=[\mathbf{P}_{\mathcal{L}}][\dot{\mathbf{z}}]$
assures that the displacement constraints of the
control program are momentarily frozen,
with $[\mathbf{C}][\dot{\mathbf{x}}]=[\mathbf{0}]$.
As a practical matter,
the conditions of Eq.~(\ref{eq:Ins1}) are met
when the matrix
$[\mathbf{P}_{\mathcal{L}}]^{\text{T}}[\widehat{\mathbf{H}^{i}}][\mathbf{P}_{\mathcal{L}}]$
has one or more negative eigenvalues,
where $[\widehat{\mathbf{H}^{i}}]$ is the symmetric
part of matrix $[\mathbf{H}^{i}]$, or 
$[\widehat{\mathbf{H}^{i}}]=\frac{1}{2}([\mathbf{H}^{i}]+[\mathbf{H}^{i}]^{\text{T}})$.
Vectors $[\dot{\mathbf{z}}]$ that satisfy the inequality
in Eq.~(\ref{eq:Ins1}) form a cone in $\mathbb{R}^{6N}$,
the \emph{cone of instability}
\cite{Darve:2000a}.
Applying the equation, however, presents two difficulties.
First, the cone formed from eigenvectors that correspond
to negative eigenvalues is a subset of the larger
cone of instability \cite{Lerbet:2020a}.
Second,
determining unsustainability requires establishing
the cone of instability and then ascertaining whether
it intersects the cone $\Omega^{i}$.
These difficulties are avoided by approaching the problem
as an LCP.
%
%
\par
In the context of the \LCP\ of Problem~2, the
following conditions
for unsustainability are derived in
\ref{sec:setting}:
\begin{linenomath}
\begin{subequations}\label{eq:instabProb2a}
\begin{align}
  \exists
  \left[\dot{\mathbf{z}}\right],
  \left[\dot{\boldsymbol{\lambda}}\right]:
  \;\;
  &\left[\mathbf{H}_{\boldsymbol{\lambda}\mathbf{x}}\right]
   \left[\mathbf{P}_{\mathcal{L}}\right]
   \left[\dot{\mathbf{z}}\right]
   +
   \left[\mathbf{H}_{\boldsymbol{\lambda\lambda}}\right]
   \left[\dot{\boldsymbol{\lambda}}\right] \ge 0\\
  &\left[\dot{\boldsymbol{\lambda}}\right] \ge 0\\
  &\left[\dot{\boldsymbol{\lambda}}\right]^{\text{T}}
   \left(
     \left[\mathbf{H}_{\boldsymbol{\lambda}\mathbf{x}}\right]
     \left[\mathbf{P}_{\mathcal{L}}\right]
     \left[\dot{\mathbf{z}}\right]
     +
     \left[\mathbf{H}_{\boldsymbol{\lambda\lambda}}\right]
     \left[\dot{\boldsymbol{\lambda}}\right]
   \right) = 0\\
  &\ddot{W}_{\text{int}} =
     \left[\dot{\mathbf{z}}\right]^{\text{T}}
     \left(
       \left[\mathbf{P}_{\mathcal{L}}\right]^{\text{T}}
       \left[\mathbf{H}_{\mathbf{xx}}\right]^{\text{T}}
       \left[\mathbf{P}_{\mathcal{L}}\right]
       \left[\dot{\mathbf{z}}\right]
       +
       \left[\mathbf{P}_{\mathcal{L}}\right]^{\text{T}}
       \left[\mathbf{H}_{\mathbf{x}\boldsymbol{\lambda}}\right]
       \left[\dot{\boldsymbol{\lambda}}\right]
     \right)
     < 0
\end{align}
\end{subequations}
\end{linenomath}
where $[\dot{\mathbf{z}}]\in \mathbb{R}^{6N}$,
$[\dot{\boldsymbol{\lambda}}]\in \mathbb{R}^{M_{\lambda}}$.
One recognizes the first three conditions,
(\ref{eq:instabProb2a}a)--(\ref{eq:instabProb2a}c),
as a linear complementarity problem, although
different than the original \LCP\ of Eq.~(\ref{eq:form2}),
with the new problem written as
$\text{LCP}(
     [\mathbf{H}_{\boldsymbol{\lambda}\mathbf{z}}]
     [\mathbf{P}_{\mathcal{L}}]
     [\dot{\mathbf{z}}],\,
     \mathbf{H}_{\boldsymbol{\lambda\lambda}})$.
As an additional condition for unsustainability,
a solution $[\dot{\boldsymbol{\lambda}}]$ of this LCP must have a negative
$\ddot{W}_{\text{int}}$, as in (\ref{eq:instabProb2a}d),
for the particular $[\dot{\mathbf{z}}]$ to which
$[\dot{\boldsymbol{\lambda}}]$ applies.
\par
The search for solutions of Eq.~(\ref{eq:instabProb2a}) is simplified
when matrix $[\mathbf{H}_{\boldsymbol{\lambda\lambda}}]$
is diagonal and positive.
If so,
signs are preserved when the
$\text{LCP}(
     [\mathbf{H}_{\boldsymbol{\lambda}\mathbf{z}}]
     [\mathbf{P}_{\mathcal{L}}]
     [\dot{\mathbf{z}}],\,
     \mathbf{H}_{\boldsymbol{\lambda\lambda}})$
is replaced with the
$\text{LCP}([\mathbf{H}_{\boldsymbol{\lambda\lambda}}]^{-1}
     [\mathbf{H}_{\boldsymbol{\lambda}\mathbf{z}}]
     [\mathbf{P}_{\mathcal{L}}]
     [\dot{\mathbf{z}}],\,
     \,\mathbf{I}\,)$,
and because $[\,\mathbf{I}\,]$ is positive definite,
the LCP has a unique solution for every
vector $[\dot{\mathbf{z}}]$ (rule~2).
This fortunate property enables enumerative
algorithms for identifying all unsustainable modes
for a granular assembly.
A presumed positive and diagonal
$[\mathbf{H}_{\boldsymbol{\lambda\lambda}}]$
applies with most contact models,
as a \emph{diagonal}
$[\mathbf{H}_{\boldsymbol{\lambda\lambda}}]$
implies locality in the contacts' active mechanisms
(i.e. a contact's active mechanism depends on neither
the contact's other mechanisms nor those of other
contacts), and a \emph{positive}
$[\mathbf{H}_{\boldsymbol{\lambda\lambda}}]$
implies positive values of the quantities
$\mathbf{g}^{pq}_{\text{s}}\cdot\mathbf{h}^{pq}_{\text{s}}$
in Eq.~(\ref{eq:gdotl})
(in a continuum setting, this is akin to non-opposing yield and
flow directions).
Based upon this assumption,
\ref{sec:setting} presents an enumerative algorithm
for finding
all unsustainable modes.
\par
The inequality $[\dot{\boldsymbol{\lambda}}]\ge 0$
exposes the directional nature of unsustainability
in frictional systems, in a more overt manner
than the unsustainability criteria for Problem~1
(Eq.~\ref{eq:Ins1}).
For example,
if a single unsustainable mode exists,
with $[\dot{\mathbf{x}}]=[\dot{\boldsymbol{\eta}_{1}}]$
and $[\dot{\boldsymbol{\lambda}}]=[\dot{\boldsymbol{\lambda}_{1}}]$,
then vectors
$\alpha[\dot{\boldsymbol{\eta}_{1}}]$ and
$\alpha[\dot{\boldsymbol{\lambda}_{1}}]$
with $\alpha>0$
are also a solution of the 
$\text{LCP}(
     [\mathbf{H}_{\boldsymbol{\lambda}\mathbf{z}}]
     [\mathbf{P}_{\mathcal{L}}]
     [\dot{\mathbf{z}}],\,
     \mathbf{H}_{\boldsymbol{\lambda\lambda}})$
and satisfy the condition 
$\ddot{W}_{\text{int}}< 0$.
Darve and coworkers \cite{Sibille:2007a,Prunier:2009a}
developed the concept of an
\emph{instability cone} for continuum constitutive operators:
for a particular stress state,
an instability cone is
the locus of
stress directions $d\boldsymbol{\sigma}$ in which the
second-order work is negative.
For discrete granular systems,
the unsustainable
modes~--- solutions of Eq.~(\ref{eq:instabProb2a})~--- form
polyhedral cones in the
displacement space of $\mathbb{R}^{6N}$, emanating
from the origin $[\dot{\mathbf{x}}]=[\mathbf{0}]$.
In the above example of
a discrete system with a single unsustainable mode,
the instability cone is a simple ray in the $\mathbb{R}^{6N}$ space
of particle movements.
For discrete systems with multiple unsustainable modes
(i.e., solution directions),
multiple detached instability cones are possible,
in the sense that cones might only share the origin but
are otherwise disjoint.
\par
A hierarchy applies to the pathologies of neutral equilibrium
and of unsustainability,
based upon the following matrix property:
the real parts of the eigenvalues of a matrix $[\mathbf{H}]$
are bounded by the smallest and largest eigenvalues of
its symmetric part.
As such, for systems
that undergo a smooth, continuous transition of
stiffness during a loading program,
unsustainability is encountered
before neutral equilibrium:
a negative eigenvalue of the symmetric matrix
$[\widehat{\mathbf{H}}]$ is encountered before
a zero eigenvalue appears
(along with a zero determinant) with the full matrix $[\mathbf{H}]$.
For a smooth stiffness operator, Challamel et al. \cite{Challamel:2010a}
have shown
that the second-order work criterion for a non-conservative
system coincides with neutral equilibrium with one homogeneous
constraint, and Lerbet et al. \cite{Lerbet:2012b}
have generalized this result to systems with $n$ constraints.
This hierarchy of second-order work and 
neutral equilibrium does not necessarily apply to
non-smooth inelastic systems,
such as those with incremental non-linearities.
The eigenvectors of the full stiffness
matrix and of its symmetric counterpart
are not equal,
and it is possible that the negative eigenvalues
of the symmetric stiffness $[\widehat{\mathbf{H}}^{i}]$
correspond to eigenvectors that
lie \emph{outside} the stiffness's domain $\Omega^{i}$;
whereas, the full matrix $[\mathbf{H}^{i}]$ can have
a zero eigenvalue with a different eigenvector, but one
that lies within $\Omega^{i}$.
Although this possibility is unlikely, 
meaning that unsustainability will likely precede neutral equilibrium,
evaluating the two pathologies in the context of an LCP
resolves the difficulty of non-smooth transitions between
regions $\Omega^{i}$.
That is, approaching a granular system
in the context of Problem~2 rather than Problem~1 simplifies
the evaluation of unsustainable modes
(e.g., the algorithm in \ref{sec:setting}).
\section{\normalsize Examples of Granular Systems}\label{sec:examples}
Three examples are presented, some with multiple variations.
All examples were solved as Problem~2 LCPs, using
methods of Sections~\ref{sec:p12final} and~\ref{sec:pathologies}
and the appendices.
The author's code for these examples is available,
as described in the section ``Data availability''
at the end of the paper.
\subsection{Example~1: three-particle system}\label{sec:three}
\begin{figure}
  \centering
  \includegraphics{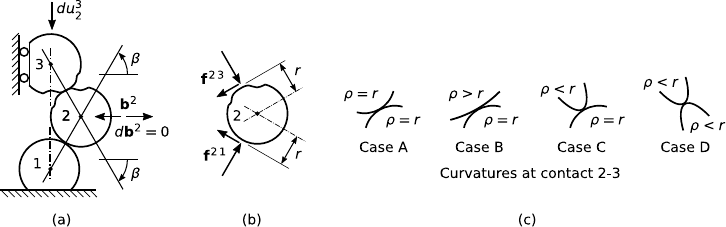}
  \caption{Three-particle system: (a)~particle arrangement,
            displacement constraints, downward loading,
            and external forces;
            (b)~normal and tangential contact forces on
            the middle particle, with contacts at the friction limit
            and $|\mathbf{f}^{2,1}|=|\mathbf{f}^{2,3}|$;
            and (c)~four cases of contact curvatures at the
            contact between particles~2 and~3.
            \label{fig:threeparticles}}
\end{figure}

A seemingly simple two-dimensional
three-particle system with two contacts,
shown in Fig.~\ref{fig:threeparticles}a,
illustrates seven of the eight
pathologies in the previous section 
as well as many of the rules in Section~\ref{sec:exist},
when applied to frictional systems.
The particles are labeled 1, 2, and 3, from bottom to top.
The control program $[\mathbf{C}]$ places constraints on the movements
of particles~1 and~3, and the control parameters assign
their directions of movement:
particle~1 is fixed while the system 
is loaded by displacing
particle~3 downward toward particle~1,
located directly below~3.
The system
has certain symmetries:
the centers of the two particle-pairs are oriented at angle $\beta$,
with inclinations to the left and right, as
shown in the figure;
the contacts' normal vectors are aligned with their
contact vectors (i.e., $\mathbf{n}^{pq}$ is aligned
with $\mathbf{r}^{pq,p}$);
and all 
contact vectors have the same length $r$
(i.e., $|\mathbf{r}^{pq,p}|=|\mathbf{r}^{pq,q}|=r$).
\par
At time $t$, both contacts are assumed at the friction limit,
$|\mathbf{f}^{\,\text{t}}|= \mu f^{\,\text{n}}$,
and the tangent forces $\mathbf{f}^{\,\text{t}}$ on particle~2 act
in the directions shown in
Fig.~\ref{fig:threeparticles}b.
Both contacts
obey the standard linear-frictional model,
with normal stiffness $k$,
tangential stiffness $\alpha k$,
and friction coefficient $\mu$ (Section~\ref{sec:linfrict}).
The system is in equilibrium, requiring external
forces to induce the friction limit:
a horizontal force $\mathbf{b}^{2}$ on the middle
particle, and opposite forces $-\frac{1}{2}\mathbf{b}^{2}$
on both top and bottom particles.
The magnitude of these forces is a function of
$\mu$ and $\beta$.
\par
During increment $dt$,
the middle particle is free to move
with zero incremental force, $d\mathbf{b}^{2}=0$,
while the top particle is constrained to move downward, with
no rotation and no horizontal shifting.
These constraints are readily applied with
matrix $[\mathbf{C}]$ and vector $[\dC]$.
The applied load increments $[\dP]$ are zero,
although reaction forces $[d\mathbf{y}]$ ensue on the
top and bottom particles,
due to their constraints.
\par
To illustrate a range of behaviors and pathologies,
three ranges of $\beta$ and four cases of the particles'
curvatures are modeled (Fig.~\ref{fig:threeparticles}a).
With the steepest Case~1 angles,
$\text{tan}^{-1}(1/\mu)<\beta<\pi/2$, the initial external
force $\mathbf{b}^{2}$ must pull particle~2
toward the right, prior to $dt$, to bring the
contacts to the friction limit.
For Case~2, with the single angle
$\beta=\text{tan}^{-1}(1/\mu)$, the contacts
are at the frictional limit without any external $\mathbf{b}^{2}$;
and for Case~3 with the shortest assembly,
$0 <\beta < \text{tan}^{-1}(1/\mu)$, force $\mathbf{b}^{2}$
must push toward the left.
With the $\mathbf{b}^{2}$ force held constant on particle~2
(i.e. $d\mathbf{b}^{2}=\mathbf{0}$),
angles in Case~1 produce hardening behavior:
the incremental reaction force on particle~3, $d\mathbf{y}^{3}$,
is downward, such that the downward force
increases as particle~3 is pushed downward.
However, the Case~3 angles exhibit softening behavior,
with $d\mathbf{y}^{3}$ being upward,
causing the downward force on particle~3 to be reduced as
the particle is pushed downward.
\par
For the contact between particles~1 and~2,
the curvatures are disk-like:
the radii of curvature $\rho$ equal the radial length $r$.
Between particles~2 and~3, however,
four contact curvatures are modeled
(Fig.~\ref{fig:threeparticles}c).
In case~A, both particles~2 and~3 are disk-like, with $\rho=r$;
in case~B, particle~3 has a flatter surface,
with $\rho > r$, whereas particle~2 is disk-like;
in case~C, particle~3 is sharper, with $\rho< r$,
and particle~2 is disk-like;
and in case~D, both~2 and~3 have sharper surfaces,
with the same $\rho < r$.
\par
Results for the twelve variations of
the three-particle system are summarized in
Table~\ref{table:threeparticles},
based upon the following parameters:
$r=1$, $k=1$, $\alpha = 1$, $\mu=0.5$, and $f^{\text{n}}=0.001$,
with the latter meaning that the initial indentation of the 
particles is 0.001 times their radius.
\begin{table}
  \caption{Results of the three-particle system of
           Fig.~\ref{fig:threeparticles}.
           \label{table:threeparticles}}
  \centering
  \small
  \begin{threeparttable}
  \begin{tabular}{clccccccc}
    \toprule
    & & $[\Mbar]$ & \multicolumn{2}{c}{Solutions (number)\tnote{2}} & Unstable $|$ & Unsustain. & Path- & Path- \\
    Case              & 
    $\beta$, curvature case &       
    classes\tnote{1} &             
    Type I  &             
    Type II &             
    likely paths\tnote{3} &             
    modes\tnote{4}  &             
    sensit.\tnote{5} &             
    incongr.\tnote{6} \\           
    \midrule
    1 & $\text{atan}(1/\mu)<\beta<\pi/2$ &   & & & &  & & \\
    & \quad A                          & 
      D, NSD & \CircCirc\ (1)& \BulletBullet\ ($\infty$)& \CircCirc$\,|\,$\BulletBullet&\BulletBullet& \checkmark& -\\
    & \quad B                          & 
      D, NSD & \CircCirc\CircBullet\ (2)& -& \CircCirc$\,|\,$\BulletBullet& \BulletBullet& - & -\\
    & \quad C                          & 
      D, NSD & \CircCirc\BulletCirc\ (2)& -& \CircCirc$\,|\,$\BulletBullet& \BulletBullet& - & -\\
    & \quad D                          & 
      I, ND & \CircCirc\CircBullet\BulletCirc\BulletBullet\ (4)& -& \CircCirc\CircBullet\BulletCirc$\,|\,$\BulletBullet& \BulletBullet& - & -\\
    \midrule
    2 & $\beta = \text{atan}(1/\mu)$ &   & & & &  & & \\
    & \quad A                          & 
      D, NSD & -& -& -& \BulletBullet& -& \checkmark\\
    & \quad B                          & 
      D, NSD & -& -& -& \BulletBullet& -& \checkmark\\
    & \quad C                          & 
      D, NSD & -& -& -& \BulletBullet& -& \checkmark\\
    & \quad D                          & 
      D, ID & -& -& -& \BulletBullet& -& \checkmark\\
    \midrule
    3 & $0<\beta<\text{atan}(1/\mu)$ &   & & & & & & \\
    & \quad A                         &
      R, D, PSD & -& \BulletBullet\ ($\infty$)& -& -& \checkmark& -\\
    & \quad B                          & 
      R, D, PSD & \BulletCirc\ (1)& -& -& -& -& -\\
    & \quad C                          & 
      R, D, PSD & \CircBullet\ (1)& -& -& -& -& -\\
    & \quad D                          & 
      R, I, ID & \BulletCirc\CircBullet\BulletBullet\ (3)& -& \CircBullet\BulletBullet$\,|\,$\BulletCirc& -& -& -\\
    \bottomrule
  \end{tabular}
  \begin{tablenotes}
    \item[1] 
           Matrix classes of $[\Mbar]$:
           R-matrix (R),
           P-matrix (P), N-matrix (N), P\textsubscript{0}
           or N\textsubscript{0} degenerate matrix (D),
           non-degenerate matrix that is neither N or P (I),
           and matrix that is
           positive definite (PD), positive semi-definite (PSD),
           indefinite (ID), negative semi-definite (NSD), or
           negative
           definite (ND).
    \item[2] Bifurcations: solutions (and their number) of the \LCP, as
           combinations of sliding and non-sliding for the lower
           contact (between particles~1 and~2) and for the upper
           contact (between particles~2 and~3):
           \CircCirc = both contacts non-sliding;
           \CircBullet = bottom contact sliding, top contact non-sliding;
           \BulletCirc = bottom contact non-sliding, top contact sliding;
           and \BulletBullet = both contacts sliding.
    \item[3] For solutions in the previous two columns,
             the solutions are classified as either unstable or
             ($|$) likely (stable).
    \item[4] Unsustainable modes,
           combinations of sliding/non-sliding contacts (see note~2).
    \item[5] Type~II solutions that exhibit path-sensitivity.
    \item[6] Conditions that exhibit path-incongruity.
  \end{tablenotes}
  \end{threeparttable}
\end{table}
The matrix classes of $[\Mbar]$ are given in the third column,
showing that the matrix spans multiple classes:
ranging from R-matrices to non-R matrices,
ranging from P-matrices to N-matrices,
and ranging from positive-definite to negative-definite.
In all cases, matrix $[\Mbar]$ is R\textsubscript{0},
such that there are no instances of
Type~III bifurcation or of a neutral equilibrium
pathology (rule~20 and Sections~\ref{sec:bif} and~\ref{sec:NE}).
\par
For Cases~1 and~2 of (steeper) angles $\beta$,
the matrix entries are all negative,
so that $[\Mbar]$ is not an S-matrix,
and the \LCP\ is not feasible for some loadings $[\dQbar]$
(rule~1, and hence the path incongruity with Case~2).
With Case~2,
matrix $[\Mbar]$ is also a Z-matrix
(rule~5),
and because the case is path-incongruent for the one loading
$[\dQbar]$, it is path-incongruent for all loadings.
As such, Case~2 presents a peculiar situation:
the system is in equilibrium in the current configuration
(although the configuration is unsustainable),
but there is no equilibrium solution during subsequent
downward downward movement of the upper particle~3.
That is, Case~2 is locked, and no solution $[\dLambda]$
exists that will maintain the constant 
zero side force $[\mathbf{b}^{2}]=[\mathbf{0}]$,
with $[d\mathbf{b}^{2}]=[\mathbf{0}]$.
Further downward loading will cause the system to enter
the dynamic domain until equilibrium is restored.
\par
Only the two Cases~3.B and~3.C have unique solutions with no
stiffness pathologies, and because the
$[\Mbar]$ is a P-matrix, a unique solution exists for
all $[\dC]$ and $[\dP]$ loadings (rule~3).
For Case~3 and all four contact curvatures, 
matrix $[\Mbar]$ is an R-matrix and satisfies rule~6,
with either status assuring
that one or more solutions exist for all curvatures
(rules~4 and~6).
\par
Multiple solutions (bifurcations) occur among 6
of the 12 situations, and the numbers and types of their solutions
depend upon the curvatures of the particles at the
contacts.
The bifurcations include systems with
two, three and four Type~I solutions in
which the \LCP\ solutions are isolated in
$\boldsymbol{\lambda}$-space.
Analogs of the system's Type~I solutions are
those in Figs.~\ref{fig:ExampleLCPs}a--c.
For example, in Case~1.C, $[\Mbar]$ is an N-matrix
of the second kind with $[\dQbar]>0$, thus assuring
two isolated solutions (rules~8 and~10, with its analog
in Fig.~\ref{fig:ExampleLCPs}c, case~A).
\par
Two cases, 1.A and 3.A, exhibit Type~II bifurcation, with each case
having a non-isolated, continuous convex set of solutions in
$\boldsymbol{\lambda}$-space,
consistent with rules~12--15
(Fig.~\ref{fig:ExampleLCPs}d
is an analog of the case I.A in
Table~\ref{table:threeparticles}).
These Type~II bifurcations are also degenerate solutions of
the \LCP, and each bifurcation
constitutes a continuous convex set of solutions:
from one extreme-point in which the top contact slides and
the bottom contact is elastic, to the other extreme-point
in which the bottom contact slides and the top contact
is elastic.
Any convex linear combination of the two extremes is also
a non-isolated solution, forming
a one-dimensional subspace in the $\dLambda$ space, $\mathbb{R}^{2}$
(rule~14).
\par
Although multiple solutions are available in many of the
situations,
certain solution paths are favored;
whereas, the other paths
are unstable by the criterion of Eq.~(\ref{eq:PI2}).
These unstable paths are indicated in the sixth column
of Table~\ref{table:threeparticles},
with the remaining solution being the
stable, preferred path.
\par
The path instability criterion can also be applied
to the Type~II bifurcations
of Cases~1.A and~3.A.
In these cases,
slip can occur exclusively
in either of the two contacts,
but slip can also occur simultaneously in both
contacts and in any combination.
No single path among this range of non-isolated
paths is preferred, as all have the same $\mathcal{I}_{2}$,
indicating that each is equally likely, at least, as far as can be
discerned from a second-order formulation of
the problem (i.e., that of Section~\ref{sec:formulate1}).
The situation is similar
to that of Case~D in Fig.~\ref{fig:ExampleLCPs}d, having
a range of solutions along the line segment between
points (0,1) and (1,0).
Note that Case~1.A also has a single Type~I isolated solution
in which both contacts are elastic, but this solution has a larger
$\mathcal{I}_{2}$ than the non-isolated solutions,
thus making the elastic path unstable.
\par
These same two cases with Type~II bifurcations also
exhibit the path-sensitive pathology,
in which slight changes in the system (surface curvatures,
contact stiffnesses or friction coefficients, etc.)
produce large, abrupt changes in the system's response.
In summary, a continuous range of relative amounts of sliding at
the two contacts can occur in Cases~1.A and~3.A,
such that the relative amount of sliding is 
accidental,
and slight changes in the system can favor sliding
at one or the other contact.
\par
For the angles $\beta$ of Cases~1 and~2,
the systems suffer unsustainability:
a change in the system can ensue,
with a spontaneous gain in kinetic energy (and, hence, instability),
without a change in loading.
The particular unsustainable mode is a slip of both contacts,
propelling particle~2 toward the right;
and (because of the asymmetry of the two contacts
in cases~B, C, and~D) a simultaneous
rotation of the particle.
For each of the four curvature variations of Case~1,
the system is not only unsustainable and in a pending condition
of collapse, but the amount and direction of movement and rotation
is indeterminate (the latter resulting from the
system's Type~II bifurcation).
\par
For Case~2, this unsustainable condition occurs even though
the system is in equilibrium in the current (unloaded) configuration
and no solutions exist for the intended subsequent loading.
Finally, for the angle $\beta$ of Case~2,
no solutions exist that are consistent with
downward loading of the particles and a constant
lateral force $\mathbf{b}^{2}$~--- an example of path incongruity~---
even though the system is in equilibrium under the
current (unloaded) condition.
Moreover,
the current system is in a tenuous condition of pending collapse, as
the system is also unsustainable at angle $\beta$,
and the single unsustainable mode will prompt the middle particle
to move toward the right, causing the system to
pass to the smaller angles $\beta$ of Case~3.
\par
The situation is more favorable with Case~3.
For all four curvature variations,
collapse is not pending, and with curvatures~B and~C,
the outcome is a single, predictable solution.
With curvature~A, the system remains stable, but because
of its Type~II bifurcation, the relative amounts and directions
of movement and rotation are indeterminate and are sensitive
to the infinitesimal changes in the particles' arrangement.
With curvature~D, the outcome is predictable (with a single
likely path), although other equilibrium paths are available
but are unstable.
\subsection{Example~2: tetragonal system of spheres}\label{sec:spheres}
\begin{figure}
  \centering
  \includegraphics{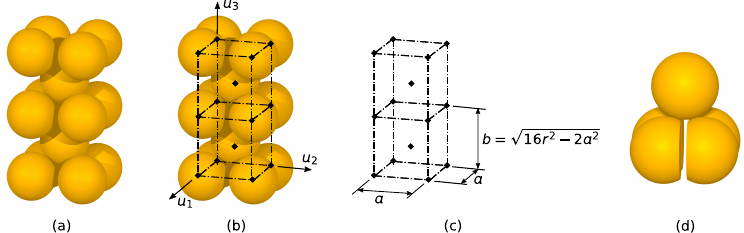}
  \caption{Fourteen spheres in a body-centered tetragonal
           system: (a)~arrangement, (b)~centers, (c)~dimensions, and
           (d)~five lower spheres, showing truncations necessary to
           avoid overlaps.
           \label{fig:BCC}}
\end{figure}

Fourteen frictional
spheres with radius $r$ are arranged as a stacked
system of two body-centered tetragonal units (Fig.~\ref{fig:BCC}).
The system is vertically compressed in the manner of standard
drained
triaxial loading of sands,
by allowing lateral movements while
maintaining constant lateral pressure
on the system's sides.
The system consists of three horizontal square planes, with
four ``planar'' spheres at each square's corners,
and two
``central' spheres that are sandwiched between the planes.
The planar spheres do not touch each other,
but each central sphere contacts eight planar spheres
(i.e., four in the plane above, and four in the plane below),
so that the system has a total of 16 contacts.
\par
At time $t$,
all contacts are at the friction limit,
$|\mathbf{f}^{\,\text{t}}|= \mu f^{\,\text{n}}$, with
tangential forces $\mathbf{f}^{\,\text{t}}$
in directions consistent with the planar spheres being
splayed outward from the system's central axis, when the
system is vertically compressed.
As such, all contacts are active and $M_{\lambda}=16$.
External forces are required on the 12 planar
spheres to equilibrate their contact forces.
Lateral components of these external forces
(components $b_{1}^{p}$ and $b_{2}^{p}$) are
used in finding an equivalent external ``membrane''
pressure on the assembly's sides, formed by $u_{1}$--$u_{3}$
and $u_{2}$--$u_{3}$ planes that pass through the centers of the
outer, planar particles.
As with conventional drained triaxial compression testing,
this membrane pressure remains constant.
\par
In increment $dt$,
vertical compression is simulated by applying an
\emph{average} downward
movement to the four upper planar spheres.
This control program can be imagined as a horizontal plane
(a ``cap'') that passes through the centers of the four
top spheres, with the center of the cap displaced
downward.
The four spheres' movements were also constrained against a
vertical warping of this upper plane.
In addition to applying the downward movement,
three alternative types of constraints on the upper spheres
(i.e., control programs)
were considered:
whether the four spheres (and the imagined cap)
were allowed to shift laterally in
$u_{1}$ and $u_2$ directions;
whether the spheres were allowed to tilt the upper plane about
the $u_{1}$ and $u_2$ directions;
and whether the four spheres could collectively translate
to twirl the cap about the $u_{3}$ axis.
These control programs, along with additional constraints
described in the next paragraph
were implemented with
the system's $[\mathbf{C}]$ constraint matrix.
\par
The four bottom spheres form the system's base.
Vertical movements of their centers were disallowed,
and the average of their horizontal
movements was constrained to zero, thus preventing a
collective shifting (note that such shifting was allowed
with other particles, so that a constraint on the base particles
is needed to prevent horizontal drift of the entire system).
A rigid-body twirling of the system about the $u_{3}$ axis
was also prevented, with a row in $[\mathbf{C}]$
that contains a
constraining sum for the base particles,
$\mathbf{e}_{3}\cdot\sum{\mathbf{u}\times d\mathbf{u}}=0$.
(Note that such twirling can be mistaken for an unsustainable
mode, since the resulting second-order product 
$[d\mathbf{b}]^{\text{T}}[d\mathbf{u}]$ will be negative.)
\par
Vertical compression of the system
causes the sides' areas,
originally of dimensions $a$ and $2b$, to change,
and the lateral external forces on the planar particles
were accordingly changed, thus
simulating a constant controlled membrane pressure.
This small effect is modeled with
a geometric stiffness $[\mathbf{H}^{\text{g-4}}]$
to induce changes in loading due to dimensional changes
(i.e., term $[\partial\mathbf{p}/\partial\mathbf{x}][\dX]$
in Eq.~\ref{eq:dbG}).
\par
Attention was also given to constraints on the spheres' rotations.
Without rotational constraints on the top or bottom spheres,
moment equilibrium is not possible, since tangential (frictional)
contact forces occur at their contacts with the central spheres.
As such,
rotations of these eight spheres were prevented
(note that in triaxial testing of sands, particles that
touch platens are also met with a greater resistance to
rotation than are interior particles).
Because of symmetry, contact forces on the
middle four planar particles are in
equilibrium, and
no external rotation constraints were applied to these spheres or
to the two central spheres.
\par
For relatively squat assemblies, with $b/a<\sqrt{2}$,
the planar particles do not touch;
but for taller assemblies with $b/a>\sqrt{2}$,
the planar particles will overlap.
To prevent unrealistically large overlaps, we imagine that
the planar spheres of taller assemblies are truncated where they
would otherwise overlap, leaving a gap, so that
contacts only occur with the central spheres
(Fig.~\ref{fig:BCC}d).
\par
Rather than finding the full stress-strain behavior of
the system,
snapshots are investigated at specific ratios $b/a$,
with the assumption that all 16 contacts are initially active
(at the friction limit).
Different behaviors are expected in
three ranges of height-width ratios.
For ratios $b/a<1$, the assembly tends to contract
(with volume decreasing) during downward compression,
while the system also softens (with a decrease in the downward
cap load).
With $1<b/a<\sqrt{2}/\mu$, the assembly tends to dilate while the
system softens.
With $b/a>\sqrt{2}/\mu$, the assembly tends to both
dilate and harden.
The last two ranges are investigated herein.
\par
Four cases are considered (Table~\ref{table:BCC}):
two height-width ratios, $b/a=$ 2.0 and 3.0;
and for each ratio, two types of cap constraints:
(a) with the cap constrained from tilting, shifting, and twirling,
and (b) with unconstrained freedom of these motions.
The following contact parameters were assigned:
$k=1$, $\alpha = 1$, $\mu=0.5$, and $f^{\text{n}}=0.001$.
With these parameters, the shorter assembly tends to dilate
and soften; whereas, the taller assembly
tends to dilate and harden.
\par
In all four cases, $M_{\lambda}$ is 16,
such that the $2^{M_{\lambda}}$ trials of
a full enumerative investigation required only modest
computer time.
Matrices $[\Mbar]$ have the following characteristics:
\begin{itemize}
\item
In all cases, $[\Mbar]$ is indefinite,
being neither positive-definite nor
negative-definite.
\item
In all cases, $[\Mbar]$ is degenerate and is neither a P-matrix,
P\textsubscript{0}-matrix,
N-matrix, nor
N\textsubscript{0}-matrix.
\item
In all cases, the $[\Mbar]$ is neither a Z-matrix nor satisfies rule~6.
\item
For the shorter assembly ($b/a=2$, cases~1 and~2),
$[\Mbar]$ is an R matrix; but for the taller
assembly ($b/a=3$, cases~3 and~4),
$[\Mbar]$ is non-R.
\item
For the shorter assembly,
the matrix is an S-matrix,
but it is not an S-matrix for the taller assembly.
\item
In all cases, $[\Mbar]$ is an R\textsubscript{0}-matrix.
\end{itemize}
%
%
For the shorter assembly ($b/a=2$),
we conclude that $[\Mbar]$ can be a non-Q matrix,
such that solutions are not guaranteed for all
loading parameters $[\dQbar]$
(i.e., possible path-incongruity),
but that multiple solutions (i.e., bifurcations)
are possible for other $[\dQbar]$ (rules~2--6).
The taller assembly, however,
is a Q-matrix, assuring solutions (possibly multiple)
for all $[\dQbar]$.
Because $[\Mbar]$ is degenerate, non-isolated solutions
(and, hence, continuous
Types~II bifurcations) as well as isolated Type~I bifurcations
are possible in all cases (rules~10--13).
Finally,
because $[\Mbar]$ is in class~R\textsubscript{0},
the assemblies cannot have Type~III unbounded bifurcations
and cannot encounter neutral equilibrium
(rules~16--20, Section~\ref{sec:NE}).
\par
Table~\ref{table:BCC} gives the number of solutions~--- both
isolated and non-isolated~--- for each of the four cases,
corresponding to
Type~I (discrete) and Type~II (continuous-bounded) bifurcations.
Each solution is a viable response of the spheres to
vertical compression, a response in the form of movements
and rotations, as well as of sliding or non-sliding
at their contacts.
The large numbers
of solutions (i.e. over one thousand) are due, in part, to symmetries
inherent in the systems, but
the large numbers also result from the unusually dense proximity of
active contacts that are all on the verge of sliding~--- with
$M_{\lambda}=M$, and with 16
active contacts among only 14 spheres~--- enabling
$2^{16}$ possible combinations of sliding/non-sliding
contacts among
the small cluster of particles.
\par
The non-isolated but bounded
Type~II bifurcations, which occur with the taller
assembly, are particularly interesting,
as each of these solutions is, itself, a continuous set of viable
solutions.
Each continuous-bounded solution is present in $\dLambda$-space
as a lower-dimensional convex region in which every point
is a solution of the \LCP.
The algorithm in \ref{sec:enum} identifies the
extreme points that form the convex hull of each of the many
Type~II solutions.
The simplest of these
solutions are one-dimensional line segments in $\dLambda$-space,
and many instances of such solutions are realized in
the taller assembly.
Some solutions, however, are more complex, with some being a
convex region of dimension 4 having 16 extreme points,
embedded in the 16-dimensional $\dLambda$-space.
\par
Although each case has numerous solutions (paths)
that are available during vertical compression,
in each case,
particular paths $[\dX^{0}]$ are energetically favored,
having the minimum second-order rate of internal entropy,
$\mathcal{I}_{2}([\dX^{\alpha}])$ of Eq.~(\ref{eq:I2}).
In this equation, the inner product
$[\dY^{\alpha}]^{\text{T}}[\dX^{\alpha}]$
is the downward reaction force increment exerted
on the four spheres by the cap, $[\dY^{\alpha}]$,
multiplied by the cap's downward movement $[\dX^{\alpha}]$.
The forces $[\dP^{\alpha}]$ in the
inner product $[\dP^{\alpha}]^{\text{T}}[\dX^{\alpha}]$
are changes in the side forces due to the
constant chamber pressure multiplied by
changes in the side areas,
as computed with the $[\mathbf{H}^{\text{g-4}}]$ geometric stiffness
and Eq.~(\ref{eq:Hg4}),
$[\dP^{\alpha}]=-[\mathbf{H}^{\text{g-4}}][\dX^{\alpha}]$.
Because this latter effect is small,
the stable response is the one that
offers the least resistance to the downward-moving cap.
\par
Features of the stable solutions are given
in the final four columns of Table~\ref{table:BCC}.
In all cases, multiple solutions share the same minimum value
of $\mathcal{I}_{2}$, which is due, in part, to symmetries;
in some cases, however, solutions with the same minimum value
are not symmetric equivalents.
When solutions share the same minimum
$\mathcal{I}_{2}$,
no single solution is favored,
and the outcome is accidental.
For the shorter assembly ($b/a=2$, Cases~1 and~2),
the stable minimum-$\mathcal{I}_{2}$
solutions are non-degenerate,
so that the solutions are isolated (Type~I, from rule~11);
for the taller assembly (Cases~3 and~4),
the minimum-$\mathcal{I}_{2}$
solutions are degenerate and satisfy rule~13, so they are
continuous Type~II solutions,
with each solution actually being a range of solutions.
For Cases~3 and~4, the numbers of these distinct continuous
convex sets of solutions are shown in the table.
Points within each continuous convex set of a Type~II solution
also share the
same $\mathcal{I}_{2}$: not only is the outcome of a particular convex set
accidental, but
the solution (point) within the set is also accidental.
\par
One obvious possible solution is for all spheres
to not rotate, and for all contacts to behave similarly,
by uniformly either sliding or elastically unloading.
This idealized homogeneity has been a common assumption for
strength and stiffness theories of infinite regular
arrays of spheres \cite{Rowe:1962a,Thornton:1979a,Chang:1989c}.
With this finite system of 14 spheres,
zero-rotation solutions do exist in all four cases,
but these solutions either are unstable or do not imply
a uniform sliding or elastic unloading among the contacts.
In Case~1, the preferred, stable solution is a
zero-rotation solution, but only 4 of the 16 contacts slip,
while the other 12 behaved elastically.
Case~2 also has zero-rotation solutions,
but none of these are stable solutions, and they all
exhibit a mix of slipping and elastic contacts.
Cases~3 and~4 have a zero-rotation solution in which all contacts
elastically unload,
but this solution is the \emph{least} favored,
having the largest $\mathcal{I}_{2}$ among the more than one thousand
solutions.
\par
Two other
pathologies~--- path-sensitivity and instability of
equilibrium~--- are also present in the
four cases, with both pathologies occurring in
multiple modes.
A path-sensitive solution $[\dLambda]$
is one that can vanish or abruptly (discontinuously)
change in number or direction with a small continuous change in the
data $[\Mbar]$ or $[\dQbar]$~---
changes that can result from perturbations in a system's geometry
or in the contacts' conditions (Section~\ref{sec:sensitivity}
and the algorithm in \ref{sec:sensalg}).
A sensitive solution can be either isolated or non-isolated:
in Cases~1--3, all path-sensitive solutions
are also non-isolated (Type~II bifurcations);
but in Case~4, both isolated and non-isolated
(Types~I and~II) solutions
occurred among the path-sensitive solutions,
meaning that any of the isolated solutions
or any solution among continuous
ranges of Type~II solutions can be favored,
depending on infinitesimal changes in the input conditions.
In Cases~1 and~2, path-sensitivity is absent in
the favored, stable solution path,
but each of these cases has multiple (more than a thousand)
path-sensitive solutions,
many with $\mathcal{I}_{2}$ values that are close to the minimum,
stable value.
Although these solutions are less-favored when spheres
are perfectly arranged,
these path-sensitive solutions
are available, and
in the event of small
perturbations to the system, can become favored solutions.
\par
Modes of 
unsustainability
were identified using the algorithm in \ref{sec:setting}.
Such modes were found
in all four cases,
with each case exhibiting multiple unsustainability directions,
signaling the possibility of
accelerated motion
upon subsequent loading in these directions,
even though the system is in equilibrium in
the current (unloaded) condition.
\begin{table}
  \caption{Results for the triaxial compression of 
           a tetragonal system of 14 spheres
           (Fig.~\ref{fig:BCC}).
           \label{table:BCC}}
  \centering
  \small
  \setlength{\tabcolsep}{5.5pt}
  \begin{threeparttable}
    \begin{tabular}{lcccccccc}
      \toprule
      & \multicolumn{4}{c}{Numbers of pathologies} &
        \multicolumn{4}{c}{Stable solutions\tnote{4}}\\
      \cmidrule(r){2-5}
      \cmidrule(r){6-9}
           & \multicolumn{2}{c}{Bifurcations\tnote{1}} &         & 
           & & & & \\
      \cmidrule(r){2-3}
      Case & Type I & Type II & \multirow[b]{-2}{1.0cm}{Path- sensit.\tnote{2}}
                           & \multirow[b]{-2}{1.5cm}{\centering Unsustain. modes\tnote{3}}
           & Number & Degen. & Isol.
                              & \multirow[b]{-2}{1.0cm}{Path- sensit.} \\
      \midrule
      Shorter, $b/a = 2.0$ &&&&&&&&\\
      \quad(1)~Constrained cap 
           & 11 & 1012 & 1012 & 10,586 & 2 & - & \checkmark & -\\
      \quad(2)~Unconstrained cap & 179 & 1660 & 1660 &38,719& 4 &-&\checkmark&-\\
      \midrule
      Taller, $b/a = 3.0$ &&&&&&&&\\
      \quad(3)~Constrained cap & 184 & 1700 & 1700 & 709& 66 & \checkmark&-&\checkmark\\
      \quad(4)~Unconstrained cap & 421 & 1152 & 1425 & 65& 49 & \checkmark& -& \checkmark\\
      \bottomrule
    \end{tabular}
    \begin{tablenotes}
       \item[1] Numbers of bifurcation solutions (Section~\ref{sec:bif}).
                Type~I solutions are isolated solutions.
                Type~II solutions, with each solution being a
                range of solutions.
       \item[2] Numbers of the solutions that are path-sensitive
                (Section~\ref{sec:sensitivity}).
       \item[3] All four cases present unsustainability
                (Section~\ref{sec:instequil}).
                Column gives the numbers of modes of unsustainability.
       \item[4] Of the Types~I and~II solutions,
                a subset are path-stable (Section~\ref{sec:PI1}).
                The four columns give their numbers and
                characteristics: whether the stable solutions
                are degenerate (Deg.), Isolated (Isol.),
                and path-sensitive (Path-sensit.).
    \end{tablenotes}
  \end{threeparttable}
\end{table}
\par
In summary, all four cases (shorter and taller assemblies,
constrained and unconstrained cap)
present multiple unstable solutions
(path instability,
just as Euler and Shanley columns present unstable solutions),
but all four cases also have multiple stable solutions.
The particles' movements in each case are unpredictable,
as any of the stable solutions is equally likely.
With the shorter assembly, these solutions are separate (isolated)
patterns of movement;
whereas, the taller assembly presents continuous (non-isolated)
ranges of movement patterns,
with the likelihood of any particular pattern being
sensitive to the assembly's initial conditions.
Even though the systems are in equilibrium in both their
initial and displaced conditions,
each system is replete with unsustainability modes,
meaning that accelerated motions are possible
in the directions of these modes,
even as the system undergoes stable movement.
\par
As a final variant,
Case~2 was altered by flattening the 14 spheres at their 16 contacts,
by increasing the radii of curvature $\rho$ by a factor of 2.0
while maintaining the \emph{original} radial distances $r$.
With the particles
becoming flattened spheres at the contacts,
this alteration was made to investigate
the effect of the system's geometric stiffness on its response
(Sections~\ref{sec:equilp}--\ref{sec:GeomH}).
The impact on Case~2~--- the shorter assembly with an unconstrained
cap~--- was substantial,
and, as with the three-particle example of the previous section,
the 14 spheres demonstrated a strong geometric influence.
Only 39 solutions, all of Type~I, remained,
compared with the 179 Type~I and 1660 Type~II solutions
of the unaltered system;
none of the solutions was path-sensitive, compared with 1660
for the unaltered system;
and the unsustainable modes
were reduced in number from 38,710 to 16,915.
Both the unaltered and altered systems exhibited softening during
the downward loading,
but the rate of softening was much less with the flattened spheres:
the softening rate (of load reduction) of the altered system was only 40\%
that of the unaltered system.
This variant illustrates that the particles' shapes at their
contacts (i.e., their curvatures $\mathbf{K}^{pq,p}$,
as in Eq.~(\ref{eq:drKPKQ}) and~(\ref{eq:deltan}))
can greatly influence an assembly's stability,
sustainability, path-sensitivity, bifurcation susceptibility,
and bulk stiffness.
\subsection{Example~3: biaxial compression of irregular assembly of disks}%
\begin{figure}
  \centering
  \includegraphics{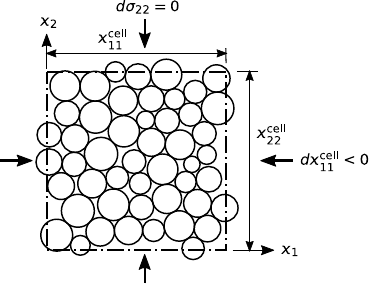}
  \caption{Assembly of 49 disks, loaded with horizontal compression
           and constant vertical stress.
           \label{fig:Assembly}}
\end{figure}

\label{section:disks49}
The final example is slow, quasi-static
loading of a small 2D assembly of 49 frictional disks,
with the purpose of exploring particle-scale mechanisms
and pathologies that lead to failure
(Fig.~\ref{fig:Assembly}).
The loading was accomplished in a series of increments,
with the particles' response in each increment determined by solving
a linear complementarity problem (LCP), as with Problem~2.
Although the author has used a similar method with
much larger assemblies,
doing so in a reasonable time requires treating each
loading increment as a Problem~3 quadratic program (QP),
thus foregoing a full enumerative analysis that can identify all
bifurcations and other pathologies at each increment
(see Table~\ref{table:summary1}).
The small number of 49 particles enabled a full enumerative
analysis in reasonable computation time.
The simulation method differs from
discrete element (DEM), discontinuous deformation analysis (DDA),
contact dynamics (CD), and other methods that model dynamic,
momentum-transfer processes
and rely upon the particles' accelerations to impel their
rearrangement.
With the method herein,
simulating true quasi-static loading,
the particles move from one equilibrium
state to another, in response to increments of the loading
path.
\par
The assembly is contained within periodic boundaries,
having dimensions $x_{11}^{\text{cell}} \times x_{22}^{\text{cell}}$
and shearing offset $x_{12}^{\text{cell}}$.
Biaxial compression is implemented
with mixed boundary conditions: reducing the horizontal
dimension in increments $dx_{11}^{\text{cell}}<0$ while
maintaining a rectangular
shape ($dx_{12}^{\text{cell}}=0$)
and constant vertical stress ($d\sigma_{22}=0$).
The average Cauchy stress $\boldsymbol{\sigma}$,
referenced to the deformed configuration,
is determined with the Love-Weber
equation, and the stress increment $d\boldsymbol{\sigma}$
ensues from the equation's differential:
\begin{equation}\label{eq:stress}
  \sigma_{ij} = \frac{1}{V}\sum_{pq} r_{i}^{pq,p} f_{j}^{pq},\quad
  d\sigma_{ij} = -\frac{dV}{V}\frac{1}{V}\sigma_{ij}
    + \frac{1}{V}\sum_{pq} dr_{i}^{pq,p} f_{j}^{pq}
    + \frac{1}{V}\sum_{pq} r_{i}^{pq,p} df_{j}^{pq}
\end{equation}
%
where, as described below, the second equation is the basis of
a matrix expression of stress increments that can be
appended to stiffness $[\mathbf{H}]$ so that
$\sigma_{22}$ is maintained constant.
All incremental quantities in Eq~(\ref{eq:stress}\textsubscript{2})
depend on the particle displacements $[\dX]$, boundary
displacements $[d\mathbf{x}^{\text{cell}}]$, and
slip increments $[\dLambda]$:
the volume strain
$dV/V=\sum_{i}{(dx^{\text{cell}}_{i}/x^{\text{cell}}_{i})}$;
the contact displacements
$d\mathbf{r}^{pq,p}=
\delta\mathbf{r}^{pq,p} + d\boldsymbol{\theta}^{p}\times\mathbf{r}^{pq,p}$,
with $\delta\mathbf{r}^{pq,p}$
given by Eqs.~(\ref{eq:delrpq})--(\ref{eq:drKPKQ});
and the force increments
$d\mathbf{f}^{pq}=\mathfrak{d}\mathbf{f}^{pq}
+\delta\hat{\mathbf{f}}^{pq}
+d\boldsymbol{\theta}^{p}\times\mathbf{f}^{pq}$, with
increment $\delta\hat{\mathbf{f}}^{pq}$ given
by Eqs.~(\ref{eq:drKPKQ}), (\ref{eq:dfhat}),
and~(\ref{eq:deltan}).
In the simulation, each force increment
$\mathfrak{d}\mathbf{f}^{pq}$
obeys the
standard linear-frictional model of Section~\ref{sec:linfrict}
and
derives from the contact deformation
$\delta\mathbf{x}^{\text{def},\,pq}$ and the frictional
slip $d\lambda_{\text{s}}^{pq}$
(Eq.~\ref{eq:dfForm2}).
A contact's deformation $\delta\mathbf{x}^{\text{def},\,pq}$
depends on the movements
of particles $p$ and $q$ (Eq.~\ref{eq:dudef});
but for those contacts that straddle periodic boundaries,
deformation
$\delta\mathbf{x}^{\text{def},\,pq}$ also depends upon
the boundary movement $d\mathbf{x}^{\text{cell}}$.
Equation~(\ref{eq:stress}\textsubscript{2}) is assembled
in the matrix form
\begin{equation}
  \left[
      \arraycolsep=3pt\def\arraystretch{1.0}
      \begin{array}{@{\extracolsep{0ex}}c;{1pt/1pt}c;
                      {1pt/1pt}c@{\extracolsep{\fill}}}
        \mathbf{H}_{\boldsymbol{\sigma}\mathbf{x}}
        &
        \mathbf{H}_{\boldsymbol{\sigma}\mathbf{x}^{\text{cell}}}
        &
        \mathbf{H}_{\boldsymbol{\sigma}\boldsymbol{\lambda}}
      \end{array}
  \right]_{3\times (3N + 3 + M_{\lambda})}
  \left[
    \arraycolsep=1pt\def\arraystretch{1.0}
    \begin{array}{l@{\extracolsep{\fill}}}
        \dX\,
        \\\hdashline[1pt/1pt]
        \dX^{\text{cell}}\,
        \\\hdashline[1pt/1pt]
        \dLambda\,
    \end{array}
  \right]
  =
  \left[
    d\boldsymbol{\sigma}
  \right]_{3\times 1}
\end{equation}
where, in 2D,
vector $[d\boldsymbol{\sigma}]$ holds the
three stress increments $d\sigma_{11}$, $d\sigma_{22}$,
and $d\sigma_{12}$;
the $6N$ displacements (3D) of $[\dX]$ are replaced with $3N$
displacements (2D); and
vector $[\dX^{\text{cell}}]_{3\times 1}$ holds three
resizing increments:
elongations $dx_{11}^{\text{cell}}$ and $dx_{22}^{\text{cell}}$,
and shearing displacement $dx_{12}^{\text{cell}}$.
\par
To manage the mixed boundary conditions,
vectors $[\dX^{\text{cell}}]$
and $[d\boldsymbol{\sigma}]$ are appended to $[\dX]$
and $[\dP]$, and
the stiffness submatrices of Eqs.~(\ref{eq:form2b})
and~(\ref{eq:hxxdx})--(\ref{eq:dQbar}) now connote
combinations of multiple stiffness effects:
\begin{equation}\label{eq:BCs}
  \begin{gathered}
    \left[\dX\right]
    \leftarrow
    \left[
      \arraycolsep=1pt\def\arraystretch{1.0}
      \begin{array}{c}
          \dX
          \\\hdashline[1pt/1pt]
          \dX^{\text{cell}}
      \end{array}
    \right]
    ,\quad
    \left[\dP\right]
    \leftarrow
    \left[
      \arraycolsep=1pt\def\arraystretch{1.0}
      \begin{array}{c}
          \dP\,
          \\\hdashline[1pt/1pt]
          d\boldsymbol{\sigma}\,
      \end{array}
    \right]
    \\
    \left[\mathbf{H}_{\mathbf{xx}}\right]
    \leftarrow
    \left[
      \arraycolsep=3pt\def\arraystretch{1.0}
      \begin{array}{@{\extracolsep{0ex}}c;{1pt/1pt}c@{\extracolsep{\fill}}}
        \mathbf{H}_{\mathbf{xx}}
        & \mathbf{H}_{\mathbf{xx}^{\text{cell}}}
        \\\hdashline[1pt/1pt]
        \mathbf{H}_{\boldsymbol{\sigma}\mathbf{x}}
        & \mathbf{H}_{\boldsymbol{\sigma}\mathbf{x}^{\text{cell}}}
      \end{array}
    \right]
    ,\quad
    \left[\mathbf{H}_{\mathbf{x}\boldsymbol{\lambda}}\right]
    \leftarrow
    \left[
      \arraycolsep=1pt\def\arraystretch{1.0}
      \begin{array}{c}
        \mathbf{H}_{\mathbf{x}\boldsymbol{\lambda}}
        \\\hdashline[1pt/1pt]
        \mathbf{H}_{\boldsymbol{\sigma}\boldsymbol{\lambda}}
      \end{array}
    \right]
    ,\quad
    \left[\mathbf{H}_{\boldsymbol{\lambda}\mathbf{x}}\right]
    \leftarrow
    \left[
      \arraycolsep=3pt\def\arraystretch{1.0}
      \begin{array}{@{\extracolsep{0ex}}c;{1pt/1pt}c@{\extracolsep{\fill}}}
        \mathbf{H}_{\boldsymbol{\lambda}\mathbf{x}}
        & \mathbf{H}_{\boldsymbol{\lambda}\mathbf{x}^{\text{cell}}}
      \end{array}
    \right]
  \end{gathered}
\end{equation}
In a similar manner, rows are added to $[\mathbf{C}]$
and $[\dC]$ to supply a resizing of the periodic cell's
dimensions, $dx^{\text{cell}}_{11}$ and $dx^{\text{cell}}_{12}$.
Because these dimensions
are controlled,
the corresponding stress responses,
$d\sigma_{11}$ and $d\sigma_{12}$,
are computed as $d\mathbf{y}$ reaction forces, as in Eqs.~(\ref{eq:hxxdx})
and~(\ref{eq:dy23}).
\par
The 49 disks were polydisperse, with mean, median, maximum,
and minimum radii of 1.0, 1.0, 1.26, and 0.63, respectively.
(Herein, dimensional units are set aside, since only the
two units of length and force are required with the inertia-free
assumption.)
The contact parameters were as follows:
normal contact stiffness
$k=1$, tangential/normal stiffness ratio $\alpha=1$,
and friction coefficient $\mu = 0.5$.
The initial assembly had 92 contacts and a roughly isotropic
fabric, with mean stress
$\frac{1}{2}(\sigma_{11}+\sigma_{22})=-0.0025$ and
an average overlap of the disks of about $5\times 10^{-3}$ times their
mean radius.
\par
Loading was advanced in small increments,
$d\varepsilon_{11}=-5\times 10^{-6}$.
With each increment, the following calculations were
performed: the various matrices
$[\mathbf{H}_{\mathbf{xx}}]$,
$[\mathbf{H}_{\mathbf{x}\boldsymbol{\lambda}}]$, etc.
were assembled;
matrix $[\Mbar]$ and vector $[\dQbar]$ were computed;
the values $[\dLambda]$ and $[\dX]$ were found
by solving either the QP of Problem~3 when $[\Mbar]$ was PD
or, otherwise, the LCP 
was solved using
an enumerative algorithm
(\ref{sec:enum}); 
and the particles' positions and orientations were advanced to
$[\mathbf{x}]+[\dX]$.
When increments $[\dX]$ produced fresh contacts,
their normal forces were
set to stiffness $k$ times the disks' overlaps
in their new positions $[\mathbf{x}]+[\dX]$,
with zero initial tangential force.
Likewise, contact forces were set to zero for
those contacts that disengaged during $[\dX]$.
These alterations
are included in the $[\Mbar]$ used in the next increment,
and they contribute to deviations from equilibrium.
\par
Even though quasi-static conditions are intended,
small deviations from equilibrium occur with each
loading increment, from several sources:
newly formed contacts, newly disengaged contacts,
changes in the assembly's geometry of an order higher
than the second-order estimates of sines and cosines,
and shape contours of an order higher than second-order curvatures.
The average imbalance of force on a particle was about
$2\times 10^{-7}$ times the mean contact force.
Although small,
any unbalanced force would be assigned as $[-\dP]$ in the
subsequent increment.
%
\par
\begin{figure}
  \centering
  \includegraphics[scale=1.00]{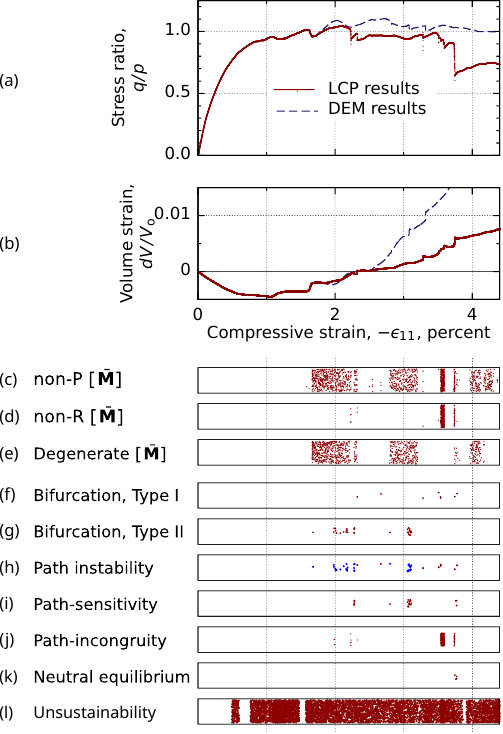}
  \caption{Results of biaxial compression of 49 disks:
           (a)~ratio $q/p$ is of the deviator stress,
           $q=\sigma_{11}-\sigma_{22}$, and the mean stress,
           $p = \frac{1}{2}(\sigma_{11}+\sigma_{22})$;
           (b)~volumetric strain;
           (c)--(l)~jitter plots of various conditions.
           The vertical axes of jitter plots are used to spread
           the specks to distinguish these individual events.
           In (h), red specs are path instabilities; 
           blue specs are indeterminate paths.
           \label{fig:AssemblyResults}}
\end{figure}

Results of biaxial compression are presented in 
Fig.~\ref{fig:AssemblyResults},
which shows the deviator stress
and volume change during loading
(results of DEM simulations are also shown,
and these are discussed in Section~\ref{sec:eff}).
The assembly's behavior is similar to that seen in both
physical and virtual tests of disk assemblies
\cite{Oda:1980a},
and the modest fluctuations of stress~--- sudden
stress-drops followed by more gradual rises~--- are to be expected
for assemblies of this size
\cite{Kuhn:2019a}.
The number of contacts $M$ was 75--90;
the average number of
active (sliding) contacts $M_{\lambda}$ was about 11,
with a maximum of~24;
and, on average, about 13\% of contacts were active.
Throughout the loading,
new contacts are created while existing contacts disengage,
and the subset of active contacts is incessantly reallocated.
\par
The lower part of the figure gives the occurrences~--- as specks
at the occurring strains~--- of
various conditions, including the stiffness pathologies
of Section~\ref{sec:pathologies}.
The vertical axis of the specks has no significance,
other than to jitter (spread) the specks so that they do not appear as
connected lines.
\par
Figs.~\ref{fig:AssemblyResults}c and~\ref{fig:AssemblyResults}d
show the strains at which matrix $[\Mbar]$ was non-P and non-R.
A solution of the \LCP\ is not assured in either case, although
the non-R category is more restrictive (rules~2--4).
A unique solution is certain for P-matrices,
and this situation prevails until the strain of 1.6\%,
which occurs well before the peak stress at strain 2.1\%.
(The weaker condition of being a PD-matrix
held until a strain of 1.2\%.)
Beyond the strain 1.6\%, the $[\Mbar]$ matrices were P-matrices
for only 78\% of strain increments, allowing possible bifurcation
or an absence of solutions
in the remaining 22\%.
The non-R condition was far less frequent than the non-P condition,
and the first non-R occurrence was at strain 2.2\%, shortly after
the peak stress.
A non-R matrix enables the possible absence of solutions in
some loading directions $[\dQbar]$, possibly leading to
path-incongruity
(note that being a non-R matrix is not conclusive,
since $[\Mbar]$ can be a Q-matrix without being
an R-matrix, rule~4).
Although results are not shown in the figure,
the criteria of Z-matrices and of rule~6 were
less helpful in determining the existence of
solutions,
as $[\Mbar]$ satisfies neither criterion at 99\%
of the strains.
%
\par
The jitter plot of 
Fig.~~\ref{fig:AssemblyResults}e shows strains at which matrix
$[\Mbar]$ is degenerate,
a condition that is present at about 8\% of the strain increments.
A degenerate matrix is necessary (but not sufficient)
for several pathologies:
Types~II and~III bifurcation (non-isolated solutions),
neutral equilibrium, and path-sensitivity.
However, because degeneracy is not a sufficient condition
for these pathologies, they occur at much fewer of the increments
(rules~10, 12, 13, 18, and Section~\ref{sec:sensitivity}).
A non-degenerate but non-P matrix can also have multiple
solutions of Type~I
for a loading direction $[\dQbar]$,
and because all $[\dQbar]$ have the same parity in their solutions,
other directions can have an even greater number of solutions (rule~11).
Also note that
a matrix is degenerate if even a single principal minor is zero,
but the simulation exposed
instances of $[\Mbar]$ with several thousands of its minors
being zero, with each such minor opening the possibility
of the pathologies mentioned above.
\par
Bifurcation occurred at 40 strains, located in the
jitter plots of
Figs.~\ref{fig:AssemblyResults}f and~\ref{fig:AssemblyResults}g.
Among these occurrences, 8 were of multiple isolated Type~I
solutions, and 32 were continuous-bounded Type~II solutions.
(Although it is possible to have both types at the same
strain, this situation was not encountered.)
No continuous-unbounded Type~III solutions occurred.
Types~I and~II were distinguished by noting whether
matrix $[\Mbar]$ is non-degenerate and matrix
$[\Mbar_{I_{2}^{\ast}I_{2}^{\ast}}]$
is non-singular for the Type~I bifurcations;
whereas,  $[\Mbar]$ is degenerate for the Type~II
bifurcations, and
$[\Mbar_{I_{2}^{\ast}I_{2}^{\ast}}]$ is singular,
except at the extreme points
(rules~13--15).
The isolated Type~I solutions were in the form of a
few active contacts taking
alternating conditions of continued
sliding or elastic unloading.
At strain 3.3\% the system encountered four isolated solutions.
At this strain, the system had 84 contacts,
of which 12 were active, and Type~I
bifurcation occurred with three of the active contacts
taking four alternative paths of either
sliding or elastically unloading.
\par
Type~II non-isolated (but bounded) bifurcation happens when an
infinity of solutions are available as a continuous
range of sliding or unloading among a subset of the active
contacts.
For the simulation, each non-isolated bifurcation
was a convex one-dimensional subset (a line segment) of solutions
within $\mathbb{R}^{M_{\lambda}}$
(rule~14).
For example,
at strain 2.8\%, the system had 78 contacts, of which 12 were active.
Two of these active contacts could adopt a range of conditions:
the two extreme-points at which one contact was sliding
sliding and the other unloading,
or with both contacts sliding in amounts
that span the continuum of the two extremes.
\par
Path instability
is possible at bifurcation strains
(Section~\ref{sec:instequil}).
For each of the eight Type~I bifurcations, one of the paths
had a lower second-order entropy, $\mathcal{I}_{2}$,
than the other paths,
making these other paths unstable (red specks in 
Fig.\ref{fig:AssemblyResults}f).
With the 32 Type~II bifurcations,
the $\mathcal{I}_{2}$ was slightly
smaller at one extreme-point than at the other.
Although the larger $\mathcal{I}_{2}$ might be interpreted
as unstable, the difference between the two values~--- the smallest
and largest~$\mathcal{I}_{2}$~--- differed
by less than 0.003\% of their mean value, and
in some cases, less than 0.00001\%.
These small differences are on the order of the
small imbalances in contact forces that were the result
of the simulation's
higher-order (i.e., higher than second-order) geometric
effects, as described above.
The author interprets the continuous-bounded Type-II bifurcations
as not engendering a path instability that would favor
a minimum-$\mathcal{I}_{2}$ path and
nudge the system onto that path.
Rather, each solution in the continuous set of solutions
is equally preferred and equally likely,
making the subsequent path indeterminate and accidental~--- at
least, insofar as can be determined with a second-order analysis.
A similar conclusion was reached with the examples of
Sections~\ref{sec:three} and~\ref{sec:spheres},
but although the conclusion in those examples
might be ascribed to symmetry of the particles' arrangements,
the same conclusion pertains to the irregular assembly of this
section.
These cases of an indeterminate (accidental) path, although not a
path instability, are marked with blue specks in
Fig.~\ref{fig:AssemblyResults}h.
\par
The pathology of
path-sensitivity is the sensitivity of
a solution $[\dLambda]$ to the underlying data
$[\Mbar]$ and $[\dQbar]$, with both being derived from the particles'
arrangement and the
contacts' conditions (Section~\ref{sec:sensitivity}).
Twenty-two instances of path-sensitivity occurred in the
simulation (Fig.~\ref{fig:AssemblyResults}i),
and these instances coincided with some of the
Type~II bifurcations,
supporting the accidental nature of the response when a
Type~II bifurcation is encountered.
\par
Perhaps most paradoxical are the 350 instances of
path-incongruity,
in which no solution path, $[\dLambda]$,
is available with the
given control program $[\mathbf{C}]$ and
control parameters $[\dQbar]$
(Fig.~\ref{fig:AssemblyResults}j and Section~\ref{sec:incongruity}).
When an absence of solutions was encountered during the simulation,
we treated the situation as a Problem~3 quadratic program (QP)
and used the path $[\dLambda]$
that minimized the objective function $z(\dLambda)$
in Eq.~\ref{eq:form3}a,
even though this path might not meet the
conditions of Eqs.~\ref{eq:form3}b and~\ref{eq:form3}c.
In defense of this makeshift remedy,
all instances of path-incongruity coincided with
unsustainability, in which
the current ``locked'' state was unsustainable even in the absence of any
further loading (Fig.~\ref{fig:AssemblyResults}l,
as described below).
Similar situations were encountered in the
two earlier examples,
and one can imagine that the paradox of having no paths forward
is resolved by small movements,
prompted in the unsustainable directions,
that nudge the system beyond its current locked condition.
The instances of path-incongruity are shown to
be clustered around four strains (Fig.~\ref{fig:AssemblyResults}j),
and at three of three of these clusters,
the system suffers a drop in stress, which is followed
by an recovery (increase) of stress.
%
\par
Only 4 instances of neutral equilibrium occurred during the
simulation, and these happened
at a strain of about 3.75\%, well past the peak
stress (Fig.~\ref{fig:AssemblyResults}k).
Neutral equilibrium is a state in which movement directions
$[\dLambda]$ are available with zero loading,
$[\dQbar]=[\mathbf{0}]$ (Section~\ref{sec:NE}).
The condition corresponds to a non-R\textsubscript{0}
matrix $[\Mbar]$, when
an unbounded solution set is available
(rules~4 and 16--20, noting that $t$ in Eq.~\ref{eq:R}
is unbounded).
The four instances happened immediately after
a large drop in stress, which coincided with a large increase in volume.
\par
Occurrences of unsustainability are shown
in Fig.~\ref{fig:AssemblyResults}l.
This condition happens when the particle arrangement
and the contact forces enable movements
$[\dX]$ and $[\dLambda]$ that are energetically available
in the absence of further loading, $[\dP]=[\dC]=[\dQbar]=\mathbf{0}$,
as precipitated by displacement modes having negative
second-order work $\ddot{W}_{\text{int}}$
(Section~\ref{sec:instequil} and \ref{sec:setting}).
%
Figure~\ref{fig:AssemblyResults}l shows that
unlike the other pathologies,
unsustainability was not an occasional, infrequent
event, but was common during the simulation.
The first occurrence was at strain 0.49\%,
during strain hardening, when the assembly was still compressing
(i.e., before dilation), and well before the peak stress.
Three gaps are seen in Fig.~\ref{fig:AssemblyResults}l,
which occurred during periods of increasing stress,
but unsustainable modes were otherwise available during
periods of both increasing and decreasing stress.
After the first occurrence, at strain 0.49\%,
unsustainability was present at 75\% of the
strain increments, making this pathology the norm rather than
the exceptional.
The enumerative algorithm of \ref{sec:setting}
identified the unsustainable modes at each strain increment,
finding over 30 different modes at some strains.
\par
One might question why the
negative second-order work
that is available in unsustainable instances,
which, in a dynamic setting,
represents a quadratic increase in kinetic energy,
is not resolved in sudden energetic movements of particles
in our quasi-static analysis.
This matter is addressed in Section~\ref{sec:expaths}.
%
%
\section{\normalsize Discussion and Conclusions}\label{sec:conclusions}
\subsection{Summary of derivations}\label{sec:derivations}
The paper establishes a framework
for the quasi-static loading
of frictional granular systems,
posed as a linear complementarity problem (LCP).
In this manner, a system's response
is reduced to a primal form,
focusing only
on those contacts that have reached the friction limit.
As an LCP, one has available a rich record, developed over the past 50 years,
of those conditions that assure the existence and uniqueness of a solution,
or contrarily, allow for pathologies, such as multiple solutions,
an absence of solutions, or unstable or sensitive solutions.
These LCP conditions
are cataloged in the paper (Section~\ref{sec:exist}),
providing criteria on $[\bar{\mathbf{M}}]$ and
$[d\bar{\mathbf{q}}]$
for evaluating quasi-static granular systems.
These means are of greatest value
for systems with multiple active contacts,
each adding to the number of incremental non-linearities
and exponentially increasing the complexity, as
measured by the number of available solution cones
(tensorial zones).
\par
Besides developing an LCP approach,
the paper adopts a kinematic constraint matrix $[\mathbf{C}]$
to impose the control program,
a method that
has advantages over the more common techniques
of partitioning and rearranging the stiffness equation
$[\mathbf{H}][\dX]=[\dP]$ \cite{Nova:1994a,Wan:2017a}
(a technique in mechanics that is equivalent to Jordan exchange
in mathematical programming \cite{Murty:1972a})
or of introducing Lagrange multipliers \cite{Lerbet:2020a}.
Besides permitting non-homogeneous constraints
(with non-zero $[\dC]$)
and treating the force and kinematic loadings,
$[\dP]$ and $[\dC]$,
in a similar manner (Eqs.~\ref{eq:form1b}b
and~\ref{eq:dQbar}), using
a constraint matrix $[\mathbf{C}]$
greatly simplifies calculating the essential data,
$[\Mbar]$ and $[\dQbar]$, of the \LCP.
\par
The paper also includes
internal geometric effects on a system’s stiffness,
effects that can result from curvatures
of the particles at their contacts
and from follower forces.
Although geometric effects are of second-order
(engendered in a quadratic fit of the particles’ contours at contacts),
it is not unusual for particles to become tenuously arranged
during failure,
and in these circumstances,
such geometric effects can determine a system’s progress
and stability.
In the first example (Section~\ref{sec:three}),
the particles' contours at a single contact affect the number and types
of the system’s solutions.
In the second example, involving 14 tightly packed spheres,
the rate of softening was controlled by geometric effects.
The author suspects that a similar influence applies
when granular media are loaded near the peak stress
and during subsequent softening.
In a continuum setting,
geometric effects are usually associated with the shifting and rotation
of a region's boundaries; whereas, material effects are viewed
as internal within the region.
With systems of discrete grains, however, geometric effects
are also internal within a system and are manifested
at the macro-scale as the material's continuum response.
\subsection{Summary of stiffness pathologies in relation to the examples}%
\label{sec:expaths}
The paper catalogs eight aberrant conditions (stiffness pathologies),
each with an unambiguous definition and which
can be most efficiently evaluated when
a granular system is expressed as an LCP.
As an example, the pathology of neutral equilibrium is difficult
to ascertain when the stiffness matrix 
has a null space of multiple dimensions,
requiring one to evaluate whether the null space intersects the
solution cone for which the stiffness applies.
When placed in the context of an LCP, however,
the question is settled by
whether matrix $[\Mbar]$ is in class R\textsubscript{0}
(Section~\ref{sec:NE}).
\par
Another aberration, path-sensitivity,
is a weak form of Lyapunov instability
(or lack of Lipschitz-like continuity).
To the author’s knowledge,
this pathology is first defined herein
and is only exposed when the problem is posed as an LCP
(Section~\ref{sec:sensitivity}).
This form of instability disregards
notions of energy or second-order work
but instead indicates that a system's response
has a discontinuous, non-smooth reliance on a granular system's
internal arrangement and on its loading.
Path-sensitivity
was manifest in all three of the paper's examples.
\par
Using methods available to LCPs,
the paper defines three types of bifurcations.
If, however, one is limited to evaluating
stiffness matrices without access to the LCP,
it becomes difficult to distinguish or even to recognize the three types
(i.e., with Eq.~\ref{eq:dxdyGBD}),
and, to the author’s knowledge,
Type~II bifurcation
solutions~--- those with a continuous but bounded range~--- are
not yet acknowledged in the mechanics literature.
The three examples in Section~\ref{sec:examples} reveal an
oddity of these Type~II solutions:
when Type~II solutions are encountered, no single stable path
is preferred within the bounded range of solution paths,
leading the author to characterize the outcome as accidental.
That is,
the system can pursue any of a continuous range of solutions,
with no preference for a particular solution within the range.
\par
Unsustainability (instability of equilibrium),
a pathology in which modes of negative second-order work are
available,
occurred in all three examples,
and in the case of biaxial compression of 49 disks
(Section~\ref{section:disks49}),
this pathology was a frequent presence during
strain-hardening and throughout post-peak loading.
With this Example~3,
the unsustainability pathology is problematic,
as one could question the role of second-order work in assessing
a possible quadratic increase
in kinetic energy in an otherwise quasi-static system.
First, unsustainability should be seen as
a latent quality that offers only the potential
of abrupt movements (in various directions),
a potential that can either be realized
or remain unfulfilled during
continued loading in a particular direction.
Second, effects of a higher-order can restore equilibrium
and mitigate sudden movement.
Third,
although kinetic surges are obviated with the quasi-static assumption,
the frequent presence of modes of negative second-order work
demonstrates
that both before and during failure,
particles take on tenuous, insecure arrangements,
while the imposed deformation is carrying them
from one precarious state to another~---
even during quasi-static loading.
Finally,
instances of path-incongruity~---
a separate pathology that occurs
when no solution paths are available~---
were consistently accompanied by
unsustainability,
and path-incongruity usually
coincided with a brief drop in stress.
Because path-incongruity is also a tenuous condition,
the tendency exists in otherwise quasi-static systems for
dynamic movements
to bring the system toward an arrangement
that becomes solvable during subsequent loading
and to bring about a recovery of stress.
\par
The author concludes that the quasi-static assumption
is valid for granular systems,
even when unsustainable modes are available.
Even though such modes are a pervasive presence during
the loading of the large system in Example~3,
unsustainability modes are considered potential local adjustments,
which occur while the system
advances from one equilibrium state to another.
\par
Only three of the eight pathologies are encountered
in the continuum mechanics literature, in relation
to a material's constitutive operator
$\mathcal{M}(\mathbf{x},d\boldsymbol{\varepsilon})$.
With rate-independent continua, the operator is assumed
homogeneous of degree~1 in $d\boldsymbol{\varepsilon}$,
at a point $\mathbf{x}$, with
$d\boldsymbol{\sigma}
=\mathcal{M}(\mathbf{x},d\boldsymbol{\varepsilon})\cdot
d\boldsymbol{\varepsilon}$.
In one pathology,
associated with shear bands and compaction bands,
a \emph{localized bifurcation} in direction $\mathbf{n}$ occurs when
the Rice criterion is met:
when the determinant of the acoustic tensor
$\mathbf{n}\cdot\mathcal{M}\cdot\mathbf{n}$
vanishes in direction $\mathbf{n}$
\cite{Rudnicki:1975a,Rice:1976a}.
This condition is an isolated Type~I bifurcation
(Section~\ref{sec:bif}), allowing two isolated solutions,
$d\boldsymbol{\varepsilon}^{1}$ and $d\boldsymbol{\varepsilon}^{2}$,
for the increment $d\boldsymbol{\sigma}$.
Another form of material failure,
the \emph{loss of controllability} at a limit point, is associated
with the vanishing of $\text{det}(\mathcal{M})$ when
the operator is constrained by a control program
\cite{Nova:1994a,Darve:2004d}.
This condition has its counterpart in the neutral equilibrium
associated with a matrix $[\Mbar]$ when the matrix
is non-R\textsubscript{0}
(Section~\ref{sec:NE}).
Note, however, that although the simple linear equation
$d\boldsymbol{\sigma}
=\mathcal{M}\cdot d\boldsymbol{\varepsilon}$
loses both existence and uniqueness when $\text{det}(\mathcal{M})=0$,
the two qualities are not coupled in the same way for an LCP,
thus requiring distinctions among the pathologies of
path-incongruence, neutral equilibrium, and Types~II and~III bifurcations.
Finally,
the \emph{loss of Hill material stability} (sustainability)
is associated with a
a non-positive value of the quadratic form
$d\boldsymbol{\varepsilon}\cdot\mathcal{M}\cdot d\boldsymbol{\varepsilon}$,
a criterion that has been proposed
at both the macro- and micro-scales
of granular materials and is associated with diffuse failure
\cite{Hill:1958a,Darve:1998a,Nicot:2007c}.
This condition is the continuum counterpart
of unsustainability in discrete systems,
discussed in Section~\ref{sec:instequil}.
%
\par
The three pathologies discussed in the previous
paragraph warrant the attention
given to them in the literature of continua, as they signal distress
and even collapse.
The other five pathologies, however, are candidly exposed when a
discrete granular system
is treated as an LCP.
All of these aberrant conditions are likely also available
in continua, particularly when more than one plastic
mechanism (the counterpart of $M^{\lambda}$) is present:
for example, in multi-mechanism plasticity, in boundary value
problems in which the plastic limit is reached at more than a single
region within a continuum,
or when multiple localization modes are available to a continuum.
The author suggests that these other pathologies will also be revealed
when analyses of continua are cast as linear complementarity
problems.
\subsection{Computation performance and DEM comparison}\label{sec:eff}
Section~\ref{section:disks49} presented the results of LCP simulations
of a small assembly of 49 disks.
The small number of particles permitted a full
enumerative analysis
of pathologies and a resolution of multiple solutions
during the assembly's loading.
The simulation was conducted with GNU Octave,
a high-level interpreted language, and the uncompiled
code required a run-time of 11~days, using an Intel i5-1135G7 processor.
Because of such long run-times,
full enumerative LCP (Problem~2)
computations are only viable for small assemblies
or for analyzing small sub-assemblies within a larger assembly.
\par
Forgoing a complete enumerative analysis,
the author also conducted the simulation as a sequence of
quadratic programs (Problem~3, QPs).
Although the simulation could compute stress and volume change,
this approach was unable to identify multiple solutions
(bifurcation),
to scrutinize stiffness pathologies, or to determine whether the
$[\Mbar]$ matrices were P-matrices, R-matrices, or degenerate.
Using strain increments of $5\times 10^{-6}$
(i.e., 9000 increments to reach strain
$-\varepsilon_{11}=0.045$),
the simulation required a more practicable 72~minutes,
and the average force-imbalance on a particle was about
$2\times 10^{-7}$ times the mean contact force.
\par
DEM simulations were also conducted with the same assembly,
using the author's optimized and compiled Fortran
OVAL/DEMPLA computer program.
Attaining a force-imbalance of $2\times 10^{-7}$
required much smaller strain increments of
$2\times 10^{-9}$ (22.5 million increments, see \cite{Suzuki:2014a}).
The run-time was 22~minutes.
Considering that optimized and compiled Fortran is expected
to run several times faster (perhaps by an order of magnitude)
than unoptimized and interpreted Octave,
the efficiency of the QP simulation is similar to that of the DEM
simulation.
Comparisons with DDA simulations were not conducted,
since previous comparisons of DDA and DEM show that DEM
has a substantial time advantage
\cite{Khan:2010a,Samiei:2013a}.
\par
The results in Fig.~\ref{fig:AssemblyResults} show that
the LCP and DEM methods give almost identical results
for pre-peak strains, but at strains greater than 1.8\%
the results diverge.
At the 1.8\% strain, the $[\Mbar]$ matrix begins to become non-P
and degenerate, meaning that path-incongruity becomes possible
(because of the non-P status) and types~II and~III bifurcations
become possible (because of degeneracy).
The DEM simulation gives a higher deviator stress $q$
and a greater dilation rate.
The LP solution is also seen to capture small perturbations
in stress and volume change; whereas, DEM tends to smooth over
these brief, erratic episodes.
\par
In regard to the choice of simulation methods, 
the LCP/QP method does require storing and inverting
large, $6N\times 6N$, matrices, which could make the method
untenable for large granular assemblies.
DEM simulations are typically conducted with much larger
strain increments, since the consequently
larger force imbalances are deemed
acceptable approximations of the quasi-static limit.
For these reasons,
the LCP/QP methods developed in the paper should primarily 
be considered
as an alternative to DEM simulations when the quasi-static condition
must be approximated with greater fidelity, or when one
must investigate bifurcations and other stiffness pathologies
within a large assembly (or masonry structure)
or within smaller sub-assemblies.
\appendix
\section*{\normalsize Appendices}
\section{\normalsize Enumerative algorithm for solving an LCP}\label{sec:enum}
\begin{figure}
  \centering
  \small
  \includegraphics[scale=1.00]{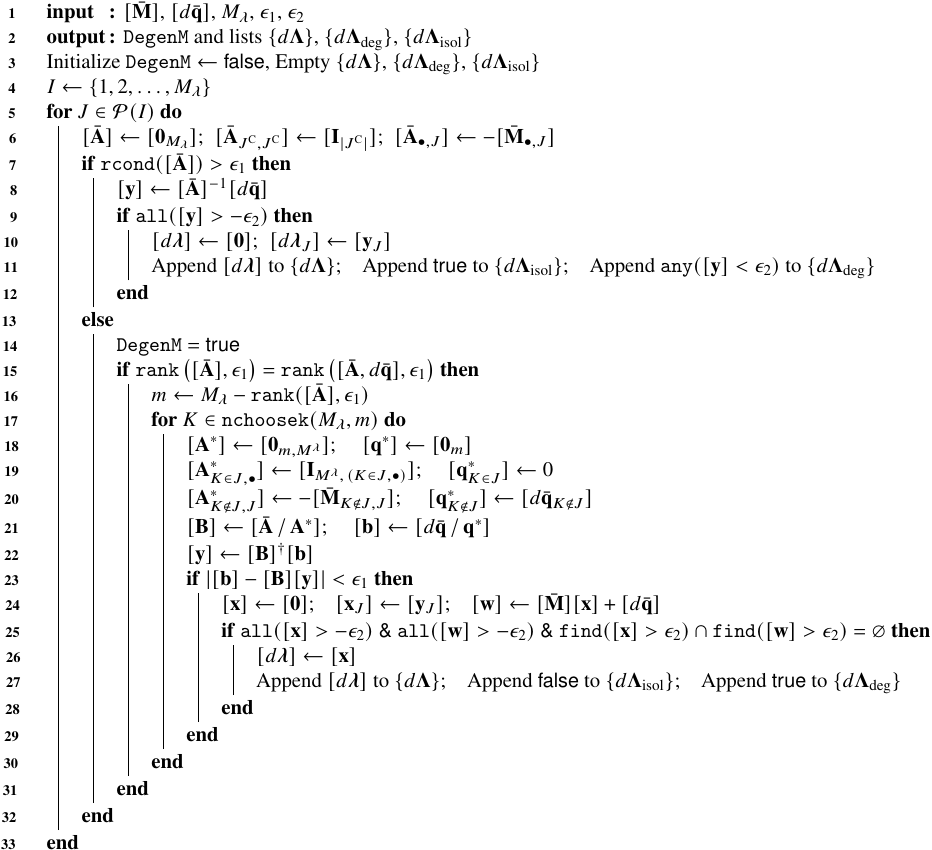}
  \caption{Enumerative algorithm for finding all solutions of
    \LCP\ and for determining whether the solutions are
    degenerate and/or isolated.
    Set $J$ is a subset of $I$;
    and set $J^{\text{C}}$ is the complementary
    subset, $J^{\text{C}}=I\setminus J$.
    A matrix $[\mathbf{0}_{K}]$ is a square matrix of zeros of scalar
    size $K$;
    and matrix $[\mathbf{I}_{K}]$ is the identity matrix of size $K$.
    Integers $|J|$ and $|J^{\text{C}}|$ are the cardinalities
    of sets $J$ and $J^{\text{C}}$.
    Functions similar to those of Matlab/Octave are used:
    $\mathtt{all}()$, $\mathtt{any}()$, $\mathtt{rcond}()$,
    $\mathtt{rank}()$, $\mathtt{nchoosek}()$, and $\mathtt{find}()$.
    \label{fig:solvealg}}
\end{figure}

The algorithm in
Fig.~\ref{fig:solvealg} finds all solutions of
an \LCP. 
The algorithm is based upon one described
by Murty \cite{Murty:1972a}\S 2.10 and
supplemented
with the findings of Jansen and Tijs \cite{Jansen:1987a}
to manage degenerate matrices $[\Mbar]$, which
give rise to non-isolated solutions.
Murty's approach is to consider each of the $2^{M_{\lambda}}$
combinations of indices $I=\{1,2,\ldots,M_{\lambda}\}$
(i.e., each subset $J\subseteq I$
of the power set $\mathcal{P}(I)$, line~5),
and to form its complementary matrix $[\Abar]$
(line~6 and Eq.~\ref{eq:comp}).
With each $[\Abar]$, one seeks the vertex in
$\dLambda$-space at the intersection of $M_{\lambda}$ planes:
the planes $([\Mbar][\dLambda]+[\dQbar])_{i}=0$, $i\in J$,
and the planes $[\dLambda]_{i}=0$, $i\in J^{C}$,
where $J^{C}$ is the complement of index set $J$.
If an $[\Abar]$ is non-singular (line~7),
a non-negative solution of the equation
$[\Abar][\mathbf{y}]=[\dQbar]$ yields a solution of the
\LCP\ (lines~8--11) that is isolated (line~12, rule~13) and
possibly degenerate (line~13).
\par
If $[\Abar]$ is singular (line~15), matrix $[\Mbar]$
is degenerate (line~16),
and solutions of $[\Abar][\mathbf{y}]=[\dQbar]$
can only exist when $[\dQbar]$ lies in the column space of
$[\Abar]$.
That is,
solutions can only exist when the rank of $[\Abar]$
and of the appended matrix $[\Abar, \dQbar]$ are equal
(line~17).
In this case, the dimension of the convex non-isolated set
of solutions (if it exists)
is given by rule~14 and Eq.~(\ref{eq:dimnoiso}).
The algorithm finds the rank deficit $m$ of matrix $[\Abar]$
(line~18),
and supplements the conditions of $[\Abar][\mathbf{y}]=[\dQbar]$
and $[\mathbf{y}]>0$
with $m$ additional conditions, as $m$ rows appended
to $[\Abar]$ and $[\dQbar]$.
The additional rows, $[\mathbf{A}^{\ast}]$ and $[\mathbf{q}^{\ast}]$,
represent additional planes (equalities)
$[\mathbf{A}^{\ast}][\mathbf{y}]=[\mathbf{q}^{\ast}]$,
taken from $m$ of the following:
the planes $([\Mbar][\dLambda]+[\dQbar])_{i}=0$, $i\in J^{C}$
(line~22),
and the planes $[\dLambda]_{i}=0$, $i\in J$
(line~21).
The algorithm queries all subsets of $I$ of cardinality $m$,
(line~19)
and appends the corresponding $m$ equations
$[\mathbf{A}^{\ast}][\mathbf{y}]=[\mathbf{q}^{\ast}]$
to $[\Mbar]$ and $[\dQbar]$, creating
the new equations $[\mathbf{B}][\mathbf{y}]=[\mathbf{b}]$
(line~23).
A solution is approximated by applying the Moore-Penrose
projection $[\mathbf{B}]^{\dagger}$
(line~24),
and if the result $[\mathbf{y}]$ satisfies all
$M^{\lambda}+m$ equations
(line~25),
complementarity is assessed
(lines~26--27).
If affirmative,
a solution $[\dLambda]$ has been found
(lines~28--31).
%
The full set of solutions for a singular $[\Abar]$
are the extreme-points of
(and form the hull of)
the convex set of non-isolated
solutions that is associated with subset $J$
(rule~15).
\par
The algorithm uses two tolerances:
a relative tolerance $\epsilon_{1}$ for testing
the reciprocal condition number ``$\mathtt{rcond}()$''
and rank (lines~7 and~17, since both tests involve
ratios of singular values),
and an absolute tolerance $\epsilon_{2}$,
based upon the machine
precision, for testing complementarity (lines~9 and~21).
\section{\normalsize Path-sensitivity algorithm}\label{sec:sensalg}
\begin{figure}
  \centering
  \small
  \includegraphics[scale=1.00]{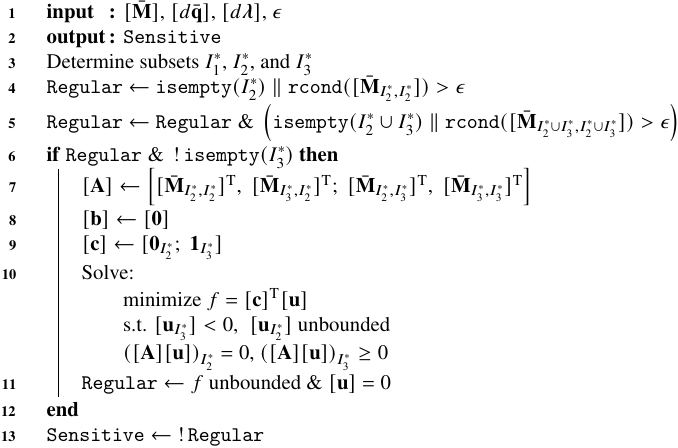}
  \caption{Algorithm for determining whether a solution
    $[\dLambda]$ of the \LCP\ is
    path-sensitive. \label{fig:sensalg}}
\end{figure}

Whether a solution $[\dLambda]$ is non-regular,
thus affirming path-sensitivity, depends
upon the partition of its components into the
subsets $I_{1}^{\ast}$, $I_{2}^{\ast}$, and $I_{3}^{\ast}$
of Eq.~(\ref{eq:sens1}) and upon test
of Eq.~(\ref{eq:sens2}).
This test involves three separate criteria,
engendered in the three parts of Eq.~(\ref{eq:sens2}\textsubscript{3}),
the last involving the solution of a system of linear equalities
and inequalities.
Fig.~\ref{fig:sensalg} presents an algorithm for determining
path-sensitivity.
The system of equalities and inequalities is solved as
a linear program, but only when the first two criteria
have shown $[\dLambda]$ to be regular (non-sensitive).
The algorithm applies a small tolerance $\epsilon$ to
the reciprocal condition number
$\mathtt{rcond}(\;)$ of a matrix.
\section{\normalsize Unsustainability in a complementarity setting,
         with algorithm}\label{sec:setting}
%
Departing from an assumption of quasi-static processes,
the thermodynamic approach of
Ba{\u{z}}ant \cite{Bazant:1989b,Bazant:1991a},
Petryk \cite{Petryk:1991a}, and 
Nicot et al. \cite{Nicot:2007c,Nicot:2012a,Nicot:2012c}
is adopted, in which unsustainability (instability of equilibrium)
is signaled by conditions that imply a second-order
gain of kinetic energy,
for a system
initially in equilibrium and
in the absence of further loading.
Alternatively, Nicot et al. \cite{Nicot:2007c}
show that under these conditions,
the system can encounter both the current and another
state without undergoing any change in the control
parameters, by following a dynamic process, thus
making the current state unsustainable.
\par
The first-order rate of change of kinetic energy at 
time $t$,
$\dot{E}(t)$, is the difference in the work rates $\dot{W}$
of the external forces $\mathbf{p}$
and internal forces $\boldsymbol{\mathfrak{f}}$,
\begin{align}\label{eq:Edot}
\dot{E}(t) = \dot{W}_{\text{ext}}(t) - \dot{W}_{\text{int}}(t)
  &=
  \left[\mathbf{p}(t)\right]^{\text{T}}\left[\dot{\mathbf{x}}\right]
  -
  \left[\boldsymbol{\mathfrak{f}}(t)\right]^{\text{T}} 
  \left[\dot{\mathbf{x}}^{\text{def}}\right] 
  =
  \left[\mathbf{p}(t)\right]^{\text{T}}\left[\dot{\mathbf{x}}\right]
  -
  \left[\boldsymbol{\mathfrak{f}}(t)\right]^{\text{T}}
  \left[\mathbf{B}\right]
  \left[\dot{\mathbf{x}}\right]
\end{align}
and this rate is zero when the external and internal forces
are in equilibrium.
The second-order rate is found by differentiating terms
in the final part this expression, such that
\begin{equation}\label{eq:ddotE1}
  \begin{aligned}
  \ddot{E}(t) =
    &\ddot{W}_{\text{ext}}(t) - \ddot{W}_{\text{int}}(t)
    \\
    =
    &- \left[\dot{\mathbf{x}}\right]^{\text{T}}
      \left(
          \left[\mathbf{H}^{\text{g-1}}\right]
        + \left[\mathbf{H}^{\text{g-2}}\right]
        + \left[\mathbf{H}^{\text{g-3}}\right]
      \right)
      \left[\dot{\mathbf{x}}\right]  
    +  \left[\dot{\mathbf{q}}\right]^{\text{T}}
       \left[\partial \mathbf{p}/\partial \mathbf{q}\right]^{\text{T}}
       \left[ \dot{\mathbf{x}}\right]
     - \left[\dot{\mathbf{x}}\right]^{\text{T}}
       \left[\mathbf{H}^{\text{g-4}}\right]
       \left[\dot{\mathbf{x}}\right]  \\
    &-
       \left(
         \frac{d}{d t}
         \left[
         \begin{array}{@{\extracolsep{\fill}}c@{\extracolsep{\fill}}}
           \mathfrak{d}\mathbf{f} \\[0.1ex]\hdashline[1pt/1pt]
           \mathfrak{d}\mathbf{m}
         \end{array}
        \right]^{\text{T}}
       \right)
       \left[\mathbf{B}\right]
       \left[\dot{\mathbf{x}}\right]
  \end{aligned}
\end{equation}
Details are given in \cite{Kuhn:2019b},
but briefly, the equation includes possible destabilizing
geometric effects that arise from the corotated
$\delta\hat{\mathbf{f}}$, $\delta\hat{\mathbf{m}}$,
$\delta\mathbf{r}$, and $d\boldsymbol{\theta}$
terms in Eqs.~(\ref{eq:db1})
and~(\ref{eq:delrpq})--(\ref{eq:Hg3});
the rates of the external forces $[\dot{\mathbf{p}}]$ are
assumed brought about
by the loading parameter rates $[\dot{\mathbf{q}}]$,
as in Eq.~(\ref{eq:dbG}); and
a geometric stiffness $[\mathbf{H}^{\text{g-4}}]$
is included for any position-dependent external forces,
as in Eq.~(\ref{eq:Hg4}).
The expression does not include terms with
accelerations $[\ddot{\mathbf{x}}]$ that arise by
differentiating Eq.~(\ref{eq:Edot}),
as the assumed equilibrium at time $t$ renders these terms zero.
\par
The final term in Eq.~(\ref{eq:ddotE1}) includes the rates
of contact forces, insofar as changes in force are due to
deformation and sliding at the contacts. 
It is with this term that Problem~2 departs from the
approach of Problem~1, since contact forces now depend upon
both the contact deformations $\Dxdefpq$
and upon any advancing active scalars $d\lambda^{pq}$,
as in the contact models of
Eqs.~(\ref{eq:dfForm2}) and~(\ref{eq:LdLambda}): 
\begin{equation}\label{eq:InstDf}
  \frac{d}{d t}
  \left[
    \begin{array}{@{\extracolsep{\fill}}c@{\extracolsep{\fill}}}
      \mathfrak{d}\mathbf{f} \\[0.1ex]\hdashline[1pt/1pt]
      \mathfrak{d}\mathbf{m}
    \end{array}
  \right]
  =
  \left[
    \frac{\partial\left[\mathfrak{d}\mathbf{f}\,/\,
                        \mathfrak{d}\mathbf{m}\right]}
       {\partial\,\mathbf{x}^{\text{def}}}
  \right]
  \left[\dot{\mathbf{x}}^{\text{def}}\right]
  +
  \left[
    \frac{\partial\left[\mathfrak{d}\mathbf{f}\,/\,
                        \mathfrak{d}\mathbf{m}\right]}
       {\partial\,\boldsymbol{\lambda}}
  \right]
  \left[\dot{\boldsymbol{\lambda}}\right]
  =
  \left[
    \begin{array}{@{\extracolsep{\fill}}c@{\extracolsep{\fill}}}
      \mathbf{F}^{\text{elastic}} \\\hdashline[1pt/1pt]
      \rule{0ex}{2.4ex}\mathbf{M}^{\text{elastic}}
    \end{array}
  \right]
  \left[ \mathbf{B} \right]
  \left[\dot{\mathbf{x}}\right]
  +
  \left[ \mathbf{L} \right]
  \left[\dot{\boldsymbol{\lambda}}\right]
\end{equation}
where individual ``$pq$'' contact quantities have been
gathered into matrices for the full assembly, as in
Eqs.~(\ref{eq:Felastdu}) and~(\ref{eq:gatherSl}).
Returning to Eq.~(\ref{eq:ddotE1}), and substituting
Eqs.~(\ref{eq:Himg}), (\ref{eq:Hxxel})--(\ref{eq:gatherSl}),
(\ref{eq:InstDf}),
and the relation $[\mathbf{A}]^{\text{T}}=[\mathbf{B}]$,
the second-order change in a system's kinetic energy is
\begin{equation}\label{eq:Eddot2}
  \begin{aligned}
    \ddot{E}(t)
    &=
    \ddot{W}_{\text{ext}}(t) - \ddot{W}_{\text{int}}(t)
    \\
    &=
    \left[\dot{\mathbf{q}}\right]^{\text{T}}
    \left[\partial \mathbf{p}/\partial \mathbf{q}\right]^{\text{T}}
    \left[\dot{\mathbf{x}}\right]
    -
    \left(
      \left[\dot{\mathbf{x}}\right]^{\text{T}}
      \left[\mathbf{H}_{\mathbf{xx}}\right]^{\text{T}}
      \left[\dot{\mathbf{x}}\right]
      +
      \left[\dot{\boldsymbol{\lambda}}\right]
      \left[\mathbf{H}_{\mathbf{x}\boldsymbol{\lambda}}\right]^{\text{T}}
      \left[\dot{\mathbf{x}}\right]
    \right)
  \end{aligned}
\end{equation}
 \par
A system is sustainable when the loading and
the system's stiffnesses,
$[\mathbf{H}_{\mathbf{xx}}]$ and
$[\mathbf{H}_{\mathbf{x}\boldsymbol{\lambda}}]$, produce
a negative rate $\ddot{E}$ for perturbations in all movement directions
$[\dot{\mathbf{x}}]$, thus restoring the system to equilibrium.
A system is in unstable equilibrium, when, in the
absence of further loading $[\dot{\mathbf{q}}]$, a negative
internal work $\ddot{W}_{\text{int}}$ applies in some direction(s)
$[\dot{\mathbf{x}}]$, consequently making worse any
perturbation in that direction(s).
\par
Although Eq.~(\ref{eq:Eddot2}) gives the fundamental character
of unsustainability,
a complete criterion must also respect four conditions:
the two limitations on
$[\dot{\boldsymbol{\lambda}}]$ that are given in the
second and fourth rows of Eq.~(\ref{eq:form2b}a),
the complementarity condition of Eq.~(\ref{eq:form2b}b),
and any external constraints on the system's movements.
The last limitation means that the direction of
$[\dot{\mathbf{x}}]$, in the absence of loading
$[\dot{\mathbf{q}}]$, is also restricted by
$[\mathbf{C}][\dot{\mathbf{x}}]=[\mathbf{0}]$.
A convenient means of imposing this constraint is
to replace $[\dot{\mathbf{x}}]$ with the product
$[\mathbf{P}_{\mathcal{L}}][\dot{\mathbf{z}}]$,
where projection matrix $[\mathbf{P}_{\mathcal{L}}]$
is defined in Eq.~(\ref{eq:PL}), and
$[\dot{\mathbf{z}}]$ is an arbitrary
vector.
With these provisions,
the criterion for unsustainability,
in the context of Problem~2,
is given in Eq.~(\ref{eq:instabProb2a}) in the body
of the paper.
\par
\begin{figure}
  \centering
  \small
  \includegraphics[scale=1.00]{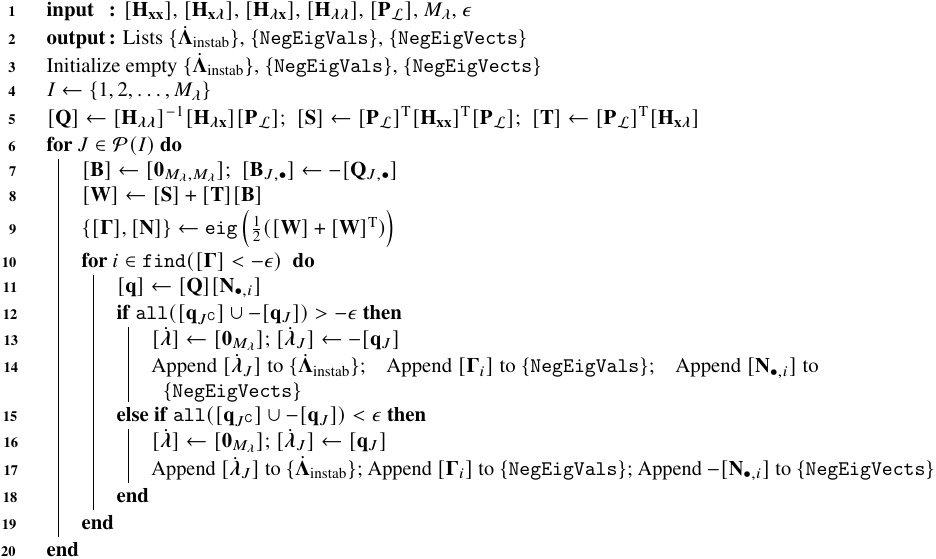}
  \caption{Enumerative algorithm for finding all unsustainable
    modes, based upon Eq.~(\ref{eq:Eddot2}).
    Matrix $[\mathbf{H}_{\boldsymbol{\lambda\lambda}}]$ is
    assumed diagonal and positive.
    Set $J$ is a subset of $I$;
    and set $J^{\text{C}}$ is the complementary
    subset, $J^{\text{C}}=I\setminus J$.
    Matrix $[\mathbf{0}_{K,K}]$ and vector $[\mathbf{0}_{K}]$
    are a square matrix or a vector of zeros of scalar
    size $K$.
    Function $\mathtt{eig}()$ is similar to that of Matlab/Octave
    and returns the eigenvalues $[\boldsymbol{\Gamma}]$
    and eigenvectors $[\mathbf{N}]$ of its argument.
    \label{fig:instabalg}}
\end{figure}

Figure~\ref{fig:instabalg} presents an enumerative algorithm for
identifying all unsustainable modes of a granular assembly.
The algorithm solves the linear complementarity problem
$\text{LCP}([\mathbf{H}_{\boldsymbol{\lambda\lambda}}]^{-1}
     [\mathbf{H}_{\boldsymbol{\lambda}\mathbf{z}}]
     [\mathbf{P}_{\mathcal{L}}]
     [\dot{\mathbf{z}}],\,
     \,\mathbf{I}\,)$
(which assumes that matrix $[\mathbf{H}_{\boldsymbol{\lambda\lambda}}]$
is diagonal and positive,
as discussed after Eq.~\ref{eq:instabProb2a})
while assuring that solutions $[\dot{\mathbf{z}}]$
and $[\dot{\boldsymbol{\lambda}}]$ produce
negative second-order work $\ddot{W}_{\text{int}}$.
The algorithm begins
by computing those matrix products in Eq.~(\ref{eq:instabProb2a})
that will be repeatedly used (line~5).
The algorithm then considers each $J$ combination of the $2^{M_{\lambda}}$
combinations of indices $I=\{1,2,\ldots,M_{\lambda}\}$,
as the power set $\mathcal{P}(I)$ in line~6.
The complementary matrix $[\Abar]$ that was used in the
method of Appendix~\ref{sec:enum} now has a simpler form:
a diagonal matrix of $-$1's and 1's,
located in the $J$ and $J^{\text{C}}$ positions,
respectively.
If we were only solving a standard LCP of form 
LCP($\mathbf{M},\mathbf{q}$),
Murty's approach, as in \ref{sec:enum},
would solve the complementary
equations $[\Abar][\mathbf{w}]=[\mathbf{q}]$ and test
whether $[\mathbf{w}]\ge 0$, and if true,
place these values into $[\boldsymbol{\lambda}]$
(see \cite{Murty:1972a}\S 2.10).
To evaluate unsustainability,
the algorithm uses an intermediate step
to seek possible
modes of negative-second order work.
To represent $[\dot{\boldsymbol{\lambda}}]$ in the second-order
$\ddot{W}_{\text{int}}$ expression of Eq.~(\ref{eq:instabProb2a}d),
one notes the assumption
that subset $[\dot{\boldsymbol{\lambda}}_{J}]\ge 0$;
the complementary subset
$[\dot{\boldsymbol{\lambda}}_{J^{\text{C}}}]= 0$;
and, because of complementarity,
$[\dot{\boldsymbol{\lambda}}_{J}]=-[\mathbf{Q}_{J,\bullet}]
[\dot{\mathbf{z}}]$.
Matrix $[\mathbf{B}]$ is created in line~7 to nullify
the $J^{\text{C}}$ rows of $[\mathbf{Q}]$ (line~7).
The product
$[\mathbf{T}][\mathbf{B}][\dot{\mathbf{z}}]$
is substituted for $[\mathbf{T}][\dot{\boldsymbol{\lambda}}]$
in Eq.~(\ref{eq:instabProb2a}d),
and the matrix $[\mathbf{W}]$ in line~8 is the basis of the quadratic
form $\ddot{W}_{\text{int}}$ (line~9).
Function $\mathtt{eig}(\,)$ solves the eigenvalues of the
symmetric part of $[\mathbf{W}]$, which are placed in vector
$[\boldsymbol{\Gamma}]$, and
the associated eigenvectors
are arranged as the columns of matrix $[\mathbf{N}]$
(line~9).
(Before beginning a lengthy eigenvalue
computation, a trial Cholesky factorization can
be used to determine whether the symmetric
part of $[\mathbf{W}]$ has any negative eigenvalues.
If not, the loop at line~10 can be skipped.)
When negative eigenvalues exist (line~10),
focus is placed upon their
corresponding eigenvectors,
which are used in computing vector $[\mathbf{q}]$,
so that the complementarity condition of
of Eq.~(\ref{eq:instabProb2a}c) can be tested (line~11).
Complementarity is satisfied when 
the conditions of either line~12 or line 17 are met
(note that either 
$[\dot{\mathbf{z}}]$ and $[\dot{\boldsymbol{\lambda}}]$
or $-[\dot{\mathbf{z}}]$ and $-[\dot{\boldsymbol{\lambda}}]$
yield the same $\ddot{W}_{\text{int}}$).
The values of $[\dot{\boldsymbol{\lambda}}]$ and of the
negative eigenvalues and their eigenvectors are appended
to lists for output (lines 14--16 and 19--21).
\\[2ex]
\small
\textbf{Data availability.}
The author's computer codes for the three examples 
are available on GitHub in
the repository
https://github.com/\-mrkuhn53/LCPgranular
\cite{Kuhn:2025b}.
The comparison with DEM
in Section~\ref{section:disks49}
was made with
the program OVAL/DEMPLA,
``Discrete Element Method for Propagation
and Liquefaction Analysis,''
available on GitHub in
the repository
https://\-github.com/mrkuhn53/dempla,
which includes
source code, documentation, and examples \cite{Kuhn:2021b}.
%
%
%
\small
%
%
\bibliographystyle{elsarticle-num}
%

\end{document}